\documentclass[trackchanges,twocolumn]{aastex7}
\usepackage{amsmath}

\newcommand{\comment}[1]{}
\usepackage{bm}
\newcommand{\Add}[1]{\textcolor{black}{#1}}
\newcommand{\Addp}[1]{\textbf{\textcolor{black}{#1}}}

\newcommand{\Erase}[1]{}

\newcommand{\M}{{\ $\mid$}}

\newcommand{\W}{{$\lambda$}}

\newcommand{\Hb}{H$\beta$}

\newcommand{\CIIIsemi}{C\,\textsc{iii}]}
\newcommand{\CIV}{C\,\textsc{iv}}

\newcommand{\OIII}{[O\,\textsc{iii}]}
\newcommand{\OIIIsemi}{O\,\textsc{iii}]}

\newcommand{\HeII}{He\,\textsc{ii}}

\newcommand{\HI}{H\,\textsc{i}}
\newcommand{\HII}{H\,\textsc{ii}}

\newcommand{\q}{$q_{\rm ion}$}
\newcommand{\MUV}{$M_\textrm{\scriptsize UV}$}

\begin{document}

\title{
Probing the Cosmic Reionization History with JWST:\\
Gunn-Peterson and Ly$\alpha$ Damping Wing Absorption at $4.5 < z < 13$
}

\correspondingauthor{Hiroya Umeda}

\author[0009-0008-0167-5129]{Hiroya Umeda}
\affiliation{Institute for Cosmic Ray Research,
The University of Tokyo,
5-1-5 Kashiwanoha, Kashiwa,
Chiba 277-8582, Japan}
\affiliation{Department of Physics, Graduate School of Science, The University of Tokyo, 7-3-1 Hongo, Bunkyo, Tokyo 113-0033, Japan}
\email[show]{ume@icrr.u-tokyo.ac.jp}

\author[0000-0002-1049-6658]{Masami Ouchi}
\affiliation{National Astronomical Observatory of Japan, 2-21-1 Osawa, Mitaka, Tokyo 181-8588, Japan}
\affiliation{Institute for Cosmic Ray Research,
The University of Tokyo,
5-1-5 Kashiwanoha, Kashiwa,
Chiba 277-8582, Japan}
\affiliation{Kavli Institute for the Physics and Mathematics of the Universe (WPI), 
University of Tokyo, Kashiwa, Chiba 277-8583, Japan}
\email{ouchims@icrr.u-tokyo.ac.jp}

\author[0009-0004-0381-7216]{Yuta Kageura}
\affiliation{Institute for Cosmic Ray Research,
The University of Tokyo,
5-1-5 Kashiwanoha, Kashiwa,
Chiba 277-8582, Japan}
\affiliation{Department of Physics, Graduate School of Science, The University of Tokyo, 7-3-1 Hongo, Bunkyo, Tokyo 113-0033, Japan}
\email{kageura@icrr.u-tokyo.ac.jp}

\author[0000-0002-6047-430X]{Yuichi Harikane}
\affiliation{Institute for Cosmic Ray Research, The University of Tokyo, 5-1-5 Kashiwanoha, Kashiwa, Chiba 277-8582, Japan}
\email{hari@icrr.u-tokyo.ac.jp}

\author[0009-0000-1999-5472]{Minami Nakane}
\affiliation{Institute for Cosmic Ray Research,
The University of Tokyo,
5-1-5 Kashiwanoha, Kashiwa,
Chiba 277-8582, Japan}
\affiliation{Department of Physics, Graduate School of Science, The University of Tokyo, 7-3-1 Hongo, Bunkyo, Tokyo 113-0033, Japan}
\email{nakanem@icrr.u-tokyo.ac.jp}

\author[0000-0002-8408-4816]{Tran Thi Thai}
\affiliation{National Astronomical Observatory of Japan, 2-21-1 Osawa, Mitaka, Tokyo 181-8588, Japan}
\email{thithai.tran@nao.ac.jp}

\author[0000-0003-2965-5070]{Kimihiko Nakajima}
\affiliation{Kanazawa University, Kakumamachi, Kanazawa, Ishikawa 920-1192, Japan}
\email{knakajima@staff.kanazawa-u.ac.jp}

\begin{abstract}
  We present a statistical analysis of Ly$\alpha$ absorption using 581 galaxies at $z=4.5$--13 observed with multiple JWST/NIRSpec spectroscopy programs, including JADES, UNCOVER, CEERS, and GO/DDT. We carefully construct composite spectra binned by redshift with homogeneous UV properties (UV magnitudes, UV slopes, and Ly$\alpha$ equivalent widths) and identify significant Ly$\alpha$ forest signals in galaxies at $z\sim5$--6, which diminish toward higher redshifts. We also find UV continuum breaks at rest-frame 1216\AA\ that soften beyond $z\gtrsim6$, confirming the effects of cosmic reionization through a self-consistent transition from Gunn-Peterson to Ly$\alpha$ damping wing absorption in galaxies. Fair comparisons of composite spectra with matched UV magnitudes and slopes across redshift reveal that UV-faint galaxies clearly show stronger Ly$\alpha$ absorption than UV-bright galaxies towards high redshift, providing insights into the topological evolution of reionization.
  We estimate Ly$\alpha$ transmission at the Gunn-Peterson trough and Ly$\alpha$ damping wing absorption by comparing the galaxy spectra to low-$z$ ($z\sim2$–5) galaxy templates that include galactic and circumgalactic absorption and Ly$\alpha$ emission. Using these measurements together with reionization simulations, we derive volume average neutral hydrogen fractions of $\langle x_{\rm HI} \rangle$ = ${0.00}^{+0.12}_{-0.00}$, ${0.25}^{+0.10}_{-0.20}$, ${0.65}^{+0.27}_{-0.35}$, ${1.00}^{+0.00}_{-0.20}$, and ${1.00}^{+0.00}_{-0.40}$ at $z\sim5$, 6, 7, 9, and 10, respectively. These values \Add{broadly} align with a reionization history characterized by a rapid transition around $z\sim7$--8, consistent with Ly$\alpha$ emitter observations. While the physical driver of this rapid reionization remains unclear, it may involve the emergence of hidden AGN populations and/or the onset of Lyman-continuum escape from galaxies.
\end{abstract}



\section{Introduction} \label{intro}
After the launch of the James Webb Space Telescope (JWST), multiple galaxy spectra are taken with high signal-to-noise rest-frame UV continuum detection. Such high-quality data have paved the way to unprecedented opportunities for understanding galaxies at the epoch of reionization (EoR), including the sciences with pure spectroscopic UV luminosity function \citep[e.g.,][]{Ha23}, UV slope measurements \citep[e.g.,][]{Saxena24,Dottorini24,Yanagisawa24}, etc. Not only the studies on high redshift galaxy itself but JWST has allowed us to use the EoR objects as background light sources to studying the nature of the intervening intergalactic medium (IGM) at the EoR. Before JWST, the only extremely bright sources such as quasar \citep[QSO; e.g., ][]{M11,B18,2018ApJ...864..142D,2019MNRAS.484.5094G,2020ApJ...896...23W} and gamma-ray burst \citep[GRB; e.g.,][]{2006PASJ...58..485T,2014PASJ...66...63T} could be used as the background light source to study the IGM and detect redward extended Ly$\alpha$ absorption called Ly$\alpha$ damping wing absorption \citep{ME98}. However, because the number densities of QSOs and GRBs are limited at the EoR, the role of abundant EoR galaxy as the background light sources is crucial to study the nature of IGM at the earliest stage of cosmic reionization. \par

After the first report on the damping wing feature detection in the galaxy spectra by \cite{CL23}, multiple reports followed Ly$\alpha$ damping wing absorption using galaxy UV-continuum \citep{Hsiao23,Williams23,Umeda24,2024A&A...689A.152D,2025A&A...693A..60H,Fu23,Asada24,2024arXiv241007377P,Witstok25,Mason25}. \cite{Umeda24} have gathered 27 spectroscopically confirmed galaxy spectra at $7<z<12$ and fitted the spectra with Ly$\alpha$ damping wing profile. From the damping wing fitting using analytical IGM absorption prescribed by \cite{ME98}, \cite{Umeda24} found an increasing IGM hydrogen neutral fraction $x_{\rm \HI}$ towards the higher redshifts, reaching $x_{\rm \HI}\sim1$ at $z\sim9$. Moreover, \cite{Mason25} developed a sophisticated methodology to measure neutral fraction via damping wing feature which incorporates the semi-numerical prescription to the sight-line IGM absorptions. Together with other methodologies using JWST galaxy spectra on understanding the nature of EoR \citep[e.g.,][]{2023ApJ...950...66K,2025arXiv250307074K,2024ApJ...976...93J,Meyer25,2025arXiv250219174R,Jones24,Nakane24,Napolitano24,Tang24,Kageura25}, Ly$\alpha$ damping wing absorption measurement using the UV-continuum of galaxies provide unique opportunity to look into the earlier stage of cosmic reionization in which visibility of Ly$\alpha$ emission is strongly suppressed. \par 

However, Ly$\alpha$ damping wing measurements do have challenges. As discussed in \cite{Heintz23} and \cite{2023arXiv230805800K}, the IGM absorption does degenerate with the damping wing absorption by the {\HI} gas in or around the galaxies. Moreover, \cite{Huberty25} present exercises on how the degenerate nature between the absorptions by IGM and local {\HI} could lead to systematics in the $x_{\rm \HI}$ measurements. While \cite{Mason25} have shown that the neutral gas column density of host absorption does not evolve significantly over the redshift from $z\sim3$, the amount of absorption by the local {\HI} gas could differ by individual galaxies. Thus, $x_{\rm \HI}$ measurements using individual galaxy spectra cannot avoid the systematic floor introduced by this degeneracy unless the absorption by local {\HI} is independently constrained. Moreover, \cite{2023arXiv230805800K} have shown that strong Ly$\alpha$ emission could dominate the spectral shapes around Ly$\alpha$ break and inferring UV continuum absorption challenging.\par 

To overcome the systematic floor introduced from the local \Add{{\HI}} gas absorption and Ly$\alpha$ emission in individual galaxy spectra, we use realistic galaxy template spectra constructed from $z\sim3$ galaxy sample to perform the spectral fitting. These realistic galaxy spectra templates already include typical host {\HI} gas absorption and Ly$\alpha$ representing observed galaxies at $z\sim3$. To compare the template spectra to the observed galaxy spectra at the EoR, we perform stacking spectra analysis using a high redshift galaxy sample constructed from JWST data. By stacking the galaxy spectra with similar spectral properties at different redshift bins, we aim to capture the redshift evolution of the Ly$\alpha$ damping wing feature encrypted to the galaxy population. By considering the stacking spectra instead of each spectrum, we mitigate the variance in the damping wing feature introduced by the local \Add{{\HI}} gas absorption for individual galaxies.\par

In this work, we first construct the stacking spectra from the public JWST NIRSpec PRISM \citep{2022A&A...661A..80J} spectra. By comparing stacking spectra for the different redshift bins, UV magnitude bins, and UV slope bins, we investigate the dependence of the Ly$\alpha$ damping wing features on the redshift and physical properties of galaxies. We also try to infer the $x_{\rm \HI}$ by fitting the stacking spectra with the model spectra constructed by applying semi-numerically predicted IGM absorption reflecting the IGM inhomogeneity to the template composite galaxy spectra based on the low redshift galax\Add{y} sample. At last, we discuss the implications on understand\Add{ing} the cosmic reionization history and the source of cosmic reionization based on our findings.\par

In this paper, we use the Planck cosmological parameter sets of the TT + TE + EE+ lowE + BAO + lensing result \citep{Planck}: $h=0.6766$, $\Omega_m=0.3103$, $\Omega_\Lambda=0.6897$, $\Omega_b h^2=0.02234$, and $Y_p=0.248$. All magnitudes are in the AB system \citep{1983ApJ...266..713O}.\par
\section{Data and Sample} \label{data}
\subsection{ERS, GTO, GO, and DDT NIRSpec Observations}
To conduct the UV continuum study, we gather NIRSpec PRISM ($R\sim100$) spectra from the publicly available datasets. The datasets used in this study have been obtained in the Cosmic Evolution Early Release Science (CEERS) observations \citep[ERS-1345; PI: S. Finkelstein;][]{Fink23}, the General Observer (GO) observations targeting a $z\sim11$ galaxy candidate \citep[GO-1433; PI: D. Coe;][]{H23}, UNCOVER \cite[GO-2561; PIs: I. Labbe \& R. Benzason;][]{Bezanson24}, and the $z>9$ bright galaxies \citep[GO-3073; PI: M. Castellano][]{Castellano24,Napolitano25}, the DDT observations targeting $z\sim$12-16 galaxy candidates \citep[DD-2750; PI: P. Arrabal Haro;][]{AH23b}, and the Guaranteed Time Observer (GTO) observations of JADES \citep[GTOs-1180, 1181, 1210, and 1286; PIs: D. Eisenstein \& N. Luetzgendorf;][]{Bun23,DEugenio25}, respectively. \Add{For the data from CEERS, GO-1433, GO-3073, and DD-2750 programs, we use the spectra reduced by \cite{Nk23}, \cite{Ha23}, and \cite{Nakane25}.} For JADES data, we use reduced NIRSpec data from the official JADES DR1 \& DR3 \citep[][]{Bun23,DEugenio25}. \footnote{\url{https://archive.stsci.edu/hlsp/jades\#section-268de08a-1ff5-430e-adfe-846e6b933f3b}} For UNCOVER data, we use reduced NIRSpec data from the official UNCOVER DR4 \citep{Price24,Furtak23}. \footnote{\url{https://jwst-uncover.github.io/\#releases}} For each objects, we adopt the magnification factor from the literature we refer to the reduced data. 

\begin{figure}[htbp]
\centering
\includegraphics[width=\linewidth]{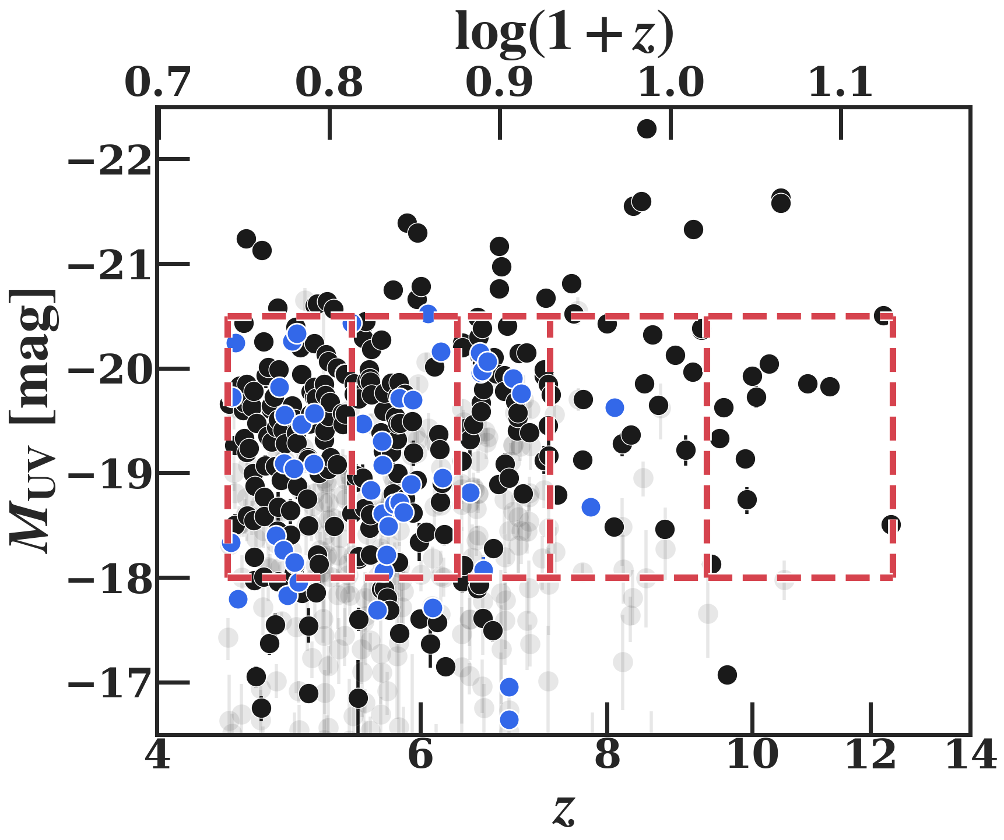}
\caption{The distribution of UV magnitude by redshift. The UV magnitude is inferred from the spectral fitting. The black (blue) circles represent the measurements for the obtained galaxy spectra with $S/N>5$ at the rest-frame 1450 {\AA} as well as the Ly$\alpha$ detection without (with) $EW_{0}>25$ {\AA}. The faint circles represent the measurements for the galaxy spectra with $S/N<5$ at the rest-frame 1450 {\AA}. Rectangular region surrounded by red dashed lines represents the parameter space used to select the ``fiducial'' subsamples.}
\label{muv_z} 
\end{figure}

\begin{figure}[htbp]
\centering
\includegraphics[width=\linewidth]{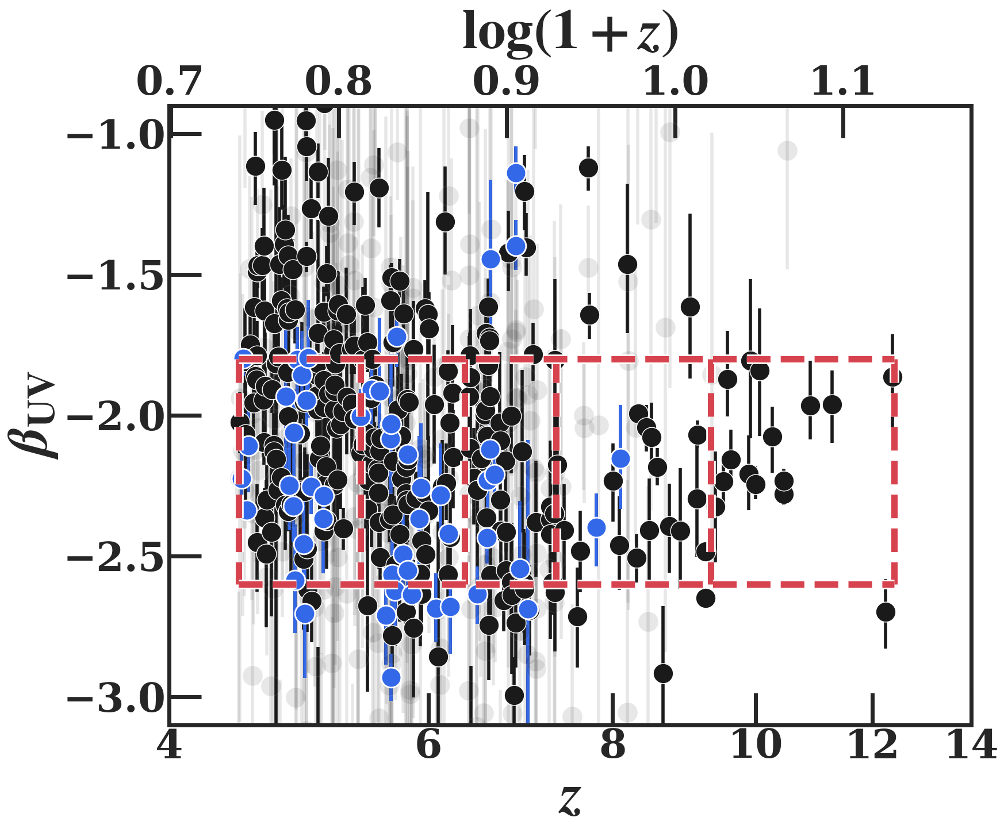}
\caption{The distribution of UV slope by redshift. The UV slope is inferred from the spectral fitting. The symbols are the same as Figure \ref{muv_z}.}
\label{beta_z} 
\end{figure}

\subsection{Sample Selection}
\subsubsection{Redshift Confirmation}
We need the precise determination of the systemic redshift from non-Ly$\alpha$ emission lines to measure the Ly$\alpha$ damping wing absorption feature. Redshift determined from the emission line is especially important as the systemic redshift determined only from the Ly$\alpha$ break systematically deviates from the systemic redshift determined from the Ly$\alpha$ \cite[e.g.,][]{Fu23,Asada24}. For this reason, we construct the galaxy sample with redshifts determined from multiple emission lines or a single emission line with a Ly$\alpha$ break feature. For JADES data, we only use the galaxy spectra with the redshift flags of A and B, corresponding to the redshift determination from multiple emission lines in grating and prism spectra, respectively. Similarly, we use the galaxy spectra with the redshift flag above the rank 2 for UNCOVER galaxy spectra so that the redshift is confirmed from multiple emission lines or the combination of Ly$\alpha$ break and an emission line. For the GO-3073 galaxy sample, we use the redshift determined from non-Ly$\alpha$ emission lines reported by \cite{Castellano24} and \cite{Napolitano25}. For CEERS, GLASS, DDT, and GO-1433 samples, we use the galaxy sample with redshift determined from non-Ly$\alpha$ emission line in \cite{Nk23} and \cite{Ha23}. \Add{While NIRSpec/PRISM do not cover strong emission lines such as {\OIII} and {\Hb} at $z>10$, we confirm that all of the $z>10$ galaxy selected into our sample have both Lyman break feature and emission line detection at the consistent wavelength positions.}\par

We further narrowed down the galaxy sample to ensure that all the spectra covers the Lyman break as well as the redward UV continuum. For this selection, we set the redshift boundary at $z=4.5$ so that the Ly$\alpha$ break can be covered in the PRISM spectra. Also, we apply additional cut based on the data quality flag in JADES and UNCOVER official catalogs. We also conduct a visual inspection to omit the galaxy spectra with significant unphysical spectral feature (e.g., negative continuum signals throughout the UV continuum).\par

\subsubsection{Final Galaxy Sample}
After these selections, we obtain the final galaxy sample with 581 galaxy spectra in the redshift range of $4.5<z<12.7$. After the selection, we obtain 353, 94, 124, 3, 1, and 6 spectra from JADES, UNCOVER, CEERS, DDT-2750, GO-1433, and GO-3073, respectively. Because we include dataset from various surveys including the one in lensed fields (e.g., UNCOVER), our sample covers wide range of the redshift \Add{and} UV magnitude. In the next section, we investigate the galaxy properties in our sample, including Ly$\alpha$ equivalent width, UV magnitude, and UV slope.
\section{Sample Properties} \label{sample}
\subsection{UV magnitude and slope} \label{mag_beta}
We first investigate the UV magnitude and slope of the galaxies in our sample. The UV magnitude and slope are crucial to understanding the intrinsic galaxy properties, such as the star formation rate, dust extinction, and the age of the stellar population. We measure the UV magnitude and slope by fitting the UV continuum of the galaxy spectra. We fit the UV continuum of the galaxy spectra by the a simple power-law function as follows:
\begin{equation}
  F_\lambda(\lambda_{\rm rest}) = F_{1450}\left(\frac{\lambda_{\rm rest}}{1450 {\rm \AA}}\right)^{\beta_{\rm UV}}
  \label{pl}
\end{equation}
Here, $F_{1450}$ and $\beta_{\rm UV}$ are the normalization factor and UV slope, respectively. We correct the magnification factors and use $F_{1450}$ to infer the UV magnitude ($M_{\rm UV}$) of objects. We consider the same fitting window as the one adopted by \cite{Saxena24}. Our \Add{and \cite{Saxena24}'s} fitting windows \Add{for UV slope measurements are modified from} the \Add{widely used} Calzetti's window \citep{Calzetti94}. We modify the mask window to consider the instrumental broadening effect on PRISM spectra. Because of low-resolution spectra, we mask the wavelength range of 1440-1590 {\AA}, 1620-1680 {\AA}
and 1860-1980 {\AA}, to avoid contributions from strong emission lines such as {\CIV\W\W}1548,1551, {\HeII\W}1640/{\OIIIsemi\W\W}1661,1666, and {\CIIIsemi\W\W}1907,1909, respectively. We fit in the wavelength range of 1350 to 2700 {\AA} to avoid the impact from strong Ly$\alpha$ damping wing absorption \citep[e.g.,][]{Heintz23}. We conduct the fitting by assuming flat prior for both of the variables. We set a loose upper/lower bound for the $\beta_{\rm UV}$ at 1 and -5, respectively. We perform Markov Chain Monte Carlo (MCMC) sampling to determine the posterior distribution for the free parameters. We use the affine invariant MCMC ensemble sampler {\texttt emcee} \citep{2013PASP..125..306F} to run the MCMC sampling throughout the analyses.  We consider the median of the posterior distribution as the best-fit value. We give the uncertainty of the best-fit value by 16/84-th percentile. In Figure \ref{muv_z} and \ref{beta_z}, we show UV magnitude and UV slope distribution per redshift, respectively. As shown in the Figure \ref{muv_z}, our galaxy sample occupies the wide range of UV magnitude from $M_{\rm UV}\sim$-23 to -16 throughout redshifts. For the UV slope, we see wider range of UV slope at the lower redshifts. The UV slope of our galaxy sample gradually converge towards bluer slope values. However, from around $z\sim9$, the UV slope measurements show a tentative trend toward the redder values, converging around \Add{$\beta_{\rm UV}\simeq-2$}. Similar trend have been quantitatively discussed by \cite{Saxena24}. While the measurements of $\beta_{\rm UV}$ are not the main topic of this work, we discuss the interpretation of UV slope evolution in terms of cosmic reionization in Section \ref{disc}. \par 

\subsection{Ly$\alpha$ emission}
\begin{figure}[htbp]
\centering
\includegraphics[width=\linewidth]{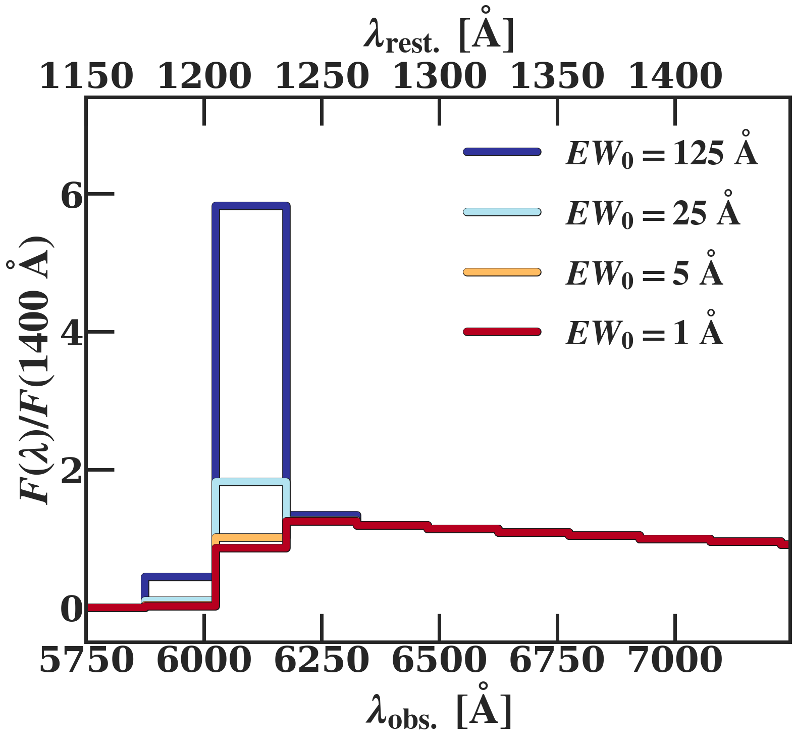}
\caption{The schematic representation of Ly$\alpha$ emission in NIRSpec/PRISM spectra at $z=6$. 
UV continuum assume the power-law with $\beta_{\rm UV}=-2.2$ with sharp drop at the rest-frame
1216 {\AA}. We assume $\delta$-function Ly$\alpha$ emission with $EW_{\rm Ly\alpha}=1$, 5, 25, and 125 {\AA}, all shown in different colors.
All the spectra are convolved to the NIRSpec/PRISM line spread function.}
\label{ew_prism} 
\end{figure}

Because strong Ly$\alpha$ emission could dominate the spectral shape around Ly$\alpha$, the inclusion of strong Ly$\alpha$ emitters in stacking sample could distort the damping wing absorption in the UV continuum \citep[e.g.,][]{Chen24}. We present example in Figure \ref{ew_prism}. In Figure \ref{ew_prism}, we show galaxy spectrum mocked by power-law with slope of -2.2 and step-function absorption at the rest-frame Ly$\alpha$. We add delta-function like Ly$\alpha$ with different rest-frame equivalent width ($EW_{\rm Ly\alpha}$) values. The spectrum is shifted to $z=6$ and convolved to the instrumental broadening according to the NIRSpec/PRISM line spread function. Throughout the study, we adopt the line spread function from the official JWST documentation \footnote{\url{https://jwst-docs.stsci.edu/jwst-near-infrared-spectrograph/nirspec-instrumentation/nirspec-dispersers-and-filters}}. We show that while weak Ly$\alpha$ emission ($EW_{\rm Ly\alpha}<5$ {\AA}), the strong Ly$\alpha$ emitters ($EW_{\rm Ly\alpha}>$25 {\AA}) do dominate the spectral feature due to the instrumental broadening. For this reason, we omit the strong Ly$\alpha$ emitters from our sample. \Add{While we do consider this selection effect via forward modeling in later analysis, we note that Ly$\alpha$ typically traces UV faint galaxies and incorporation of fainter galaxies in the future analysis would provide more information on how the Ly$\alpha$ transmission around galaxies depends on the galaxy properties and its surrounding environment (see Section \ref{sec:top}). }\par 

To omit strong Ly$\alpha$ emitters from our sample, we first measure the Ly$\alpha$ equivalent width ($EW_{\rm Ly\alpha}$) for each galaxy. For the object overlapping with the sample from \cite{Kageura25}, we adopt the measured values from \cite{Kageura25}. For the object without $EW_{\rm Ly\alpha}$ values reported in \cite{Kageura25}, we similarly measure $EW_{\rm Ly\alpha}$  as \cite{Kageura25}. \cite{Kageura25} fit individual galaxy spectra assuming the combination of gaussian profile emission line as well as simple power-law UV spectrum \Add{in the high resolution spectrum and apply instrumental broadening afterwards \citep[e.g.,][]{Chen24,Tang24,Jones24}}. We follow their procedure and fit the Ly$\alpha$ emission line by varying the amplitude, central wavelength, and width of the gaussian profile for Ly$\alpha$ emission line as well as the normalization factor and the UV slope index for the UV continuum. We determine the posterior distribution for free parameters by assuming uniform prior for the free parameter as well as the gaussian shaped likelihood as follows:
\begin{align}
  \label{Lyaline}
  F_\lambda({\lambda_{\rm obs}}) &= a\left(\frac{\lambda_{\rm obs}}{\lambda_\alpha(1+z)}\right)^{\beta} \\
  & + \frac{A}{\sqrt{2\pi}\sigma}\exp\left(-\frac{{(\lambda_{\rm obs}-(\lambda_\alpha+\Delta\lambda)(1+z))}^2}{2\sigma^2}\right).\nonumber
\end{align}
Here, $\lambda_{\alpha}$ and $\lambda_{\rm obs}$ correspond to the rest-frame Ly$\alpha$ and observed wavelengths, respectively, and $a$, $\beta$, $A$, and $\sigma$ corresponds to the free parameters. The first term of Equation \ref{Lyaline} corresponds to the flat UV continuum and the second term corresponds to the gaussian shape Ly$\alpha$ line.
For the simplicity and consistency between \cite{Kageura25}, we assume flat continuum and step-function like IGM absorption switching at the Ly$\alpha$ wavelength. \Add{While this model is oversimplified and does not consider the mixing effect from damped Ly$\alpha$ absorbers \citep[e.g.,][]{Chen24,Witstok25}, this model should still be able to detect Ly$\alpha$ line with significant contribution to the spectral shape. In terms of omitting strong Ly$\alpha$ emitters from our sample, our simplified strategy still serves the purpose.} We calculate the posterior probability distribution for EW by propagating the probability distribution for Ly$\alpha$ emission strength and the UV continuum flux. We consider the mode of the posterior distribution for Ly$\alpha$ EW as the best-fit value. We give the EW uncertainty by the 68 percentile highest posterior density interval (HPDI). If the 99 percentile (i.e., counterpart for $3\sigma$ in case of gaussian) HPDI for Ly$\alpha$ EW is consistent with 0 {\AA}, we consider Ly$\alpha$ emission to be not detected. From the following analysis, we define strong Ly$\alpha$ emitters as the one with Ly$\alpha$ EW above 25 {\AA}. In Figures \ref{muv_z} and \ref{beta_z}, we show the galaxy with Ly$\alpha$ EW above 25 {\AA} in blue color.\par

\section{Stacking Spectra} \label{stack}
\subsection{Stacking Methodology}
 We first describe the stacking methodology. We adopt the mean spectra of each galaxy subsample as the stack spectra. To obtain mean spectra, we first shift all galaxy spectra into the common rest-frame wavelength grid in a manner to conserve the original fluxes. Before we \Add{take the mean of} spectra \Add{in the sample}, we also need to correct for the wavelength dependent instrumental broadening. To do this, we calculate the line spread function for each galaxy in the rest-frame wavelength grid. Next, we calculate the maximum line spread function value at each wavelength grid among our galaxy sample and adopt the maxima \Add{at each wavelength grid point} as the ``common'' line spread function. We convolve all the galaxy spectra to match the spectral broadening by the ``common'' line spread function. After correcting for the wavelength dependent spectral broadening, we normalize all the galaxy spectra by the magnification factor and the UV magnitude determined in Section \ref{sample}. After the normalization, we finally calculate the mean spectra. We conduct 1000 steps of bootstrap resampling to estimate the average mean spectra and the covariance matrices for the mean spectra. To consider the uncertainty from the flux measurements, we also fluctuate the flux at each wavelength pixel with Gaussian noise according to the original error spectrum. We also apply the sigma clipping algorithm to omit outlying signals from the galaxy sample to avoid distortion of the mean stack spectra by unphysical features. In every step of the bootstrap, we omit the data points that deviate 2 standard deviation away from the sample mean. We recalculate the sample mean after omitting. We adopt conservative value of 2$\sigma$ because many galaxy spectra in our sample are detected at the not high signal-to-noise values (i.e., $S/N\sim5$).\par

 \subsection{Constructing Galaxy Subsample}
 \subsubsection{Fiducial Subsamples}
 We first divide the galaxy sample into five subsamples by binning according to the following redshift range: $4.5<z<5.5$, $5.5<z<6.5$, $6.5<z<7.5$, $7.5<z<9.5$, and $z>9.5$. Because we focus on these subsamples in the following analysis, we call them ``fiducial'' subsamples and give them ID as Fid1 to Fid5 from low to high redshift bins. For the sake of quality, we do not use the spectra with signal-to-noise (S/N) below 5 at the rest-frame wavelength $1300-1350$ {\AA} for the subsample. To construct the galaxy stack spectrum from the homogeneous galaxy sample, we also apply the following sample selection based on the UV magnitude cut of $-20.5<M_{\rm UV}<-18$ mag and UV slope cut of $-2.6<\beta_{\rm UV}<-1.8$. We also omit the strong Ly$\alpha$ emitters (i.e., $EW_{Ly\alpha}>25~{\rm \AA}$) determined in Section \ref{sample} from our subsamples to capture UV continuum shape curved by the damping wing absorption. After these sample selections, our \Add{five} galaxy subsamples contain 68, 47, 29, 14, and 12 galaxies, respectively. We define representative redshift values $\langle z \rangle$ of each redshift bin by the average redshift value among the galaxies in each subsample. The representative redshift for the ``fiducial'' subsamples are $\langle z \rangle=5.0$, 5.8, 7.0, 8.6, and 10.4. We summarize the properties of our subsample in Table \ref{table:stack}. \par
 \subsubsection{Subsamples Around the End of Cosmic Reionization}
 According to quasar's Ly$\alpha$ opacity measurements, the Gunn-Peterson trough \citep{GP65} signals starts to disappear around $z<6$ which signals the end of cosmic reionization \citep[e.g.,][]{Bosman22}. While the Ly$\alpha$ forest measurements require very bright sources such as quasars, we may detect the signal with multiple galaxy spectra via boosting S/N by stacking spectra. To independently confirm the timing of the disappearance of Gunn-Peterson trough signals, we create another galaxy subsamples focusing on redshift corresponding the end of cosmic reionization. We divide the galaxy sample in the redshift bin of $\Delta z=0.5$ in the redshift range of $4.5<z<6.0$. We adopt the same selections regarding the UV magnitude, slopes, and Ly$\alpha$ equivalent width as the ``Fiducial'' subsamples. We call these four subsamples as the ``End of EoR'' (EEoR) subsamples. The EEoR subsamples comprises of 22, 30, 32, and 10 objects with the average redshift of $\langle z \rangle=$4.8, 5.2, 5.7, and 6.3, respectively. We also summarize the properties of EEoR subsamples in Table \ref{table:stack}.
 \subsubsection{Subsamples by UV Magnitudes}
 To see the connection between the galaxy properties and the UV turnover features, we also create stack spectra with different UV properties. For UV properties, we construct the stack spectra by dividing the galaxy sample into UV magnitude bins. We divide the galaxy sample into ``bright'' and ``faint'' subsample. For ``bright'' and ``faint'' subsample, we apply UV magnitude selection of $-22.0<M_{\rm UV}<-20.0$ mag and $-19.0<M_{\rm UV}<-17.0$ mag, respectively. We then further divide ``bright'' and ``faint'' subsample by the redshift range of $4.5<z<6.0$ and $z>7.5$. Hereafter, we call the subsample of ``bright'' and ``faint'' for the lower (higher) redshift bin as BriL (BriH) and FaiL (FaiH), respectively. Beside the redshift and UV magnitude selections, we apply same sample selection using criteria adopted in for Fid1-5 subsamples.
 \subsubsection{Subsamples by UV Slopes}
 We also perform similar analysis with different sample selections based on UV slopes. We divide the galaxy sample into ``blue'' and ``red'' subsample. For the ``blue'' (``red'') subsample, we apply UV slope selection of $-3.0<\beta_{\rm UV}<-2.4$ ($-2.0<\beta_{\rm UV}<-1.4$). We then further divide ``blue'' and ``red'' subsample by the redshift range of $4.5<z<6.0$ and $z>7.5$. Hereafter, we call the subsample of ``blue'' and ``red'' for the lower (higher) redshift bin as BluL (BluH) and RedL (RedH), respectively. Beside the redshift and UV slope selections, we apply same sample selection using criteria adopted in for Fid1-5 subsamples.
 \begin{deluxetable}{ccccccc}
  \tablecolumns{7}
  \tabletypesize{\scriptsize}
  \tablecaption{Characteristics of Subsamples
  \label{table:stack}}
  \tablehead{
  \colhead{ID} & 
  \colhead{$N$} &
  \colhead{$\langle z \rangle$} & 
  \colhead{$z_{\rm min}$} &  
  \colhead{$z_{\rm max}$} & 
  \colhead{$\langle M_{\rm UV} \rangle$~[mag]} & 
  \colhead{$\langle \beta_{\rm UV} \rangle$} \\
  (1) & 
  (2) & 
  (3) &
  (4) & 
  (5) & 
  (6) & 
  (7)
  }
  \startdata 
  \hline
  \multicolumn{7}{c}{\bf Fiducial Subsamples} \\ \hline
  \multicolumn{7}{c}{$4.5<z<13.0$} \\
  \multicolumn{7}{c}{$-20.5<M_{\rm UV}<-18.0$} \\
  \multicolumn{7}{c}{$-2.6<\beta_{\rm UV}<-1.8$}\\
  \hline
  Fid1 & 52 & 5.0 & 4.512 & 5.446 & -19.2 & -2.1\\ 
  Fid2 & 42 & 5.8 & 5.503 & 6.372 & -19.3 & -2.2\\ 
  Fid3 & 23 & 7.0 & 6.548 & 7.483 & -19.6 & -2.2\\ 
  Fid4 & 11 & 8.6 & 7.509 & 9.433 & -19.6 & -2.4\\ 
  Fid5 & 11 & 10.4 & 9.570 & 12.47 & -19.4 & -2.0\\
  \hline
  \multicolumn{7}{c}{\bf End of EoR Subsamples} \\ \hline
  \multicolumn{7}{c}{$4.5<z<6.0$} \\
  \multicolumn{7}{c}{$-20.5<M_{\rm UV}<-18.0$} \\
  \multicolumn{7}{c}{$-2.6<\beta_{\rm UV}<-1.8$}\\
  \hline
  EEoR1 & 22 & 4.8 & 4.512 & 4.999 & -19.2 & -2.1\\ 
  EEoR2 & 30 & 5.2 & 5.013 & 5.446 & -19.3 & -2.1\\ 
  EEoR3 & 32 & 5.7 & 5.503 & 5.944 & -19.3 & -2.2\\ 
  EEoR4 & 10 & 6.3 & 6.058 & 6.372 & -19.1 & -2.2\\ 
  \hline
  \multicolumn{7}{c}{\bf Bright/Faint Subsamples} \\ \hline
  \multicolumn{7}{c}{$4.5<z<6.0$ or $7.5<z<13$} \\
  \multicolumn{7}{c}{$-22<M_{\rm UV}<-20$ or $-19<M_{\rm UV}<-17$} \\
  \multicolumn{7}{c}{$-2.6<\beta_{\rm UV}<-1.8$}\\
  \hline
  FaiL & 36 & 5.3 & 4.554 & 5.937 & -18.4 & -2.2 \\
  FaiH & 7 & 9.5 & 7.589 & 12.47 & -18.3 & -2.2 \\ 
  BriL & 14 & 5.2 & 4.637 & 5.877 & -20.5 & -2.0 \\ 
  BriH & 11 & 9.2 & 7.781 & 10.61 & -20.9 & -2.2 \\ \hline
  \multicolumn{7}{c}{\bf Blue/Red Subsamples} \\ \hline
  \multicolumn{7}{c}{$4.5<z<6.0$ or $7.5<z<13$} \\
  \multicolumn{7}{c}{$-20.5<M_{\rm UV}<-18.0$} \\
  \multicolumn{7}{c}{$-2.0<\beta_{\rm UV}<-1.4$ or $-3.0<\beta_{\rm UV}<-2.4$}\\
  \hline
  RedL & 63 & 5.2 & 4.549 & 5.988 & -19.5 & -1.8 \\ 
  RedH & 9 & 10.05 & 7.892 & 12.47 & -19.4 & -1.8 \\ 
  BluL & 10 & 5.4 & 4.643 & 5.989 & -19.1 & -2.5 \\ 
  BluH & 8 & 8.7 & 7.589 & 9.433 & -19.6 & -2.5 \\
  \hline
  \enddata
  \tablecomments{
  (1): ID for the galaxy subsample for stack spectra. 
  (2): Number of galaxies subsample. 
  (3): Average redshift of galaxies included in the composite spectra. 
  (4): Minimum redshift of galaxies in the subsample. 
  (5): Maximum redshift of galaxies in the subsample. 
  (6): Average rest-frame UV magnitude of the subsample.
  (7): Average UV slope of the subsample.
  }
  \end{deluxetable}

 \subsection{Stacking Spectra Results}
 \subsubsection{Spectral Evolution at the EoR}
 In Figure \ref{stack}, we show our stack galaxy spectra in different color. We can see the spectral flattening at the rest-frame 1216 {\AA} towards the high redshift while the redder UV continuum shape stays similarly throughout the redshift. The softening break feature as seen in Figure \ref{stack} as well as seen in \cite{Umeda24} suggest increasing Ly$\alpha$ damping wing absorptions due to increasing $x_{\rm \HI}$ with redshifts. We do see that the $\langle z \rangle$=8.6 and 10.4 subsample show significant UV turnover which can be attributed to the increase in IGM absorption towards the higher redshifts.\par
 \begin{figure*}[htbp]
 \centering
 \includegraphics[width=\linewidth]{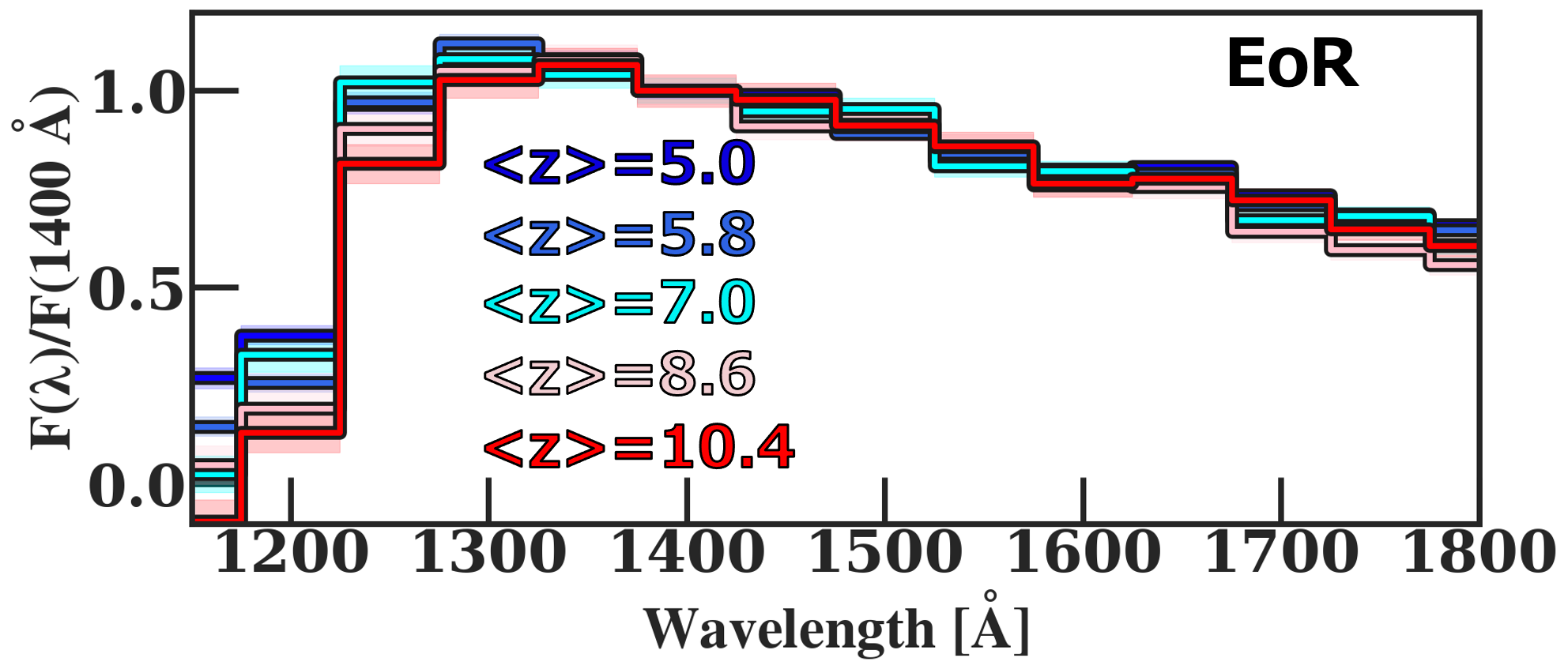}
 \caption{The stacked spectra of the fiducial galaxy sample from redshift 4.5 to 13. The stacked spectra are shown in the rest-frame wavelength. The redshift of the galaxy sample is shown in the upper right corner of each panel. The stack spectra shown here are constructed from the Fid1-5, with UV magnitude around $M_{\rm UV}\simeq-19.5$ and UV slope around $\beta_{\rm UV}\simeq-2.2$. The dark blue, navy blue, pale blue, pink, and red colored solid lines represent the stacked spectra at $\langle z \rangle$=5.0, 5.8, 7.0, 8.6, and 10.4. The corresponding 1$\sigma$ errors are shown in the shaded regions. All the stacked spectra are normalized at 1400 {\AA}. }
\label{stack} 
\end{figure*}
 \subsubsection{Spectral Evolution Around the End of Cosmic Reionization}
\begin{figure*}[htbp]
\centering
\includegraphics[width=\linewidth]{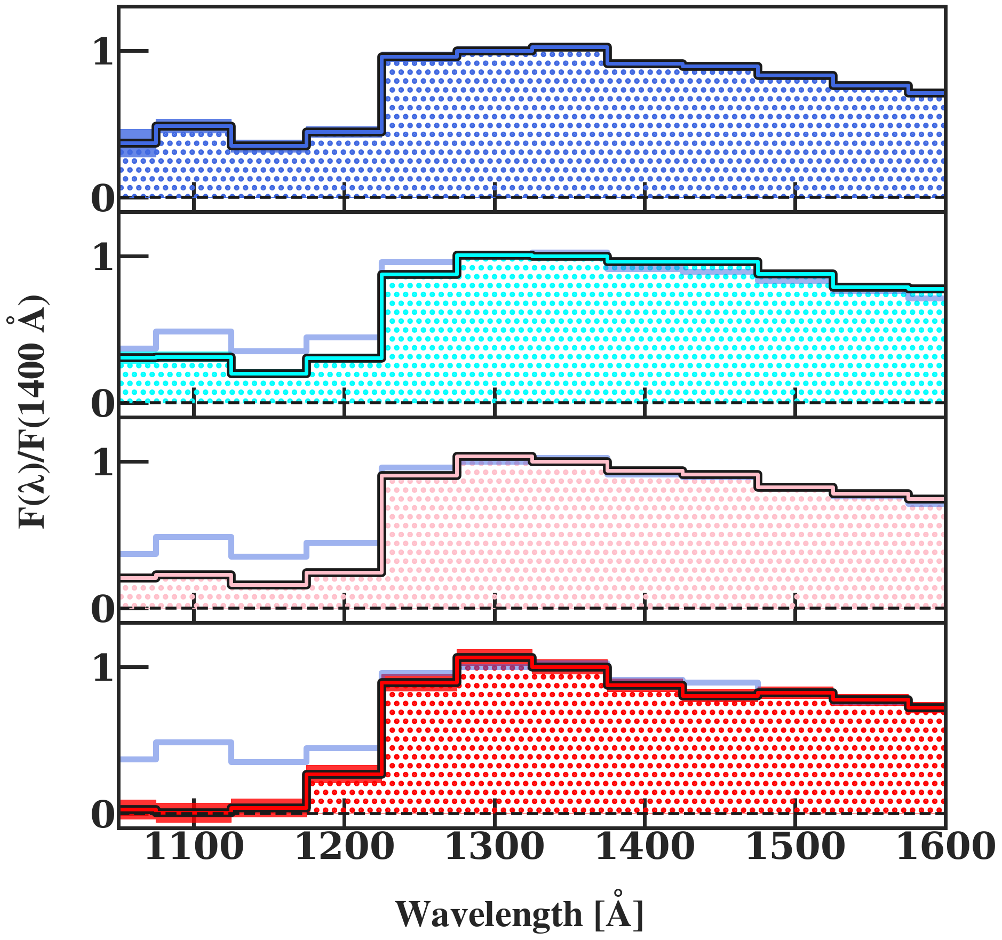}
\caption{The stacked spectra of galaxy subsample around the end of cosmic reionization. \Add{Panels from the top to bottom show the stacked spectra stacked spectra at $\langle z \rangle$=4.8, 5.2, 5.7, and 6.4 in navy blue, pale blue, pink, and red colored solid lines, respectively. \Addp{The darker shades represent 1$\sigma$ uncertainty of the stack spectra}.\Addp{The dotted infill highlights the area associated with positive flux values and is for visual emphasis only.} The faint navy blue line in all the panels except for the top one represent the stack spectra at $\langle z \rangle$=4.8 for the comparison.} The stacked spectra are shown in the rest-frame wavelength. The redshift of the galaxy sample is shown in the upper right corner of each panel. The stack spectra shown here are constructed from the EEoR1-4, with UV magnitude around $M_{\rm UV}\simeq-19.5$ and UV slope around $\beta_{\rm UV}\simeq-2.2$. The corresponding 1$\sigma$ errors are shown in the shaded regions. All the stacked spectra are normalized at 1400 {\AA}. }
\label{stack_lz} 
\end{figure*}

 In Figure \ref{stack_lz}, we show our stack galaxy spectra for EEoR subsamples in different color.  We see the diminishing signal at the Gunn-Peterson trough toward the higher redshift. The Ly$\alpha$ forest is detected above errors for the stack spectra of $\langle z \rangle$=4.8, 5.2, and 5.7 stack spectra. For $\langle z \rangle$=6.3 stack spectra, the Gunn-Peterson trough signal is consistent with non-detection within 1$\sigma$ errorbars. While Ly$\alpha$ opacity increases toward \Add{higher redshift} at the postreionization epoch as well \cite[e.g.,][]{Madau24}, the non-detection of the signal at the wavelength below 1200 {\AA} coincide well with the claim that cosmic reionization ends around $z\simeq5-6$ \cite[e.g.,][]{2006AJ....132..117F,Bosman22}. \Add{Our Ly$\alpha$ forest detection using galaxy spectra aligns with recent independent work that also report the detection of Ly$\alpha$ forest signal using galaxy spectra \citep[][]{Mason25,Meyer25}.} \par
 \subsubsection{Spectral Evolution with Different UV Brightness}
\begin{figure*}
\centering
\begin{minipage}[t]{0.43\linewidth}
    \centering
    \includegraphics[width=\linewidth]{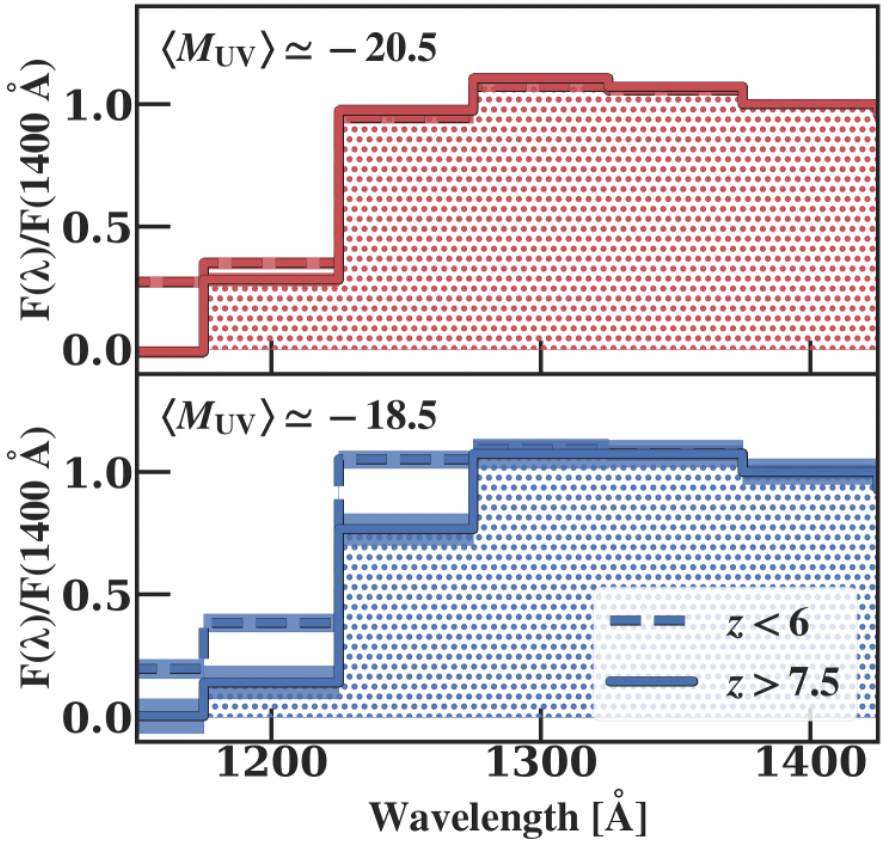}
\end{minipage}
\hfill
\begin{minipage}[t]{0.48\linewidth}
    \centering
    \includegraphics[width=\linewidth]{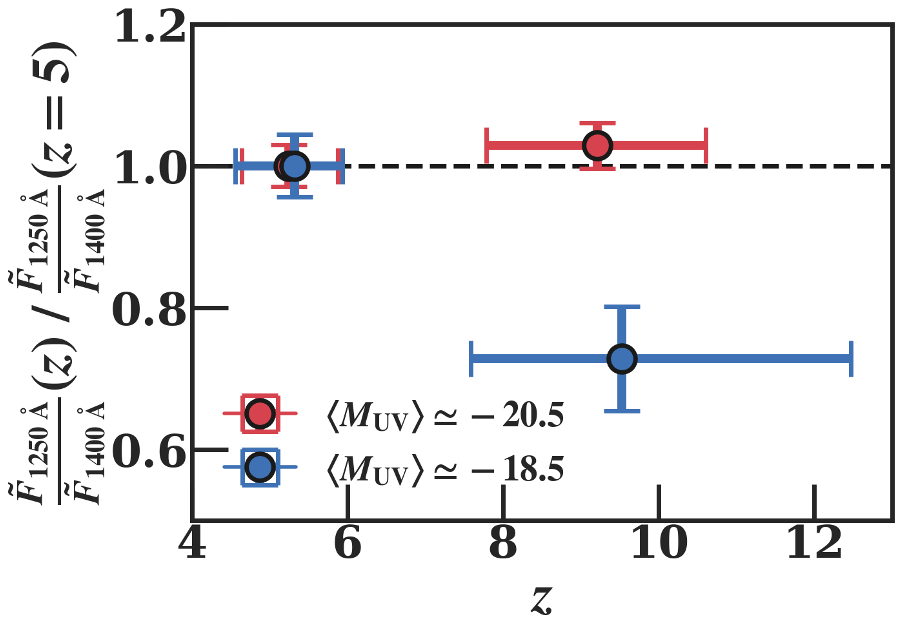}
\end{minipage}
    \caption{(Left): The stack spectra of the galaxy sample binned by the UV magnitude. The stacked spectra are shown in the rest-frame wavelength. \Add{The top and bottom panels show the stacked spectra for the bins of $-19.0<M_{\rm UV}<-17.0$ (i.e., FaiL/H subsamples) and $-22.0<M_{\rm UV}<-20.0$ (i.e., BriL/H subsamples), respectively. The solid and dashed lines correspond to the stack spectra for the redshift bin of $4.5<z<6.0$ and $z>7.5$, respectively.} The corresponding 1$\sigma$ errors are shown in the shaded regions. \Addp{The dotted infill highlights the area associated with positive flux values of $z>7.5$ stack spectra and is for visual emphasis only.} All the stacked spectra are normalized at 1400 {\AA}. (Right): The ratio between instrumentally broadened flux at the rest-frame 1400 and 1250 {\AA} for the stack spectra based on different UV magnitude bin subsamples, normalized by the value at $z\simeq5$.}
\label{muv_lya} 
\end{figure*}

  We present the stack spectra for different UV magnitudes samples in the left panel of Figure \ref{muv_lya}. We do not see strong evolution for the ``bright'' subsample, while we see the stronger Ly$\alpha$ softening for the high redshift subsample. We qualitatively assess the difference in the UV turnover feature between different UV slope subsamples by calculating the ``smoothed out'' flux ($\tilde{F}$) ratio at the rest-frame 1250 and 1400 {\AA}. We explicitly call ``smoothed out'' flux  because we are comparing fluxes based on the low spectral resolution spectra. As shown in Figure \ref{muv_lya}, we show flux ratio at the rest-frame 1250 and 1400 {\AA} for Bri/H and FaiL/H subsample stack spectra. In the right panel of Figure \ref{muv_lya}, we normalize the flux ratio for both ``bright'' and ``faint'' subsample based on the measurement at the lower redshift bin. For bright subsamples, we do not find significant difference between the redshift evolution of flux ratios beyond the errorbar. However, we find the redshift evolution beyond the errorbar for the faint subsamples. This may indicate that brighter (i.e., $M_{\rm UV}\simeq-20.5$) galaxies selectively locate in the ionized region compared to the fainter galaxies (i.e., $M_{\rm UV}\simeq-18.5$). 

 \subsubsection{Spectral Evolution with Different UV Slopes}
\begin{figure*}
\centering
\begin{minipage}[b]{0.43\linewidth}
    \centering
    \includegraphics[width=\linewidth]{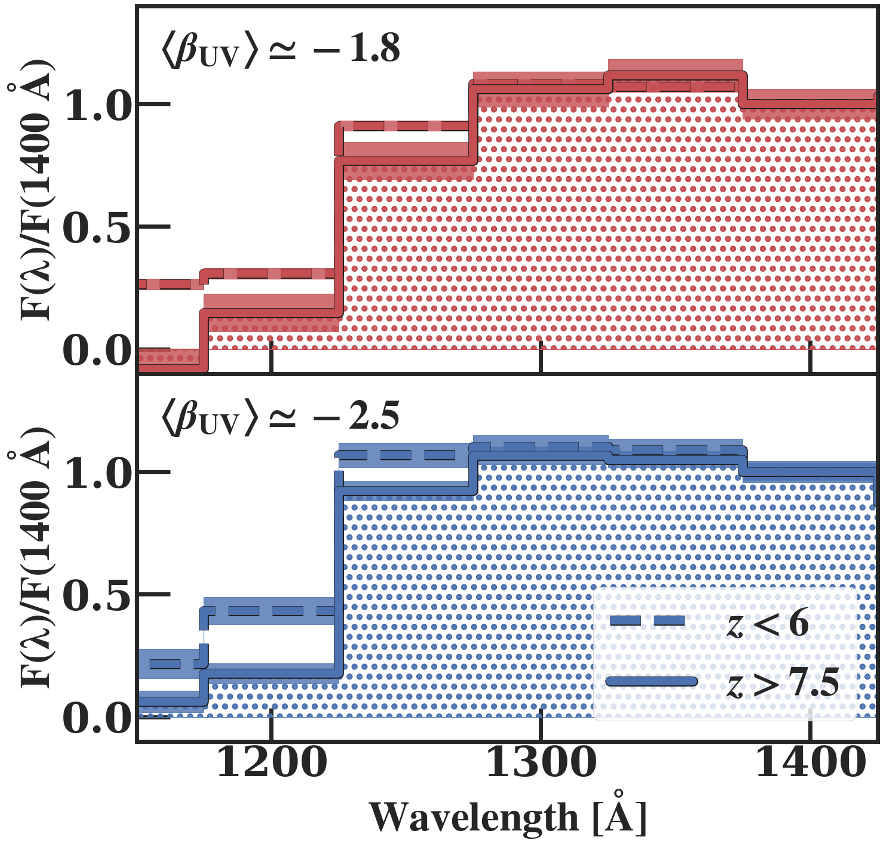}
\end{minipage}
\hfill
\begin{minipage}[b]{0.48\linewidth}
    \centering
    \includegraphics[width=\linewidth]{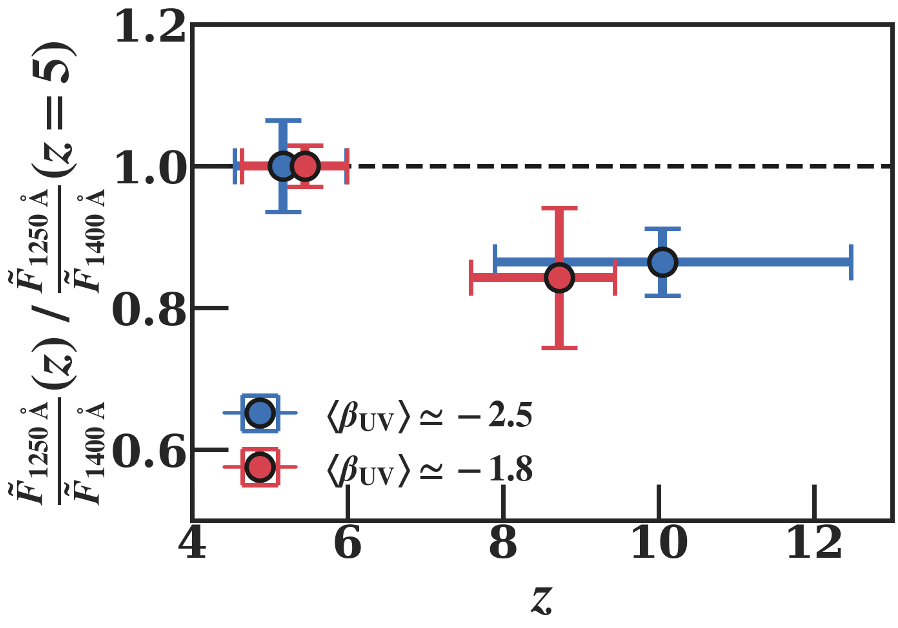}
\end{minipage}
\caption{(Left):The stack spectra of the galaxy sample binned by the UV slope. The stacked spectra are shown in the rest-frame wavelength. \Add{The top and bottom solid panels show the stacked spectra for the bins of $-2.0<\beta_{\rm UV}<-1.4$ (i.e., RedL/H subsamples) and $-3.0<\beta_{\rm UV}<-2.4$ (i.e., BluL/H subsamples), respectively. The solid and dashed lines correspond to the result for the redshift bin of $4.5<z<6.0$ and $z>7.5$, respectively.} The blue and red solid lines in each panel represent the stacked spectra at the redshift bin of $4.5<z<6.0$ and $z>7.5$, respectively. All other symbols are the same as the left panel of Figure \ref{muv_lya}. (Right):The ratio between instrumentally broadened flux at the rest-frame 1400 and 1250 {\AA} for the stack spectra based on different UV slope bin subsamples. All the symbols are the same as the right panel of Figure \ref{muv_lya}.}
\label{beta_lya} 
\end{figure*}

  We present the stack spectra for different UV slope samples in the left panel of Figure \ref{beta_lya}. In the right panel of Figure \ref{beta_lya}, we normalize the flux ratio for both \Add{``blue'' and ``red''} subsample based on the measurement at the lower redshift bin. We calculate the flux ratio at the rest-frame 1250 and 1400 {\AA}. As shown in Figure \ref{beta_lya}, we show flux ratio at the rest-frame 1250 and 1400 {\AA} for RedL/H and BluL/H. For both UV slope bin, we find decreasing flux ratios as towards higher redshift consistently between the different UV slope bin. While the bluer slope are tracer of higher ionizing photon escape fraction \citep{Zackrisson17}, the UV slope of the galaxy may not be tightly correlated with the ionizing region around the galaxy.\par

\section{Lyman Alpha Damping Wing Measurement} \label{spec_fit}
\subsection{IGM Attenuation} \label{dwfit}
\cite{Umeda24} calculate the IGM attenuation along the sightline by assuming homogeneously ionized IGM with galaxy residing at the center of fully ionized bubble sphere. We update the IGM attenuation in more realistically by considering the inhomogeneously ionized IGM \Add{\citep[e.g.,][]{MF08a,Mesinger15,Mason18,Greig18,2018ApJ...864..142D}}. To calculate the optical depth of the Ly$\alpha$ damping wing at each observed wavelength $\lambda_{\rm obs}$, $\tau(\lambda_{\rm obs})$, we adopt Equation 2 of \cite{2006PASJ...58..485T} which is based on the formulation of \cite{ME98}:

\begin{align}
\tau(\lambda_{\rm obs})&=\frac{x_{\rm \HI}\Lambda_{\alpha}\lambda_{\alpha}\tau_{\rm GP}(z_s)}{4\pi^2c}\left(\frac{1+z_{\rm obs}}{1+z_{s}}\right)^{3/2}\\
&\times\left[I\left(\frac{1+z_{b}}{1+z_{\rm obs}}\right)-I\left(\frac{1+z_{e}}{1+z_{\rm obs}}\right)\right], \nonumber
\label{DW_eq}
\end{align}
where $\Lambda_{\alpha}$, $\lambda_{\alpha}$, $c$, and $\tau_0$, are the decay constant for the Ly$\alpha$ resonance, the rest-frame Ly$\alpha$ wavelength, the speed of light, and the Gunn-Peterson optical depth, respectively. $z_{b}$ and ${z_{e}}$ are defined as the redshifts corresponding to the beginning and end of the neutral patch, respectively, whereas $z_{\rm obs}$ is a value defined as $\lambda_{\rm obs}/\lambda_\alpha-1$. $I(x)$ is an integration defined as follows:

\begin{align}
    I(x)&=\frac{x^{9/2}}{1-x}+\frac{9}{7}x^{7/2}+\frac{9}{5}x^{5/2}+3x^{3/2} \\
    &+9x^{1/2}-\frac{9}{2}\ln{\frac{1+x^{1/2}}{1-x^{1/2}}}. \nonumber
\end{align}
We calculate $\tau_{\rm GP}$ using the following formula also given by \cite{ME98}:
\begin{equation}
\tau_{\rm GP}(z)=\frac{3\lambda_\alpha^3\Lambda_\alpha n_0}{8\pi H(z)},
\label{GP_eq}
\end{equation}
where $n_0$ and $H(z)$ are the hydrogen number density and the Hubble parameter as the function of redshift. We calculate $n_0$ using the critical density of the universe, baryon density parameter, and primordial helium abundance. Because the ionization of the IGM are inhomogeneous, the multiple neutral patch along the line of sight contributes to the attenuation. To take an account of this effect, we can represent the damping wing optical depth by the summation of optical depth from the $i$-th neutral patch (i.e., $\tau_i$) in the following manner \citep[e.g.,][]{Mesinger16}:
\begin{align}
\tau(\lambda_{\rm obs})&=\sum_{i} \tau_{i}(\lambda_{\rm obs}) \\\nonumber
&=\frac{\Lambda_{\alpha}\lambda_{\alpha}\tau_{\rm GP}(z_s)}{4\pi^2c}{\left(\frac{1+z_{\rm obs}}{1+z_s}\right)}^{3/2} \\
&\times\sum_{i} x_{{\rm \HI},i}\left[I\left(\frac{1+z_{b,i}}{1+z_{\rm obs}}\right)-I\left(\frac{1+z_{e,i}}{1+z_{\rm obs}}\right)\right]. \nonumber
\end{align}
Here, $x_{\rm \HI}$ is the neutral fraction of the $i$-th neutral patch along the line of sight. Note that $x_{{\rm \HI},i}$ do not generally equal to the volume-averaged neutral fraction $\langle x_{\rm \HI} \rangle$. 
\par
\cite{2023arXiv230805800K} have shown that residual {\HI} gas in the ionized region plays an important role in suppressing the Ly$\alpha$ emission even in the already ionized region. On the other hand, transmission at the wavelength below the rest-frame 1216 {\AA} is not fully saturated at the end of EoR. To consider these various cases, we allow different Ly$\alpha$ absorption by the residual {\HI} gas in the ionized regions \Add{\citep[e.g.,][]{MesingerHaiman04,Laursen11,BoltonHaehnelt13,Mason18}}. We assume that blue Ly$\alpha$ photons are absorbed at the optical depth given by following equation:

\begin{equation}
  \tau_{\rm \HII}(z)=\frac{3\lambda_\alpha^3\Lambda_\alpha n_0x_{\rm \HI, res}}{8\pi H(z)},
  \label{resHII}
\end{equation}

For the ionized region, we consider constant residual neutral fraction of $x_{\rm \HI, res}$. Because this prescription do not account for Ly$\beta$ absorption, we only consider the flux above 1075 {\AA} to avoid the contribution from Ly$\beta$ absorption/emission. We note that definition of the ``ionized'' region becomes ambiguous at the tail of EoR as most of the Universe is already ionized. We consider these cases separately in the following analysis.\par

\subsection{IGM Sightlines} \label{cfit}
To estimate the strength of Ly$\alpha$ damping wing absorption in the galaxy continuum, we need to assume the gas density and ionization state of the gas along the sightline of the galaxy. The cosmic reionization likely proceeded in the spatially inhomogeneous manner \citep[e.g.,][]{Ishimoto22,Meyer25}. Also, the size of ionized region around the galaxies are expected to couple with  $\langle x_{\rm \HI} \rangle$ \citep[e.g.,][]{FO05,Lu23}. 
To account for inhomogeneity in the IGM, we adopt a similar approach as \cite{Mason18}. We use semi-numerical N-body radiative transfer code {\sc 21cmFASTv3} \citep{Murray20} to simulate the IGM ionization and density structure together at the EoR. In the {\sc 21cmFAST} simulation, we also generate the dark matter halo spatial distributions based on the Lagrangian 2nd-order perturbation theory with the excursion set formalism. We then allocate the galaxy with a specific UV magnitude by the the halo mass UV magnitude relation from \cite{Mason15}. \Add{After we allocate the galaxy to dark matter halos, we} obtain the sightlines for different UV magnitudes. We apply the IGM absorption for the $M_{\rm UV}=-19.0$ galaxies to all the stacked spectra \Add{to match with the typical {\MUV} magnitude for the sample (i.e., {\MUV}$=-18.0\sim-20.5$)}. \Add{We do not consider the scatter for $M_h$ and {\MUV} relation because the impact on the {\MUV} dependence to the Ly$\alpha$ visibility is small and do not affect the $x_{\rm \HI}$ inference more than 0.05 \citep{2020MNRAS.495.3602W}.}\par

For cosmic reionization scenarios, we mimic the EoS simulation \citep{Mesinger16}. For this, we prepare the simulation with boxsize of $600^3~{\rm cMpc}^{3}$ and spatial resolution of 1024 (256) pixels per side for density (ionization) field. For the ionization efficiency ($\zeta$) and minimal virial temperature ($T_{\rm vir}^{\rm min}$), we adopt $\zeta=20$ and $T_{\rm vir}^{\rm min}=2\times10^4$ K, respectively. Ionization efficiency corresponds to the amount of photon provided per the mass of halo and the minimum virial temperature correlates with the minimum halo mass which can host galaxy. This prescription corresponds to the fiducial model in the EoS simulation. To calculate various IGM sightlines for Ly$\alpha$ transmission at different $\langle x_{\rm \HI} \rangle$ and redshift, we superimpose ionization map for corresponding $\langle x_{\rm \HI} \rangle$ to the density field at the corresponding redshift. \cite{Mesinger08} find that superimposition of ionization map over the density field at different redshift do not significantly impact the inference of Ly$\alpha$ transmission \citep[cf. ][]{McQuinn07,Kageura25}.\par

Because the strong Ly$\alpha$ emitters are omitted from the stacking analysis in Section \ref{stack}, we must also omit the similar IGM sightlines which could lead to the strong Ly$\alpha$ transmission. To do this, we assign Ly$\alpha$ emission in the same manner as \cite{Mason18}. We assign Ly$\alpha$ equivalent width randomly chosen from the UV magnitude dependent EW distributions \Add{for each mock galaxy}. For the UV magnitude dependent EW distributions, we adopt the prescription adopted in \cite{Mason18}. Other properties such as Ly$\alpha$ velocity offset and the velocity dispersion are also prescribed as the same in \cite{Mason18}.
\Add{A} recent findings from large LAE surveys have discovered that characteristic EW strength does not evolve between redshift $z\sim2$ to 6 \citep{Umeda25}. Thus, we assume that the intrinsic or ``emergent'' Ly$\alpha$ EW distributions do not change beyond $z>6$. After we apply the emergent Ly$\alpha$ emission line, we calculate the transmitted Ly$\alpha$ emission line strength. If the transmitted Ly$\alpha$ emission has the rest-frame EW above 25 {\AA}, we reject the sightline from our IGM model. In this way, we try our best to mimic the selection bias introduced by omitting strong Ly$\alpha$ emitters in our stacking analysis. \par 

We run the simulation for at the redshift corresponding to $\langle z \rangle$ of each ``fiducial'' subsample. We obtain the set of IGM sightlines for $\langle x_{\rm \HI}\rangle$ from 0 to 1 with a gridsize of $\Delta x_{\rm \HI}=0.05$. For $\langle x_{\rm \HI}\rangle =0$, we simply do not apply any IGM attenuation from a neutral patch. For $\langle x_{\rm \HI}\rangle=1$, we assume all IGM is fully neutral by setting the ionization fraction at all simulation pixels to 1.

\subsection{Spectral Fitting} \label{cfit}
To quantitatively infer the $\langle x_{\rm \HI} \rangle$ values from the damping wing feature of the galaxy spectra, we perform the spectral fitting incorporating the IGM absorption. 

\subsubsection{Spectral Template} \label{temp}
\begin{figure}[htbp]
\centering
\includegraphics[width=\linewidth]{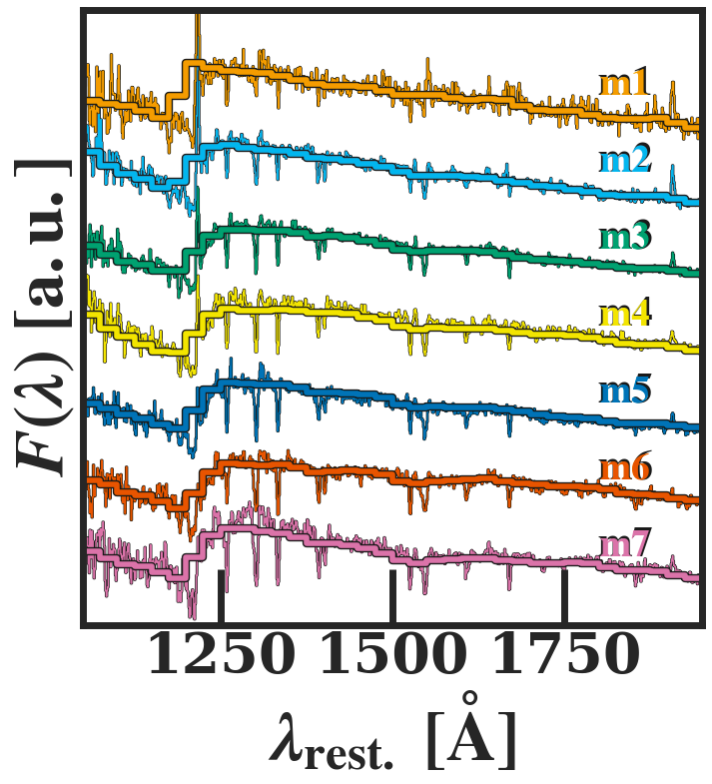}
\caption{
The composite spectra constructed by \cite{Cu19} based VANDELS survey galaxy spectra. We convolve the spectra with the official NIRSpec/PRISM line spread function assuming the object locates at $z=9$. The thin and thick lines correspond to the original composite spectra and the spectra convolved to the JWST NIRSpec/PRISM line spread function.}
\label{vandels} 
\end{figure}

 We fit the Ly$\alpha$ damping wing feature seen in the stacked spectra shown in Figure \ref{stack} using the prescription described in Section \ref{dwfit}. We fit the Ly$\alpha$ damping wing profile with different UV spectrum shapes using data from the rest frame 1075 to 1800 {\AA}. As noted in \cite{Heintz23}, the intrinsic absorption by the circum-galactic medium (CGM) also softens the Lyman break and degenerates with the Ly$\alpha$ damping wing absorption. While breaking the degeneracy between host galaxy and IGM attenuation is a difficult problem for individual galaxy spectra \cite[e.g.,][]{Huberty25}, recent study by \cite{Mason25} find that the median values of neutral {\HI} gas column density measured for the galaxy population do not evolve up to $z\sim11$ from the $z\sim3$ measurement using stacked spectra constructed from using $\sim1000$ Lyman break galaxy spectra \citep{Reddy2016}. Motivated by these findings, we incorporate the host galaxy {\HI} absorption in the spectral fitting manner by using the composite UV spectra of galaxies at moderate redshift (i.e., $2.5<z<5.0$) as the template spectra. \Add{Using spectra of low-z counterparts as templates is a common approach when inferring IGM attenuation \citep[e.g.,][]{2018ApJ...864..142D,Mason18}.} Because adding IGM absorption to the low-resolution ($R\sim100$) spectrum is technically hard, we adopt moderate-resolution ($R\sim1000$) composite spectra obtained by the VANDELS survey as template spectra \Add{from} \cite{Cu19}.\footnote{\url{https://fcullen.github.io/data/}} \cite{Cu19} constructed seven composite spectra from 681 galaxy spectra at $2.5<z<5.0$ binned by the logarithmic stellar mass in ranges of $8.16-8.70$, $8.70-9.20$, $9.20-9.50$, $9.50-9.65$, $9.65-9.80$, $9.80-10.00$, and $10.00-11.00$ in unit of the solar mass (hereafter, we call these composite spectra as m1, m2, m3, m4, m5, m6, and m7, respectively). These spectra have rest frame Ly$\alpha$ EW values below 25 {\AA}, satisfying our selection of JWST galaxy samples. \Addp{While recent study by \cite{Begley25} suggest the possible redshift evolution of nebular properties (i.e., ionizing photon production efficiency) by comparing samples from VANDELS and JWST surveys, the enhancement of ionizing photon production efficiencies are mostly attributed by the very blue ($\beta_{\rm UV}<-2.6$) galaxies in JWST sample \citep[see Fig 9 of][]{Begley25}. We want to note that our selection criteria for the fiducial subsample omit galaxy spectra with $\beta_{\rm UV}<-2.6$.} We summarize the properties of the composite spectra in Table \ref{table:vandels}. By using all available composite spectra in different mass ranges, we consider the various shapes of intrinsic UV spectra together with no additional IGM attenuation applied. We show the composite spectra shifted to $z=9$ in Figure \ref{vandels}. In Figure \ref{vandels}, we do not apply any additional IGM attenuation. We also plot the composite spectra after broadened by the JWST NIRSpec/PRISM line spread function. 

\begin{deluxetable}{ccc}
  \tablecolumns{3}
  \tabletypesize{\scriptsize}
  \tablecaption{Characteristics of VANDELS Template Spectra
  \label{table:vandels}}
  \tablehead{
  \colhead{Template ID} & \colhead{$\langle z \rangle$} & \colhead{Stellar Mass Range ${M_\odot}$} \\
  (1) & (2) & (3)
  }
  \startdata 
  m1 & 3.82  & 8.16 - 8.70\\ 
  m2 & 3.67 & 8.70 - 9.20 \\
  m3 & 3.47 & 9.20 - 9.50 \\
  m4 & 3.50 & 9.50 - 9.65 \\
  m5 & 3.48 & 9.65 - 9.80 \\
  m6 & 3.34 & 9.80 - 10.00 \\
  m7 & 3.24 & 10.00 - 11.00 \\\hline
  \enddata
  \tablecomments{(1): Template ID from \cite{Cu19} (2): the average redshift of galaxies included in the composite spectra. (3): Stellar mass range used to construct the corresponding subsample.
  }
  \end{deluxetable}
 \subsubsection{Inference Procedure} \label{infmet}
 Our fitting \Add{methodology} comprises two steps. In the first step, we fit the observed stack spectra with the template spectra masking the Ly$\alpha$ break region. More specifically, we fit the spectra using the wavelength range covered by 1075 to 1175 {\AA} and 1300 to 1800 {\AA}, masking the region affected by the Ly$\alpha$ damping wing. In this step, we apply the overall normalization to \Add{the original template spectra} and attenuation by the residual hydrogen in the ionized region with a constant $x_{\rm \HI, res}$ value given by Equation \ref{resHII}. \Add{Hereafter, we call the model template spectra after applying normalization factor and attenuation by residual hydrogen as semi-attenuated model spectra.} We set the normalization factor and Ly$\alpha$ transmission at $z=\langle z\rangle$ (i.e., $T_{\alpha}(\langle z\rangle)\equiv\exp(-\tau_{\HII}(z=\langle z\rangle))$) free. We let $T_\alpha$ as a free parameter instead of $x_{\rm \HI, res}$ because the flux is linearly dependent to the $T_\alpha$. However, we still apply the IGM attenuation from {\HII} region assuming Equation \ref{resHII} assuming a constant {$x_{\rm \HI, res}$}. We perform MCMC to sample the posterior distribution function (PDF), with uniform prior for both the normalization factor and $T_{\alpha}$. We set lower boundary for prior of $T_\alpha$ at 0 and 1, respectively. When we fit the semi-attenuated model spectra to the observed stack spectra, we assume the following likelihood function:
\begin{align}
 \label{gp_like}
 -2\ln \mathcal{L} &=\sum_{j,k}\left( F_{{\rm mod},i}(\langle x_{\rm \HI} \rangle=0) - F_{{\rm obs}}\right)_j \\ \nonumber
 &\times{C_{{\rm obs}}}^{-1}_{j,k}\\ \nonumber
 &\times\left(F_{{\rm mod},i}(\langle x_{\rm \HI} \rangle=0) - F_{{\rm obs}}\right)_k\\ \nonumber
 &+\ln2\pi|C_{\rm obs}| \nonumber
 \end{align}
 Here, $F_{\rm obs}$ and $F_{i,{\rm mod}}(\langle x_{\rm \HI} \rangle)$ represent the observed stack spectra and \Add{semi-attenuated model spectra} generated by the \Add{original m$i$ template type} and the IGM attenuation at $\langle x_{\rm \HI}\rangle$, respectively. In Equation \ref{gp_like}, we are calculating the chi-square of between \Add{the fluxes of semi-attenuated model spectra} and the observed stack spectrum in between different spectral pixels (i.e., $F_j,F_k$) weighted by the covariance matrices of the observed stack spectra (i.e., $C_{\rm obs}$).\par
 In the second step, we \Add{randomly} sample \Add{the normalization factor and $T_\alpha$} from PDF we sample in the first step to generate semi-attenuated model spectra sample. Then, for each $\langle x_{\rm \HI} \rangle$, we randomly select the IGM sightlines we produced using {\sc 21cmFAST}, apply Ly$\alpha$ damping wing absorption to the semi-attenuated model spectra, and then convolve to the ``common'' line spread function to match the spectral resolution to the observed stacked spectra. \Add{Hereafter, we call the model spectra with IGM damping wing attenuation as final model spectra.} After we conduct this step for 1000 sightlines and spectra, we calculate the mean and covariance of the final model spectra. After obtaining the mean and covariance for the \Add{final model spectra} for corresponding $\langle x_{\rm \HI} \rangle$ and \Add{the template-type} combination, we calculate the probability density for the corresponding $\langle x_{\rm \HI} \rangle$ and the template types in the following way:
 \begin{align}
 \label{dw_like}
 -2\ln P(\langle x_{\rm \HI}\rangle, i) &=\sum_{j,k}\left(\langle F_{{\rm mod},i}(\langle x_{\rm \HI} \rangle)\rangle - F_{{\rm obs}}\right)_j \\ \nonumber
 &\times[{C_{{\rm obs}}+C_{{\rm mod,}i}(z_l,\langle x_{\rm \HI} \rangle)]}^{-1}_{j,k}\\ \nonumber
 &\times\left(\langle F_{{\rm mod},i}(\langle x_{\rm \HI} \rangle)\rangle - F_{{\rm obs}}\right)_k\\ \nonumber
 &+\ln2\pi|C_{\rm obs} + C_{{\rm mod},i}| \nonumber
 \end{align}
 As similar in the Equation \ref{gp_like}, we calculate the chi-square of between the fluxes of \Add{mean final model spectra} and the observed stack spectrum in between different spectral pixels (i.e., $F_j,F_k$) weighted by the covariance matrices from both model and observed stack spectra (i.e., $C_{\rm mod}$ and $C_{\rm obs}$, respectively). We calculate the probability defined by Equation \ref{dw_like} using spectral pixels at the wavelength range of 1075 to 1800 {\AA}, with no masking around Ly$\alpha$ break. After we calculate the probability density for all \Add{original template types} and $\langle x_{\rm \HI}\rangle$ pattern, we calculate the posterior PDF for $\langle x_{\rm \HI}\rangle$ by marginalizing over the different template in the following manner:
 \begin{equation}
 P(\langle x_{\rm \HI}\rangle)=\sum_{i\in {\rm template}}\ P(\langle x_{\rm \HI}\rangle, i)
 \label{pdf_xhi}
 \end{equation}
 While we use the marginalized posterior PDF given by \ref{pdf_xhi}, we define the best-fit final model spectra by the final model spectra generated from the combination of $\langle x_{\rm \HI} \rangle$ and \Add{the original template types that yield the minimum chi-square for the representation purpose}. We calculate the probability density with all the template \Add{types} (i.e., m1 to m7) and $\langle x_{\rm \HI} \rangle$ from 0 to 1 with a grid size of $\Delta \langle x_{\rm \HI}\rangle=0.05$.

\subsection{Spectral Fitting Results} \label{res}

\subsubsection{Ly$\alpha$ Damping Wing Fitting and Inference of the IGM Neutral Fraction} \label{igminf}

\begin{figure}[htbp]
\centering
\includegraphics[width=\linewidth]{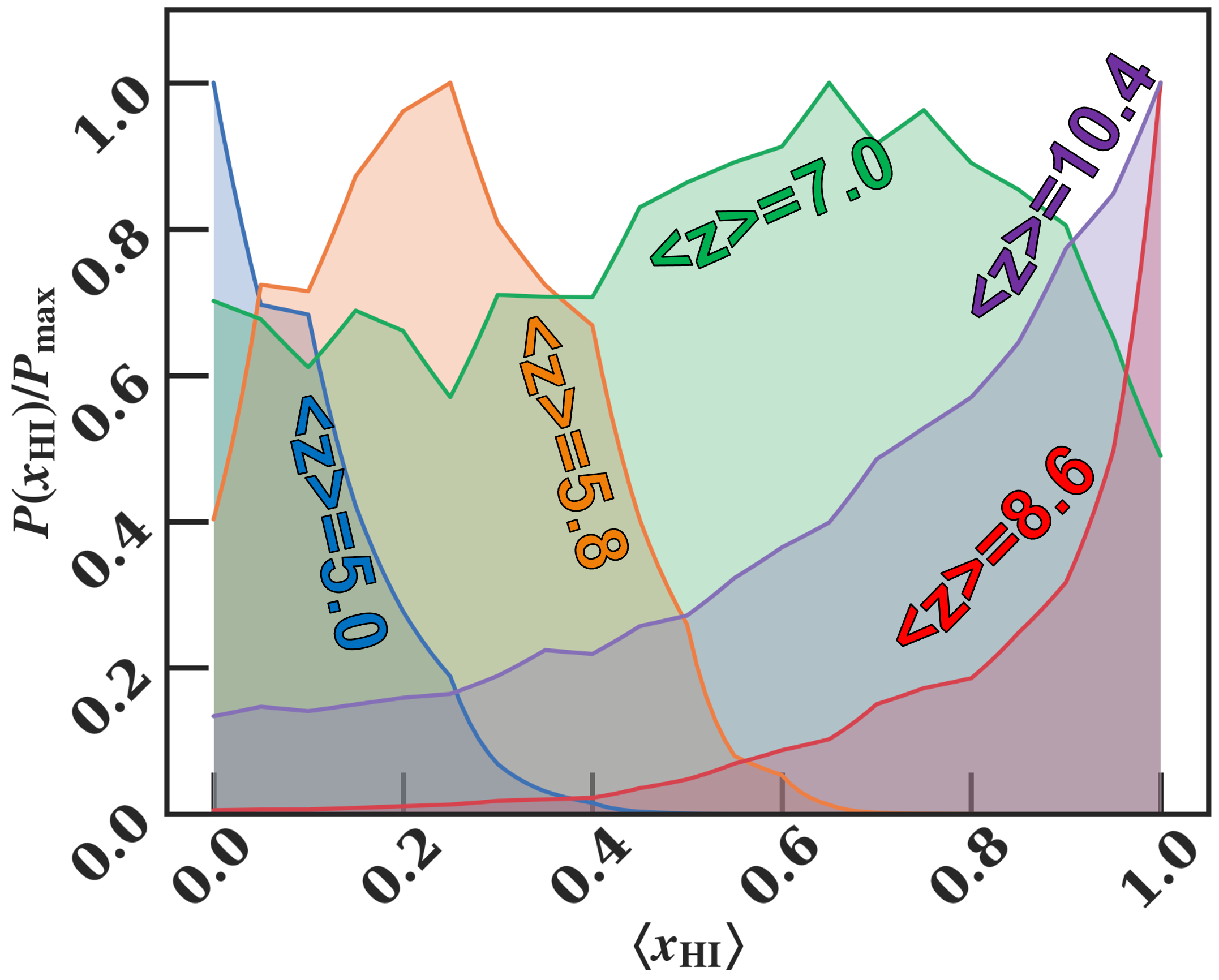}
\caption{The \Add{marginalized} probability distribution of $x_{\rm \HI}$ at the redshift of the galaxy sample. The red, orange, green, red, and purple PDFs represent the results for the galaxy sample at $\langle z \rangle$=5.0, 5.8, 7.0, 8.6, and 10.4 respectively.}
\label{pdf_faint} 
\end{figure}

\begin{figure*}[htbp]
    \centering
    \includegraphics[width=0.9\linewidth]{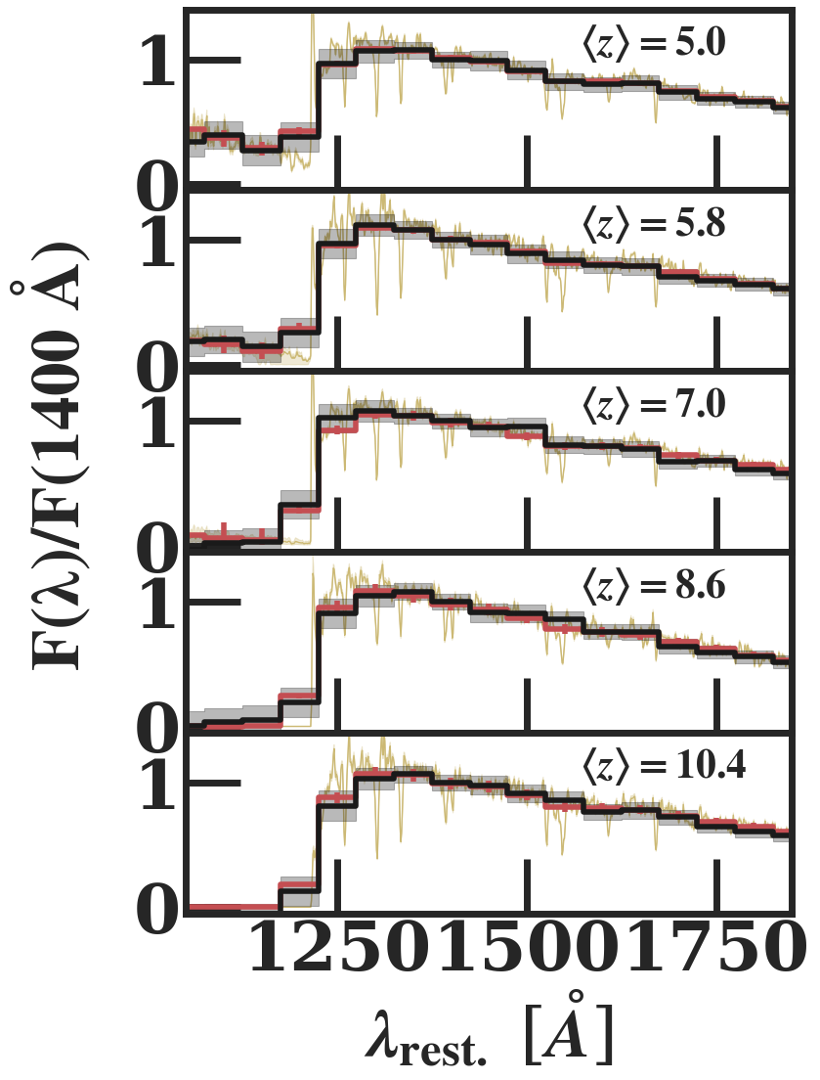}
    \caption{The comparison between the stacked spectra from the observation and the best-fit composite model spectra. The panels from the top to bottom corresponds to the results for $\langle z \rangle$=5.0, 5.8, 7.0, 8.6, and 10.4, respectively. The black solid lines (shades) represent the (1$\sigma$ error of) stacked spectra from the observation. The red solid lines represent \Addp{the mean of best-fit final model spectra. The red horizontal bars in each spectral pixel represent the fluctuation corresponding to the 16-/84-percentile range for the resampled model spectra at the best-fit template model.} \Addp{We note that this fluctuation only represents the fluctuations owing to the inhomogeneity in the IGM at the fixed $\langle x_{\rm \HI} \rangle$ and the normalization.} The yellow lines represent the \Add{final model spectra without the instrumental broadening.}}
    \label{fit_faint} 
\end{figure*}

In Figure \ref{fit_faint}, we present the fitting results for the stacked spectra at $\langle z \rangle$=5.0, 5.8, 7.0, 8.6, and 10.4. We show the comparison between the observed stacked spectra and the \Add{the best-fit final model spectra for the representative purpose}. \Add{The template types for the best-fit final model spectra at the redshift bin of $\langle z \rangle$=5.0, 5.8, 7.0, 8.6, and 10.4, correspond to m2, m4, m2, m1, and m4, respectively, and we do not find clear trend between the template type (i.e., the characteristic stellar masses of the $z\sim3$ stack spectra) and the redshift.} \Addp{The reduced chi-square values of the best-fit final model spectra above 1350 {\AA} (i.e., no IGM attenuated region are 1.0, 1.3, 0.5, 0.6, and 1.0 for $\langle z \rangle$=5.0, 5.8, 7.0, 8.6, and 10.4, respectively. These small, near unity reduced chi-square values suggest that the template spectra well represent the stack spectra without significant residuals. With different template, different $\langle x_{\rm \HI}\rangle$ are inferred. For example, for $\langle z \rangle=5.8$, the least chi-square $\langle x_{\rm \HI}$ value for m7 template is 0.35. However, the difference in chi-square values of m4 (i.e., the best-fit template) and m7 at their least chi-square $\langle x_{\rm \HI}\rangle$ is over 30 and thus it does not affect the marginalized $x_{\rm \HI}$ inference.} We also show the probability distribution of $x_{\rm \HI}$ \Add{at different redshifts marginalized over different template type} in Figure \ref{pdf_faint}. We interpolate log-scale PDF values $\langle x_{\rm \HI} \rangle$ values between the grid by linear interpolation. With interpolated continuous PDF for $\langle x_{\rm \HI} \rangle$, we derive the mode and 68 percent HPDI as the best fit and error, correspondingly. \Add{We list the the inferred $\langle x_{\rm \HI} \rangle$ from the marginalized PDFs for all redshift bin in Table \ref{table:xHI}.} As shown in the Figure \ref{pdf_faint}, we see that the $\langle z \rangle=5.0$ result is consistent with fully ionized universe, while at the $\langle z \rangle=5.8$, the PDF shows the peak around moderately neutral IGM. The $\langle x_{\rm \HI}\rangle$ PDF for $\langle z \rangle=7.0$ shows relatively \Add{unconstrained} distribution, reflecting the scatter in the IGM attenuation due to the inhomogeneous IGM at the middle stage of cosmic reionization \citep{Mesinger08}. Both of the two highest redshift bin (i.e., $\langle z \rangle=8.6,  10.4$) are consistent with fully neutral universe, \Add{matching with the strong attenuation feature around Lyman break in the corresponding stack spectra.} \Add{However, there are alternative explanations for the strong attenuation feature at the Lyman break.} One possibility is the two-photon emission seen in the strong nebular dominant galaxies \cite[e.g., ][]{Cameron24,Katz24}. Because of the intense nebular emission, strong nebular dominant galaxies show Balmer jump. \cite{Katz24} mention that many of the nebular dominated galaxies candidates found using the visual inspection for Balmer jump show the decrease of flux by more than 20\% from the rest-frame 3500 to 4200 {\AA}. We calculate the flux at 3500 and 4200 {\AA} for the ``fiducial'' subsample stack spectra and find that all stack spectra except for that of $\langle z \rangle=8.6$ show flux increase from 3500 to 4200 {\AA}. Also, even for $\langle z \rangle=8.6$ stack spectra, it only shows a tentative (i.e., $\sim5$\%) decrease in the flux with around 10\% uncertainty. While the uncertainty is still large, we do not find any clear feature suggesting strong contribution from nebular dominated galaxies. We also check whether the additional attenuation from strong damped Ly$\alpha$ absorber is required to explain the stack galaxy spectra at the highest redshift bin. For $\langle z \rangle=10.4$ best-fit stack spectra, we investigate whether we need extra strong attenuation from damped Ly$\alpha$ absorbers by adding the Voigt profile absorption with different neutral {\HI} column density $N_{\rm \HI}$. For the simplicity, we fix the center of local {\HI} gas absorption at $z=\langle z \rangle$ and the velocity dispersion of the gas at 100 km/s. We let $\log N_{\rm \HI}/{\rm cm}^{-2}$ free in the range of 17 to 24 with uniform prior. We conduct the MCMC sampling assuming the same likelihood as Equation \ref{gp_like} using the full wavelength range between 1075 to 1800 {\AA}, and obtain the median value at $\log N_{\rm \HI}/{\rm cm}^{-2}=19.2$, which is around a dex smaller than the average column density at $z\sim3$ \citep{Reddy2016}. Moreover, cumulative probability for $\log N_{\rm \HI}/{\rm cm}^{-2}>22$ is only 0.03\%, rejecting the contribution of dominant contributions from the strong damped Ly$\alpha$ absorbers with $\log N_{\rm \HI}/{\rm cm}^{-2}>22$ \citep[e.g.,][]{Heintz23,Umeda24} to the UV turnover feature of average stack spectra at $z\sim10$ \par

\begin{deluxetable}{cccccc}
  \tablecolumns{6}
  \tabletypesize{\scriptsize}
  \tablecaption{Inferred \Add{Marginalized} $\langle x_{\rm \HI} \rangle$ at different redshifts
  \label{table:xHI}}
  \tablehead{
   & \colhead{$\langle z\rangle$} & & & \colhead{$\langle x_{\rm \HI} \rangle$} &  \\
   & (1) & & & (2) &
  }
  \startdata 
   & 5.0 & & & ${0.00}^{+0.12}_{-0.00}$ & \\
   & 5.8 & & & ${0.25}^{+0.10}_{-0.20}$ & \\
   & 7.0 & & & ${0.65}^{+0.27}_{-0.35}$ & \\
   & 8.6 & & & ${1.00}^{+0.00}_{-0.20}$ & \\
   & 10.4 & & & ${1.00}^{+0.00}_{-0.40}$ & \\
  \enddata
  \tablecomments{(1): Average redshift of the subsample (2): Inferred $x_{\rm \HI}$ values correspond to the mode and 68-th percentile values.}
  \end{deluxetable}

\subsubsection{The Gunn-Peterson Trough Measurements} \label{gpmeas}
 We also measure the Ly$\alpha$ transmission around the end of cosmic reionization using the EEoR subsamples. We conduct the similar spectral fitting procedure described in Section \ref{infmet}. Before we fit the template spectra to the observed spectra, we first correct for the IGM attenuation already incorporated in the template spectra. We correct for the IGM attenuation using the IGM attenuation law from \cite{Madau95} by assuming that the template galaxy spectra are obtained at the average redshift of the sample. After we correct for the IGM attenuation in \Add{the original template spectra}, we fit the observed spectra to the model spectra via MCMC. We set the uniform prior for both normalization factor and Ly$\alpha$ transmission. To avoid the contamination from unresolved Ly$\beta$ and Ly$\alpha$ emission/absorption, we mask the range from 1175 to 1275 {\AA}. With this masking, we are measuring the Ly$\alpha$ transmission at the redshift corresponding to the rest-frame $1075-1175$ {\AA}. \par
\begin{figure}[htbp]
\centering
\includegraphics[width=\linewidth]{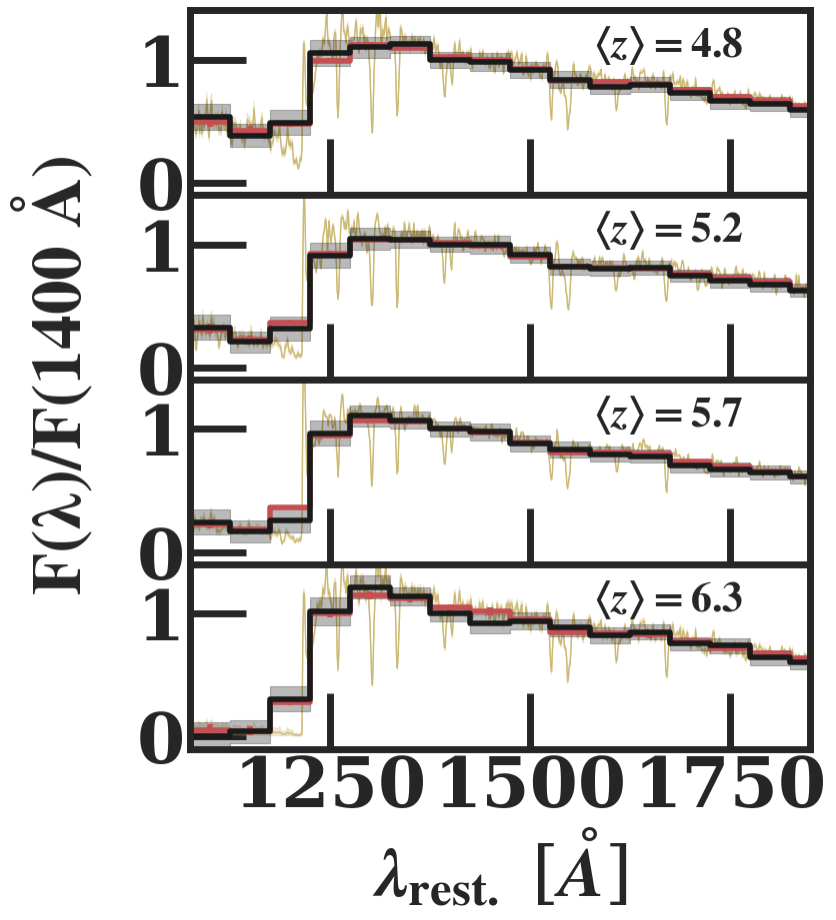}
\caption{
The comparison between the stacked spectra from the observation and the best-fit composite model spectra. The panels from the top to bottom corresponds to the results for $\langle z \rangle$=4.8, 5.2, 5.7, and 6.3, respectively. The black solid lines (shades) represent the (1$\sigma$ error of) stacked spectra from the observation. The red solid lines represent the best-fit model spectra with IGM attenuation incorporated. \Addp{The red horizontal bars in each spectral pixel represent the fluctuation corresponding to the 16-/84-percentile range for the resampled model spectra at the best-fit template model with fixed $\langle x_{\rm \HI}\rangle$.} \Addp{We note that this fluctuation only represents the fluctuation owing to the inhomogeneity in the IGM and normalization of the spectra at the fixed template.} The yellow lines represent the best-fit model with IGM attenuation before convolved to line spread function. Note that the data between the rest-frame wavelength from 1175 to 1275 {\AA} is not used in the fitting to avoid contamination by the host galaxy {\HI} absorption and Ly$\alpha$ emission.
}
\label{fit_eeor} 
\end{figure}

\begin{figure}[htbp]
\centering
\includegraphics[width=\linewidth]{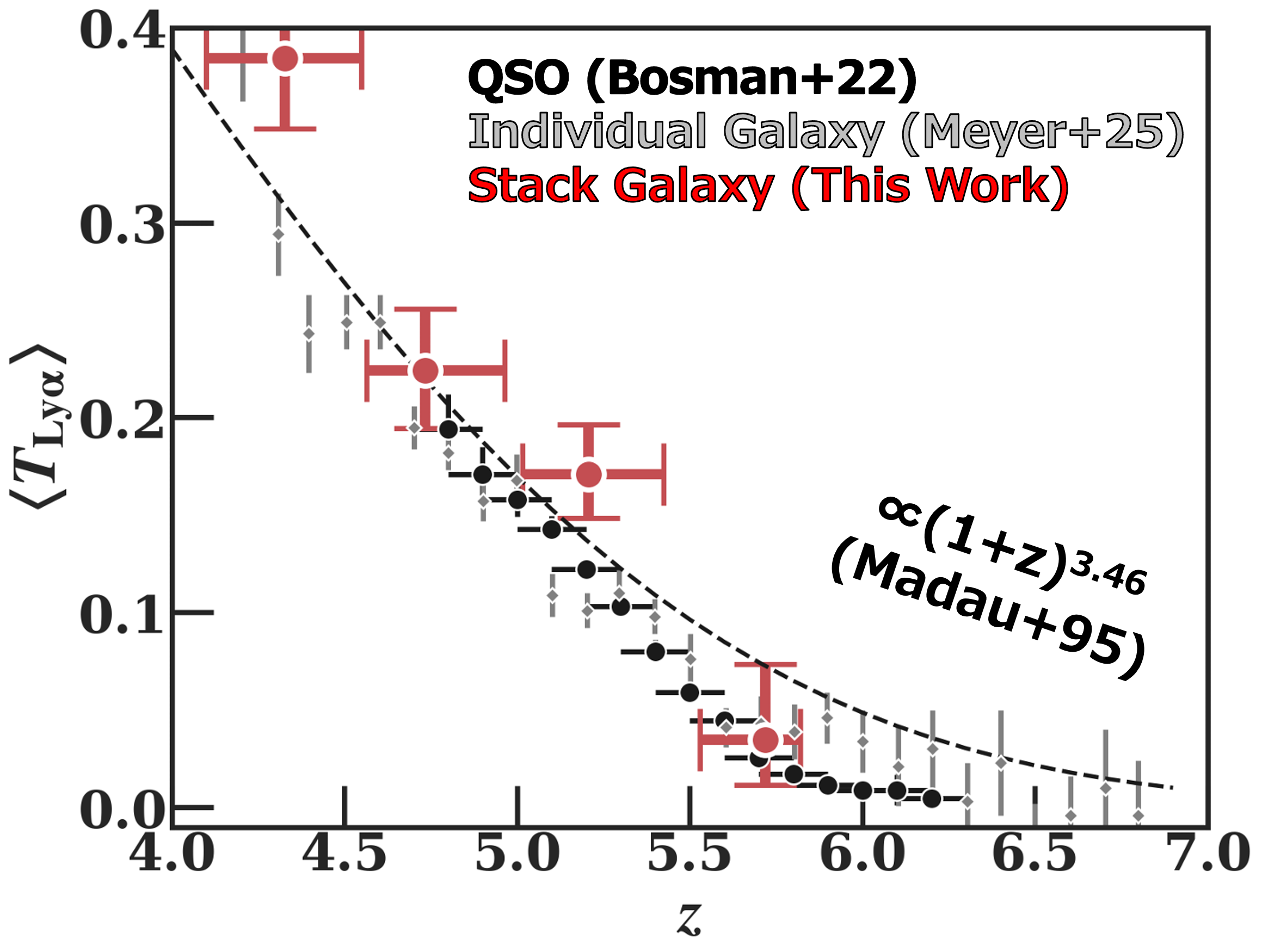}
\caption{
Ly$\alpha$ transmission at the end tail of cosmic reionization.
We present the Ly$\alpha$ transmission values derived from our fiducial stack spectra.
We also plot Ly$\alpha$ transmission values by the redshift derived from QSO's Ly$\alpha$ forest measurement from XQR-30 survey \citep{Bosman22} \Add{and by individual galaxy Ly$\alpha$ forest measurements by \cite{Meyer25} in black and grey points, respectively.}
Moreover, we plot postreionization Ly$\alpha$ transmission relation numerically calculated from \cite{Madau95}.
}
\label{lya_tail} 
\end{figure}

 
 We perform the fitting and show our best-fit spectra compared to observed stack spectra in Figure \ref{fit_eeor}. We can see that the model reproduces the observed Gunn-Peterson trough signal within errorbars at each spectral pixels. The inferred median values for $T_{\alpha}$ measured around the rest-frame $1125$ {\AA} for $\langle z \rangle=4.8$, 5.2, 5.7, and 6.3 (i.e., corresponding to the $T_\alpha$ at $z\sim4.3$, 4.7, 5.2, and 5.7) are 0.38, 0.22, 0.17, and 0.03, respectively. \Add{The best-fit template shown in the Figure \ref{fit_eeor} are m4, m3, m2, m4, and m4 for $\langle z \rangle=4.8$, 5.2, 5.7, and 6.3, respectively.} We plot our $T_\alpha$ measurements by the redshift corresponding to the rest-frame 1125 {\AA} in Figure \ref{lya_tail}.

\section{Discussion} \label{disc}
\subsection{Ly$\alpha$ Forest Signals at the End Tail of Reionization}\label{tail}
From the inferred $x_{\rm \HI, res.}$ values, we measure the Ly$\alpha$ transmission by comparing the flux at from rest-frame 1075 to 1175 {\AA} for the observed stack spectra and model spectra selected in the fitting in Section \ref{gpmeas}. We show \Add{the comparison between between the fitted model and the observed stack spectra} in Figure \ref{lya_tail}. We confirm the consistency with the Ly$\alpha$ transmission measurements precise measurements by \cite{Bosman22} using QSO spectra from XQR-30 survey. \Add{In Figure \ref{lya_tail}, we also compare the results from independent analysis by \cite{Meyer25} that derive Ly$\alpha$ transmission via SED fitting of individual galaxies.}  Our Ly$\alpha$ transmission measurements using stacking galaxy spectra as well the results based on individual fitting of galaxy spectra re-ensure that the reionization has completed by $z\sim5$ but has a tentative residual neutral hydrogen island at the tail of reionization (i.e., $z=5-6$). Moreover, our demonstration on measuring Ly$\alpha$ transmission using $z>5$ galaxies could open up new possibility in the EoR science. For example, we could take an advantage of the higher number density of galaxies compared to that of QSOs by measuring Ly$\alpha$ transmission in numbers of sightline at the specific redshift. We can infer IGM spatial structures (i.e., temperature, ionization, density) after EoR from the spatial distribution of Ly$\alpha$ transmission \Add{as hinted by \cite{Meyer25}}.\par

\subsection{Cosmic Reionization History}

\subsubsection{$x_{\rm \HI}$ Evolution}\label{xHI_evolve}
\begin{figure*}[h]
  \centering
  \includegraphics[width=\linewidth]{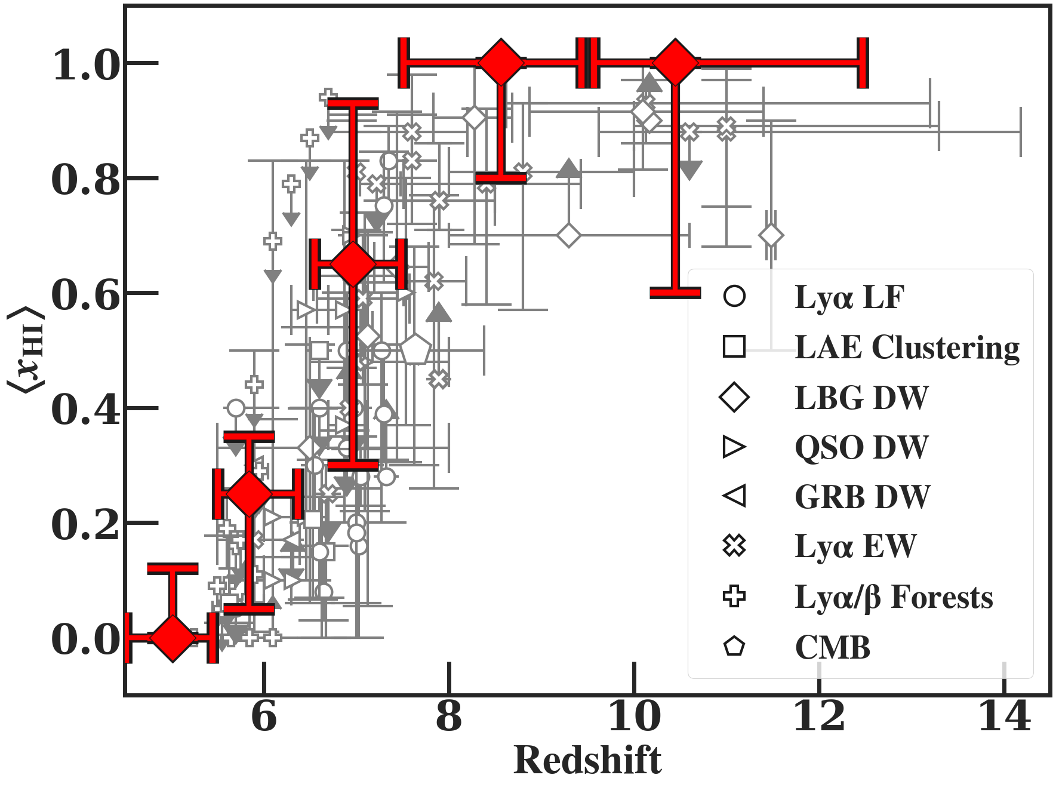}
  \caption{Redshift evolution of $x_{\rm \HI}$. The red diamonds represent our $x_{\rm \HI}$ estimates from this work based on the galaxy continuum damping wing absorption.
  Beside $x_{\rm \HI}$ estimates from this work, we also present $x_{\rm \HI}$ estimate from literature using Ly$\alpha$ luminosity function \citep[circles;][]{Ouchi10,Konno14,Zheng17,2018PASJ...70...55I,Morales21,Ning22,Wold22,Umeda25}, LAE clustering \citep[squares][]{Sobacchi15,Ouchi18,Umeda25}, Ly$\alpha$ damping wing measurement of LBGs \cite[diamonds;][]{Umeda24,CL23,Hsiao23,Mason25}, damping wing measurements of QSOs \cite[right-tipped triangles;][]{2013MNRAS.428.3058S,2018ApJ...864..142D,2019MNRAS.484.5094G,2020ApJ...896...23W,Durovcikova23}, damping wing measurements of GRBs \citep[left-tipped triangles;][]{2006PASJ...58..485T,2014PASJ...66...63T}, Ly$\alpha$ equivalent width distributions \citep[X marks;][]{2015MNRAS.446..566M,2019ApJ...878...12H,2019MNRAS.485.3947M,2020ApJ...904..144J,2020MNRAS.495.3602W,Bolan22,Br23,Mo23,Nakane24,Tang24,Jones24}, Ly$\alpha$ forests and/or Ly$\alpha$+$\beta$ dark fraction/gaps measurements \citep[pluses;][]{2006AJ....132..117F,2015MNRAS.446..566M,2023ApJ...942...59J,2022ApJ...932...76Z,Zhu24,Spina24}, and the electron scattering of CMB \citep[pentagon;][]{Planck20}.}
  \label{xHI_alldata}
\end{figure*}

We show the redshift evolution of $\langle x_{\rm \HI} \rangle$ in Figure \ref{xHI_alldata}. We find that \Add{our $\langle x_{\rm \HI} \rangle$ constraints} increases with redshift, reaching $x_{\rm \HI}\sim1$ at $z\sim9$. Our $x_{\rm \HI}$ estimates are consistent with the recent measurements from the recent Ly$\alpha$ luminosity functions and LAE clusterings measurement using Subaru/HSC imaging data \citep{Umeda25} and Ly$\alpha$ equivalent width distribution measurements using JWST data \citep[e.g.,][]{Kageura25,Nakane24,Jones24,Tang24,Napolitano24}. Also, our $x_{\rm \HI}$ values agree with the independent work of \cite{Mason25} \Add{which use individual galaxy spectra to measure IGM Ly$\alpha$ damping wing.}\par 

\subsubsection{Source of Ionization and {\sc Hii} Topology}\label{sec:top}
Topology of ionized IGM imprints the source of ionizing photons at the EoR. In another words, the IGM attenuation along the line of sight should depend on the galaxy properties such as the UV brightness. The difference in the absorption profile by the galaxy properties could possibly hint the source of IGM ionizing photons. For example, we expect less Ly$\alpha$ absorption for the brighter galaxies than the fainter ones if the ionizing photons are mainly \Add{supplied} from the \Add{overdensed environment hosting} bright galaxies. As discussed in Section \ref{sample},  we see that for the $\langle M_{\rm UV} \rangle\simeq-20.5$ bin (i.e., the brightest bin), the stacked spectra at the different redshift do not evolve. However, for the $\langle M_{\rm UV} \rangle\simeq-18.5$ bin (i.e., the faintest bin), we could capture the strong evolution of the absorption feature by redshift between $z<6$ to $z>7.5$. This feature is consistent with the picture that the ionization of IGM starts earlier around the massive dark matter halo hosting these bright galaxies \Add{at the overdensed region of the universe \citep[e.g.,][]{Mason18,Mason18b,Weinberger18,Gronke21,Smith22}}. If the objects hosted in the massive halos are the dominant source of ionizing photons, cosmic reionization proceeds late and rapidly, which is qualitatively consistent with inferred $\langle x_{\rm \HI} \rangle$ in Section \ref{xHI_evolve}. \Add{The future observations of more fainter galaxies via lensed cluster surveys or spectroscopic follow-up of gamma ray burst's bright afterglow \citep[e.g.,][]{2006PASJ...58..485T} would be helpful to further test the UV magnitude dependence of the Ly$\alpha$ visibility at the EoR.} \Add{At the same time, we need variety of spectra sample covering wider stellar and nebular property ranges to ensure the recovery of the unattenuated spectral shape of various kind of EoR objects. Not just on the galaxy property itself, but also extending the wavelength coverage is important. Incorporating the Balmer line information may enable better inference on the intrinsic Ly$\alpha$ strength, therefore allowing the IGM attenuation inference using the galaxy with strong Ly$\alpha$ emission contributions.} \Addp{While stack spectra from \cite{Cu19} are based on ground-based observations, we can combine data from ground-based and JWST to construct template spectra spanning from the rest-frame UV to optical wavelength of galaxies at post-EoR.} \par

\begin{figure}[h]
  \centering
  \includegraphics[width=\linewidth]{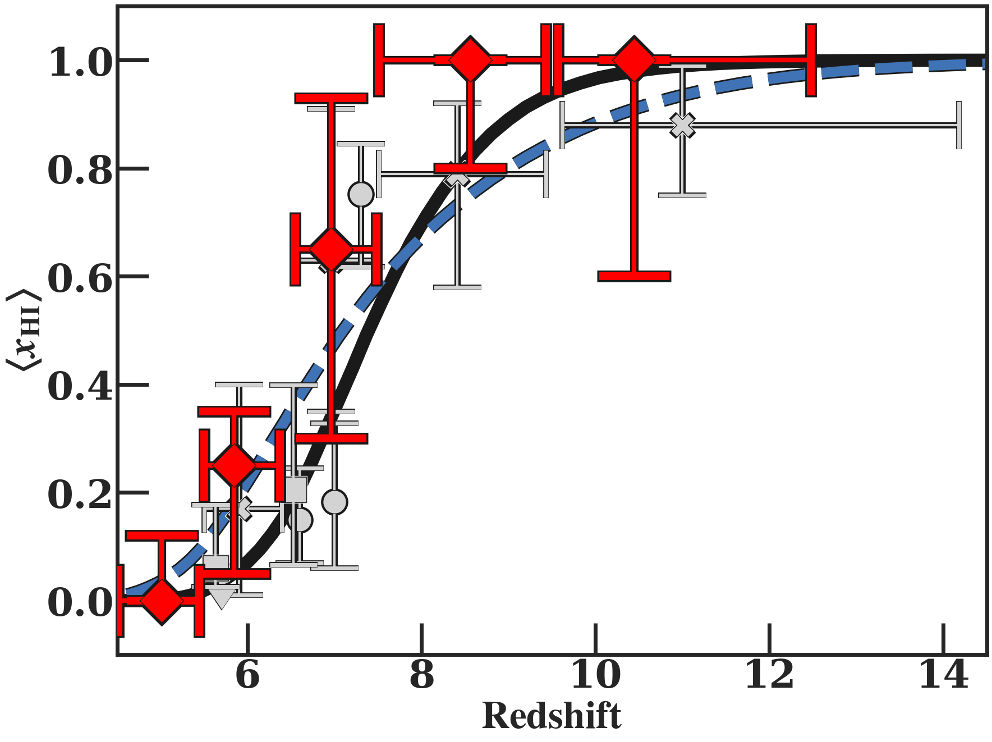}
  \caption{Comparison between cosmic reionization scenarios and $\langle x_{\rm \HI} \rangle$ estimates. The \Add{black} solid lines corresponds to the best-fit $x_{\rm \HI}$ evolution prediction from \cite{Kageura25} with $\log\zeta=2.5$ and $\log T^{\rm min}_{\rm vir}/{\rm K}=5.6.$. The blue dashed line corresponds to the faint galaxy dominant model (i.e., $\zeta=20$, $T^{\rm min}_{\rm vir}=5\times10^4~{\rm K}$). The red diamonds represent $x_{\rm \HI}$ measurements from this work based on Ly$\alpha$ damping wing absorption measurements. The grey circles and squares represent $x_{\rm \HI}$ from \cite{Umeda25} based on Ly$\alpha$ luminosity functions and LAE clustering, respectively. The grey x-marks represent the $\langle x_{\rm \HI} \rangle$ inferred from Ly$\alpha$ equivalent width distributions by \cite{Kageura25}, respectively.}
  \label{xHI_compare}
\end{figure}

\subsubsection{Physical Origin of Rapid Reionization}
As discussed in the previous sections, cosmic reionization suggested from our work and numerous other works suggest that rapid reionization.  To clearly the trend, we plot our $\langle x_{\rm \HI} \rangle$ values inferred by JWST damping wing, JWST LAE equivalent width distribution \cite{Kageura25}, and Ly$\alpha$ luminosity function and spatial clustering by the LAE sample constructed from the imaging data of ground-based Subaru telescope \cite{Umeda25} in Figure \ref{xHI_compare}. We see that the $\langle x_{\rm \HI} \rangle$ rapidly transitions from 1 to 0 at the redshift range around $z\sim7-8$, where the electron scattering measurement by \cite{Planck20} infer to be the timing where the IGM are half ionized. To interpret such a rapid and late reionization history, we overplot several redshift evolution of $\langle x_{\rm \HI} \rangle$ with different cosmic reionization scenarios discussed in \cite{Kageura25}. We overplot the \Add{``faint galaxy dominated''} cosmic reionization scenario assuming $\zeta=20$, $T_{\rm vir}^{\rm min}=5\times10^4$ K, and the clumping factor of 3 at $z\sim5$. This scenario corresponds to the situation where the galaxies as faint as $M_{\rm UV}=-10$ contributes to the ionization of IGM and that the ionizing photons are mainly distributed by the faint galaxies (i.e., $M_{\rm UV}>-17$). The ionization of IGM starts earlier than our observed constraints for the faint galaxy dominant scenario, ending up more slower cosmic reionization \Add{history}.\par 
We also compare with the \Add{``best-fit''} model inferred by \cite{Kageura25} based on the $x_{\rm \HI}$ measurements from Ly$\alpha$ forest by \cite{Bosman22}, Ly$\alpha+\beta$ dark gap measurements by \cite{2022ApJ...932...76Z}, Ly$\alpha$ luminosity function and clustering measurements by \cite{Umeda25}, and Ly$\alpha$ EW distributions by \cite{Kageura25}. Kageura's \Add{``best-fit''} model assumes $\zeta=10^{2.5}$ and $T_{\rm vir}^{\rm min}=10^{5.6}$ K. More physically speaking, Kageura's \Add{``best-fit''} model assumes that the ionization of IGM is dominated by the ionizing sources hosted by the dark matter halo with mass of $>10^{10.5}~M_{\odot}$ with ionizing photon escape fraction of 50\%. With scenario, the $x_{\rm \HI}$ evolution is more rapid and late than the faint dominant model. \Add{We find that this rapid and late reionization scenario suggested by \cite{Kageura25}'s \Add{``best-fit'' model} are consistent with our inferred $x_{\rm \HI}$ values at all redshift within errorbars.} Such an assumption is hard to interpret physically if we assume simple galaxy population. The AGNs may be the main driver of cosmic reionization, as AGN are most likely form earlier at the massive dark matter halos. However, the number density suggested from recent JWST observations suggest that the AGN cannot fully account for the ionization photon budget to complete cosmic reionization \citep[e.g.,][]{Harikane24b,Asthana24,Madau24}. Because most of the AGN discovery at $z>6$ relies on the broad Balmer line selection, the number density of TypeII AGN may be underestimated than the assumption made to assess the impact of AGN to cosmic reionization \citep{Scholtz23}. Further investigation on the AGNs at the EoR is essential to accurately predict how much AGN may contribute to the ionizing photon budget at the EoR.\par 

Another way to look at the rapid and late reionization is to think how to regulate ionizing photon supplies from galaxies at the EoR to match cosmic reionization history inferred from the observations. Massive galaxies can produce enough ionizing photons to complete cosmic reionization. The problem lies in how to adjust the timing and amount of ionizing photons escaping to the IGM. If the massive galaxies provide too much ionizing photons at the early stage of the cosmic reionization, then the reionization completes too early than suggested by $z\sim5-6$ Ly$\alpha$ forest measurements. One possible evidence for the change in the ionizing properties of the galaxies is the redshift evolution of UV slopes. As \cite{Saxena24} claim, we also could see the trend of UV slope generally turns bluer toward higher redshift until $z\sim9.5$, but turns redder at the even higher redshifts. When ionizing photons are absorbed before escaping into the IGM, the strength of nebular continuum is enhanced and the stronger contribution of nebular continuum to the galaxy spectra leads to redder UV slope. In another words, bluer spectra suggest the stronger escape of ionizing photons. With this interpretation, the redshift evolution of UV slope suggest that the ionizing photon escape from galaxies are inhibited at $z>9.5$ while the ionizing photons can efficiently escape from galaxies at $z<9.5$. Several explanation exists to explain the time evolving escape fraction. For example, \cite{Ferrara24} suggest that galaxies may experience attenuation free model after several stage of the star formation. During the attenuation-free stage, the dusts are blown away by the radiative-driven outflow and carving out ionized channels that allow efficient ionizing photon escape. Based on the study of $z\sim2$ LAE's emissivity, \cite{Matthee22} also suggest that bright LAEs with high escape fraction (i.e., $f_{\rm esc}\simeq50$\%) and ionizing photon production efficiency (i.e., $\xi_{\rm ion}=10^{25.9}~{\rm erg/s}$) could escape the the rapid cosmic reionization scenario taking place at $z\sim6-9$. \cite{2022MNRAS.510.4582N} argues that such a efficient ionization could be achieved at the early stage of star-formation where the massive stars start to die and disturb the surrounding neutral contents via feedbacks. With such a galaxy evolutionary track, we may possibly explain delayed progress of cosmic reionization after the intense star-formation at the early Universe. Another interesting point of view is to check the consistency between supernova driven ionizing photon escape. From the chemical abundance pattern of $z>9$ galaxies, \cite{Nakane25} suggest that there may be very \Add{energetic} supernova (e.g., pair-instability supernova) which could enhance the ionizing photon escape by creating ionizing channels via supernova driven outflow. In this sense, cosmic reionization history may be coupled with the chemical enrichment history of galaxies via the occurrence of supernova. Further investigation need to investigate the galaxy evolution to understand how the ionizing photons escapes are regulated.\par 
\section{conclusion}\label{conc}
In this work, we investigated $\sim600$ JWST/NIRSpec spectra of galaxies at $4.5<z<13$ from public JWST survey. We construct stacking spectra from homogeneous samples and investigate how the typical galaxy spectra evolve with increasing IGM Ly$\alpha$ attenuation toward high redshift. We summarize our findings as follows:
\begin{enumerate}
\item We infer $\langle x_{\rm \HI}\rangle$ at $z=$5.0, 5.8, 7.0, 8.6, and 10.4 using realistic galaxy template model and IGM transmission model based on semi-numerical simulations. We find that the $\langle x_{\rm \HI} \rangle$ values of ${0.00}^{+0.12}_{-0.00}$, ${0.25}^{+0.10}_{-0.20}$, ${0.65}^{+0.27}_{-0.35}$, ${1.00}^{+0.00}_{-0.20}$, and ${1.00}^{+0.00}_{-0.40}$ at $z=$5.0, 5.8, 7.0, 8.6, and 10.4, respectively. \par 
\item Our $\langle x_{\rm \HI} \rangle$ inference are consistent with recent independent $\langle x_{\rm \HI} \rangle$ inference based on Ly$\alpha$ equivalent width distribution measurement using JWST data and Ly$\alpha$ luminosity function/clustering measurements by Subaru telescopes. The inferred redshift evolution of $x_{\rm \HI}$ \Add{are also broadly consistent with }rapid and late \Add{reionization recently suggested by \cite{Kageura25}} that are hard to reproduce by simply enhancing/limiting the ionizing photon production and/or escape fraction. We may require the emergence of hidden source of ionizing photon (e.g., AGNs) or time-evolving escape fractions to explain the inferred late and rapid cosmic reionization.
\item We confirm the detection of Gunn-Peterson trough signal at $z<5.5$, indicating that IGM is almost completely ionized. We find consistency between the high precision Ly$\alpha$ transmission measurements using QSO spectra. Our measurements do not disagree with recent claim that cosmic reionization persist until $z\simeq5.3$.
\item We investigate the dependency of Ly$\alpha$ damping wing absorption feature by the galaxy properties such as $M_{\rm UV}$ and $\beta_{\rm UV}$. We find no significant difference between the damping wing feature between different UV slopes, whereas we find the fainter galaxies show stronger damping wing feature at $z>7$ than the brighter galaxies. \Add{Because brighter galaxies are more likely to trace the overdense environments and massive halos, the UV magnitude dependence of the Ly$\alpha$ transmission suggests} that the IGM around the massive halos are ionized prior to the small halos. 
\end{enumerate}

\section*{Acknowledgements}\label{ack}
We thank Frederick Davies, Sarah Bosman, Charlotte Mason, Laura Keating, Claudia Scarlata, Martin Haehnelt, Sandro Tacchella, and Hiroto Yanagisawa for the valuable discussion regading this work. This work is based on observations with the NASA/ESA/CSA James Webb Space Telescopes at the Space Telescope Science Institute, which is operated by the Association of Universities for Research in Astronomy, Inc., under NASA contract NAS 5-03127 for JWST. These observations are associated with programs ERS-1345 (CEERS), GO-1433, GO-3073, DD-2750, GO-2561 (UNCOVER), GTOs-1180, 1191, 1210, and 1286 (JADES). The authors acknowledge the CEERS, GLASS, GO-1433, DD-2750, UNCOVER, and JADES teams led by Steven L. Finkelstein, Dan Coe, Pablo Arrabal Haro, Marco Castellano, Ivo Labbe, Rachel Benzason, Nora Lutzgendorf, and Daniel Eisenstein, respectively. We thank the JADES and UNCOVER team for publicly releasing reduced spectra and catalog from JADES and UNCOVER survey, respectively. We thank Fergus Cullen and the VANDELS team for publicly releasing the composite UV spectra from VANDELS survey. This publication is based upon work supported by the World Premier International Research Center Initiative (WPI Initiative), MEXT, Japan, and KAKENHI (20H00180, 21K13953, 21H04467, 23KJ0646, 22K21349) through Japan Society for the Promotion of Science. This work was supported by JSPS Core-to-Core Program (grant number: JPJSCCA20210003). This work was supported by the joint research program of the Institute for Cosmic Ray Research (ICRR), University of Tokyo. This research was supported by FoPM,WINGS Program, the University of Tokyo.


\bibliography{main}{}

\begin{thebibliography}{}
\expandafter\ifx\csname natexlab\endcsname\relax\def\natexlab#1{#1}\fi
\providecommand{\url}[1]{\href{#1}{#1}}
\providecommand{\dodoi}[1]{doi:~\href{http://doi.org/#1}{\nolinkurl{#1}}}
\providecommand{\doeprint}[1]{\href{http://ascl.net/#1}{\nolinkurl{http://ascl.net/#1}}}
\providecommand{\doarXiv}[1]{\href{https://arxiv.org/abs/#1}{\nolinkurl{https://arxiv.org/abs/#1}}}

\bibitem[{P. {Arrabal Haro} {et~al.}(2023){Arrabal Haro}, {Dickinson}, {Finkelstein}, {Kartaltepe}, {Donnan}, {Burgarella}, {Carnall}, {Cullen}, {Dunlop}, {Fern{\'a}ndez}, {Fujimoto}, {Jung}, {Krips}, {Larson}, {Papovich}, {P{\'e}rez-Gonz{\'a}lez}, {Amor{\'\i}n}, {Bagley}, {Buat}, {Casey}, {Chworowsky}, {Cohen}, {Ferguson}, {Giavalisco}, {Huertas-Company}, {Hutchison}, {Kocevski}, {Koekemoer}, {Lucas}, {McLeod}, {McLure}, {Pirzkal}, {Trump}, {Weiner}, {Wilkins}, \& {Zavala}}]{AH23b}
{Arrabal Haro}, P., {Dickinson}, M., {Finkelstein}, S.~L., {et~al.} 2023, \bibinfo{title}{{Spectroscopic verification of very luminous galaxy candidates in the early universe},} arXiv e-prints, arXiv:2303.15431, \dodoi{10.48550/arXiv.2303.15431}

\bibitem[{Y. {Asada} {et~al.}(2024){Asada}, {Desprez}, {Willott}, {Sawicki}, {Brada{\v{c}}}, {Brammer}, {Dubath}, {Iyer}, {Martis}, {Muzzin}, {Noirot}, {Paltani}, {Sarrouh}, {Harshan}, \& {Markov}}]{Asada24}
{Asada}, Y., {Desprez}, G., {Willott}, C.~J., {et~al.} 2024, \bibinfo{title}{{Improving photometric redshifts of Epoch of Reionization galaxies: a new transmission curve with the neutral hydrogen damped Ly$\alpha$ absorption},} arXiv e-prints, arXiv:2410.21543, \dodoi{10.48550/arXiv.2410.21543}

\bibitem[{S. {Asthana} {et~al.}(2024){Asthana}, {Haehnelt}, {Kulkarni}, {Bolton}, {Gaikwad}, {Keating}, \& {Puchwein}}]{Asthana24}
{Asthana}, S., {Haehnelt}, M.~G., {Kulkarni}, G., {et~al.} 2024, \bibinfo{title}{{The impact of faint AGN discovered by JWST on reionization},} arXiv e-prints, arXiv:2409.15453, \dodoi{10.48550/arXiv.2409.15453}

\bibitem[{E. {Ba{\~n}ados} {et~al.}(2018){Ba{\~n}ados}, {Venemans}, {Mazzucchelli}, {Farina}, {Walter}, {Wang}, {Decarli}, {Stern}, {Fan}, {Davies}, {Hennawi}, {Simcoe}, {Turner}, {Rix}, {Yang}, {Kelson}, {Rudie}, \& {Winters}}]{B18}
{Ba{\~n}ados}, E., {Venemans}, B.~P., {Mazzucchelli}, C., {et~al.} 2018, \bibinfo{title}{{An 800-million-solar-mass black hole in a significantly neutral Universe at a redshift of 7.5},} \nat, 553, 473, \dodoi{10.1038/nature25180}

\bibitem[{R. {Begley} {et~al.}(2025){Begley}, {McLure}, {Cullen}, {McLeod}, {Dunlop}, {Carnall}, {Stanton}, {Shapley}, {Cochrane}, {Donnan}, {Ellis}, {Fontana}, {Grogin}, \& {Koekemoer}}]{Begley25}
{Begley}, R., {McLure}, R.~J., {Cullen}, F., {et~al.} 2025, \bibinfo{title}{{The evolution of [O III] + H{\ensuremath{\beta}} equivalent width from z ≃ 3-8: implications for the production and escape of ionizing photons during reionization},} \mnras, 537, 3245, \dodoi{10.1093/mnras/staf211}

\bibitem[{R. {Bezanson} {et~al.}(2024){Bezanson}, {Labbe}, {Whitaker}, {Leja}, {Price}, {Franx}, {Brammer}, {Marchesini}, {Zitrin}, {Wang}, {Weaver}, {Furtak}, {Atek}, {Coe}, {Cutler}, {Dayal}, {van Dokkum}, {Feldmann}, {F{\"o}rster Schreiber}, {Fujimoto}, {Geha}, {Glazebrook}, {de Graaff}, {Greene}, {Juneau}, {Kassin}, {Kriek}, {Khullar}, {Maseda}, {Mowla}, {Muzzin}, {Nanayakkara}, {Nelson}, {Oesch}, {Pacifici}, {Pan}, {Papovich}, {Setton}, {Shapley}, {Smit}, {Stefanon}, {Taylor}, \& {Williams}}]{Bezanson24}
{Bezanson}, R., {Labbe}, I., {Whitaker}, K.~E., {et~al.} 2024, \bibinfo{title}{{The JWST UNCOVER Treasury Survey: Ultradeep NIRSpec and NIRCam Observations before the Epoch of Reionization},} \apj, 974, 92, \dodoi{10.3847/1538-4357/ad66cf}

\bibitem[{P. {Bolan} {et~al.}(2022){Bolan}, {Lemaux}, {Mason}, {Brada{\v{c}}}, {Treu}, {Strait}, {Pelliccia}, {Pentericci}, \& {Malkan}}]{Bolan22}
{Bolan}, P., {Lemaux}, B.~C., {Mason}, C., {et~al.} 2022, \bibinfo{title}{{Inferring the intergalactic medium neutral fraction at z 6-8 with low-luminosity Lyman break galaxies},} \mnras, 517, 3263, \dodoi{10.1093/mnras/stac1963}

\bibitem[{J.~S. {Bolton} \& M.~G. {Haehnelt}(2013){Bolton} \& {Haehnelt}}]{BoltonHaehnelt13}
{Bolton}, J.~S., \& {Haehnelt}, M.~G. 2013, \bibinfo{title}{{On the rapid demise of Ly {\ensuremath{\alpha}} emitters at redshift z {\ensuremath{\gtrsim}} 7 due to the increasing incidence of optically thick absorption systems},} \mnras, 429, 1695, \dodoi{10.1093/mnras/sts455}

\bibitem[{S.~E.~I. {Bosman} {et~al.}(2022){Bosman}, {Davies}, {Becker}, {Keating}, {Davies}, {Zhu}, {Eilers}, {D'Odorico}, {Bian}, {Bischetti}, {Cristiani}, {Fan}, {Farina}, {Haehnelt}, {Hennawi}, {Kulkarni}, {Mesinger}, {Meyer}, {Onoue}, {Pallottini}, {Qin}, {Ryan-Weber}, {Schindler}, {Walter}, {Wang}, \& {Yang}}]{Bosman22}
{Bosman}, S. E.~I., {Davies}, F.~B., {Becker}, G.~D., {et~al.} 2022, \bibinfo{title}{{Hydrogen reionization ends by z = 5.3: Lyman-{\ensuremath{\alpha}} optical depth measured by the XQR-30 sample},} \mnras, 514, 55, \dodoi{10.1093/mnras/stac1046}

\bibitem[{S. {Bruton} {et~al.}(2023){Bruton}, {Lin}, {Scarlata}, \& {Hayes}}]{Br23}
{Bruton}, S., {Lin}, Y.-H., {Scarlata}, C., \& {Hayes}, M.~J. 2023, \bibinfo{title}{{The Universe is at Most 88\% Neutral at z=10.6},} arXiv e-prints, arXiv:2303.03419, \dodoi{10.48550/arXiv.2303.03419}

\bibitem[{A.~J. {Bunker} {et~al.}(2023){Bunker}, {Cameron}, {Curtis-Lake}, {Jakobsen}, {Carniani}, {Curti}, {Witstok}, {Maiolino}, {D'Eugenio}, {Looser}, {Willott}, {Bonaventura}, {Hainline}, {Uebler}, {Willmer}, {Saxena}, {Smit}, {Alberts}, {Arribas}, {Baker}, {Baum}, {Bhatawdekar}, {Bowler}, {Boyett}, {Charlot}, {Chen}, {Chevallard}, {Circosta}, {DeCoursey}, {de Graaff}, {Egami}, {Eisenstein}, {Endsley}, {Ferruit}, {Giardino}, {Hausen}, {Helton}, {Hviding}, {Ji}, {Johnson}, {Jones}, {Kumari}, {Laseter}, {Luetzgendorf}, {Maseda}, {Nelson}, {Parlanti}, {Perna}, {Rawle}, {Rix}, {Rieke}, {Robertson}, {Rodriguez Del Pino}, {Sandles}, {Scholtz}, {Sharpe}, {Skarbinski}, {Stark}, {Sun}, {Tacchella}, {Topping}, {Villanueva}, {Wallace}, {Williams}, \& {Woodrum}}]{Bun23}
{Bunker}, A.~J., {Cameron}, A.~J., {Curtis-Lake}, E., {et~al.} 2023, \bibinfo{title}{{JADES NIRSpec Initial Data Release for the Hubble Ultra Deep Field: Redshifts and Line Fluxes of Distant Galaxies from the Deepest JWST Cycle 1 NIRSpec Multi-Object Spectroscopy},} arXiv e-prints, arXiv:2306.02467, \dodoi{10.48550/arXiv.2306.02467}

\bibitem[{D. {Calzetti} {et~al.}(1994){Calzetti}, {Kinney}, \& {Storchi-Bergmann}}]{Calzetti94}
{Calzetti}, D., {Kinney}, A.~L., \& {Storchi-Bergmann}, T. 1994, \bibinfo{title}{{Dust Extinction of the Stellar Continua in Starburst Galaxies: The Ultraviolet and Optical Extinction Law},} \apj, 429, 582, \dodoi{10.1086/174346}

\bibitem[{A.~J. {Cameron} {et~al.}(2024){Cameron}, {Katz}, {Witten}, {Saxena}, {Laporte}, \& {Bunker}}]{Cameron24}
{Cameron}, A.~J., {Katz}, H., {Witten}, C., {et~al.} 2024, \bibinfo{title}{{Nebular dominated galaxies: insights into the stellar initial mass function at high redshift},} \mnras, 534, 523, \dodoi{10.1093/mnras/stae1547}

\bibitem[{M. {Castellano} {et~al.}(2024){Castellano}, {Napolitano}, {Fontana}, {Roberts-Borsani}, {Treu}, {Vanzella}, {Zavala}, {Arrabal Haro}, {Calabr{\`o}}, {Llerena}, {Mascia}, {Merlin}, {Paris}, {Pentericci}, {Santini}, {Bakx}, {Bergamini}, {Cupani}, {Dickinson}, {Filippenko}, {Glazebrook}, {Grillo}, {Kelly}, {Malkan}, {Mason}, {Morishita}, {Nanayakkara}, {Rosati}, {Sani}, {Wang}, \& {Yoon}}]{Castellano24}
{Castellano}, M., {Napolitano}, L., {Fontana}, A., {et~al.} 2024, \bibinfo{title}{{JWST NIRSpec Spectroscopy of the Remarkable Bright Galaxy GHZ2/GLASS-z12 at Redshift 12.34},} \apj, 972, 143, \dodoi{10.3847/1538-4357/ad5f88}

\bibitem[{Z. {Chen} {et~al.}(2024){Chen}, {Stark}, {Mason}, {Topping}, {Whitler}, {Tang}, {Endsley}, \& {Charlot}}]{Chen24}
{Chen}, Z., {Stark}, D.~P., {Mason}, C., {et~al.} 2024, \bibinfo{title}{{JWST spectroscopy of z 5-8 UV-selected galaxies: new constraints on the evolution of the Ly {\ensuremath{\alpha}} escape fraction in the reionization era},} \mnras, 528, 7052, \dodoi{10.1093/mnras/stae455}

\bibitem[{F. {Cullen} {et~al.}(2019){Cullen}, {McLure}, {Dunlop}, {Khochfar}, {Dav{\'e}}, {Amor{\'\i}n}, {Bolzonella}, {Carnall}, {Castellano}, {Cimatti}, {Cirasuolo}, {Cresci}, {Fynbo}, {Fontanot}, {Gargiulo}, {Garilli}, {Guaita}, {Hathi}, {Hibon}, {Mannucci}, {Marchi}, {McLeod}, {Pentericci}, {Pozzetti}, {Shapley}, {Talia}, \& {Zamorani}}]{Cu19}
{Cullen}, F., {McLure}, R.~J., {Dunlop}, J.~S., {et~al.} 2019, \bibinfo{title}{{The VANDELS survey: the stellar metallicities of star-forming galaxies at 2.5 < z < 5.0},} \mnras, 487, 2038, \dodoi{10.1093/mnras/stz1402}

\bibitem[{E. {Curtis-Lake} {et~al.}(2023){Curtis-Lake}, {Carniani}, {Cameron}, {Charlot}, {Jakobsen}, {Maiolino}, {Bunker}, {Witstok}, {Smit}, {Chevallard}, {Willott}, {Ferruit}, {Arribas}, {Bonaventura}, {Curti}, {D'Eugenio}, {Franx}, {Giardino}, {Looser}, {L{\"u}tzgendorf}, {Maseda}, {Rawle}, {Rix}, {Rodr{\'\i}guez del Pino}, {{\"U}bler}, {Sirianni}, {Dressler}, {Egami}, {Eisenstein}, {Endsley}, {Hainline}, {Hausen}, {Johnson}, {Rieke}, {Robertson}, {Shivaei}, {Stark}, {Tacchella}, {Williams}, {Willmer}, {Bhatawdekar}, {Bowler}, {Boyett}, {Chen}, {de Graaff}, {Helton}, {Hviding}, {Jones}, {Kumari}, {Lyu}, {Nelson}, {Perna}, {Sandles}, {Saxena}, {Suess}, {Sun}, {Topping}, {Wallace}, \& {Whitler}}]{CL23}
{Curtis-Lake}, E., {Carniani}, S., {Cameron}, A., {et~al.} 2023, \bibinfo{title}{{Spectroscopic confirmation of four metal-poor galaxies at z = 10.3-13.2},} Nature Astronomy, \dodoi{10.1038/s41550-023-01918-w}

\bibitem[{F.~B. {Davies} {et~al.}(2018){Davies}, {Hennawi}, {Ba{\~n}ados}, {Luki{\'c}}, {Decarli}, {Fan}, {Farina}, {Mazzucchelli}, {Rix}, {Venemans}, {Walter}, {Wang}, \& {Yang}}]{2018ApJ...864..142D}
{Davies}, F.~B., {Hennawi}, J.~F., {Ba{\~n}ados}, E., {et~al.} 2018, \bibinfo{title}{{Quantitative Constraints on the Reionization History from the IGM Damping Wing Signature in Two Quasars at z > 7},} \apj, 864, 142, \dodoi{10.3847/1538-4357/aad6dc}

\bibitem[{F. {D'Eugenio} {et~al.}(2024){D'Eugenio}, {Maiolino}, {Carniani}, {Chevallard}, {Curtis-Lake}, {Witstok}, {Charlot}, {Baker}, {Arribas}, {Boyett}, {Bunker}, {Curti}, {Eisenstein}, {Hainline}, {Ji}, {Johnson}, {Kumari}, {Looser}, {Nakajima}, {Nelson}, {Rieke}, {Robertson}, {Scholtz}, {Smit}, {Sun}, {Venturi}, {Tacchella}, {{\"U}bler}, {Willmer}, \& {Willott}}]{2024A&A...689A.152D}
{D'Eugenio}, F., {Maiolino}, R., {Carniani}, S., {et~al.} 2024, \bibinfo{title}{{JADES: Carbon enrichment 350 Myr after the Big Bang},} \aap, 689, A152, \dodoi{10.1051/0004-6361/202348636}

\bibitem[{F. {D'Eugenio} {et~al.}(2025){D'Eugenio}, {Cameron}, {Scholtz}, {Carniani}, {Willott}, {Curtis-Lake}, {Bunker}, {Parlanti}, {Maiolino}, {Willmer}, {Jakobsen}, {Robertson}, {Johnson}, {Tacchella}, {Cargile}, {Rawle}, {Arribas}, {Chevallard}, {Curti}, {Egami}, {Eisenstein}, {Kumari}, {Looser}, {Rieke}, {Rodr{\'\i}guez Del Pino}, {Saxena}, {{\"U}bler}, {Venturi}, {Witstok}, {Baker}, {Bhatawdekar}, {Bonaventura}, {Boyett}, {Charlot}, {Danhaive}, {Hainline}, {Hausen}, {Helton}, {Ji}, {Ji}, {Jones}, {Juod{\v{z}}balis}, {Maseda}, {P{\'e}rez-Gonz{\'a}lez}, {Perna}, {Pusk{\'a}s}, {Shivaei}, {Silcock}, {Simmonds}, {Smit}, {Sun}, {Villanueva}, {Williams}, \& {Zhu}}]{DEugenio25}
{D'Eugenio}, F., {Cameron}, A.~J., {Scholtz}, J., {et~al.} 2025, \bibinfo{title}{{JADES Data Release 3: NIRSpec/Microshutter Assembly Spectroscopy for 4000 Galaxies in the GOODS Fields},} \apjs, 277, 4, \dodoi{10.3847/1538-4365/ada148}

\bibitem[{D. {Dottorini} {et~al.}(2024){Dottorini}, {Calabr{\`o}}, {Pentericci}, {Mascia}, {Llerena}, {Napolitano}, {Santini}, {Roberts-Borsani}, {Castellano}, {Amor{\'\i}n}, {Dickinson}, {Fontana}, {Hathi}, {Hirschmann}, {Koekemoer}, {Lucas}, {Merlin}, {Morales}, {Pacucci}, {Wilkins}, {Arrabal Haro}, {Bagley}, {Finkelstein}, {Kartaltepe}, {Papovich}, \& {Pirzkal}}]{Dottorini24}
{Dottorini}, D., {Calabr{\`o}}, A., {Pentericci}, L., {et~al.} 2024, \bibinfo{title}{{Evolution of the UV slope of galaxies at cosmic morning (z > 4): the properties of extremely blue galaxies},} arXiv e-prints, arXiv:2412.01623, \dodoi{10.48550/arXiv.2412.01623}

\bibitem[{X. {Fan} {et~al.}(2006){Fan}, {Strauss}, {Becker}, {White}, {Gunn}, {Knapp}, {Richards}, {Schneider}, {Brinkmann}, \& {Fukugita}}]{2006AJ....132..117F}
{Fan}, X., {Strauss}, M.~A., {Becker}, R.~H., {et~al.} 2006, \bibinfo{title}{{Constraining the Evolution of the Ionizing Background and the Epoch of Reionization with z\raisebox{-0.5ex}\textasciitilde6 Quasars. II. A Sample of 19 Quasars},} \aj, 132, 117, \dodoi{10.1086/504836}

\bibitem[{A. {Ferrara}(2024){Ferrara}}]{Ferrara24}
{Ferrara}, A. 2024, \bibinfo{title}{{Super-early JWST galaxies, outflows, and Ly{\ensuremath{\alpha}} visibility in the Epoch of Reionization},} \aap, 684, A207, \dodoi{10.1051/0004-6361/202348321}

\bibitem[{S.~L. {Finkelstein} {et~al.}(2023){Finkelstein}, {Bagley}, {Ferguson}, {Wilkins}, {Kartaltepe}, {Papovich}, {Yung}, {Arrabal Haro}, {Behroozi}, {Dickinson}, {Kocevski}, {Koekemoer}, {Larson}, {Le Bail}, {Morales}, {P{\'e}rez-Gonz{\'a}lez}, {Burgarella}, {Dav{\'e}}, {Hirschmann}, {Somerville}, {Wuyts}, {Bromm}, {Casey}, {Fontana}, {Fujimoto}, {Gardner}, {Giavalisco}, {Grazian}, {Grogin}, {Hathi}, {Hutchison}, {Jha}, {Jogee}, {Kewley}, {Kirkpatrick}, {Long}, {Lotz}, {Pentericci}, {Pierel}, {Pirzkal}, {Ravindranath}, {Ryan}, {Trump}, {Yang}, {Bhatawdekar}, {Bisigello}, {Buat}, {Calabr{\`o}}, {Castellano}, {Cleri}, {Cooper}, {Croton}, {Daddi}, {Dekel}, {Elbaz}, {Franco}, {Gawiser}, {Holwerda}, {Huertas-Company}, {Jaskot}, {Leung}, {Lucas}, {Mobasher}, {Pandya}, {Tacchella}, {Weiner}, \& {Zavala}}]{Fink23}
{Finkelstein}, S.~L., {Bagley}, M.~B., {Ferguson}, H.~C., {et~al.} 2023, \bibinfo{title}{{CEERS Key Paper. I. An Early Look into the First 500 Myr of Galaxy Formation with JWST},} \apjl, 946, L13, \dodoi{10.3847/2041-8213/acade4}

\bibitem[{D. {Foreman-Mackey} {et~al.}(2013){Foreman-Mackey}, {Hogg}, {Lang}, \& {Goodman}}]{2013PASP..125..306F}
{Foreman-Mackey}, D., {Hogg}, D.~W., {Lang}, D., \& {Goodman}, J. 2013, \bibinfo{title}{{emcee: The MCMC Hammer},} \pasp, 125, 306, \dodoi{10.1086/670067}

\bibitem[{S. {Fujimoto} {et~al.}(2024){Fujimoto}, {Wang}, {Weaver}, {Kokorev}, {Atek}, {Bezanson}, {Labbe}, {Brammer}, {Greene}, {Chemerynska}, {Dayal}, {de Graaff}, {Furtak}, {Oesch}, {Setton}, {Price}, {Miller}, {Williams}, {Whitaker}, {Zitrin}, {Cutler}, {Leja}, {Pan}, {Coe}, {van Dokkum}, {Feldmann}, {Fudamoto}, {Goulding}, {Khullar}, {Marchesini}, {Maseda}, {Nanayakkara}, {Nelson}, {Smit}, {Stefanon}, \& {Weibel}}]{Fu23}
{Fujimoto}, S., {Wang}, B., {Weaver}, J.~R., {et~al.} 2024, \bibinfo{title}{{UNCOVER: A NIRSpec Census of Lensed Galaxies at z = 8.50{\textendash}13.08 Probing a High-AGN Fraction and Ionized Bubbles in the Shadow},} \apj, 977, 250, \dodoi{10.3847/1538-4357/ad9027}

\bibitem[{S.~R. {Furlanetto} \& S.~P. {Oh}(2005){Furlanetto} \& {Oh}}]{FO05}
{Furlanetto}, S.~R., \& {Oh}, S.~P. 2005, \bibinfo{title}{{Taxing the rich: recombinations and bubble growth during reionization},} \mnras, 363, 1031, \dodoi{10.1111/j.1365-2966.2005.09505.x}

\bibitem[{L.~J. {Furtak} {et~al.}(2023){Furtak}, {Zitrin}, {Weaver}, {Atek}, {Bezanson}, {Labb{\'e}}, {Whitaker}, {Leja}, {Price}, {Brammer}, {Wang}, {Marchesini}, {Pan}, {Dayal}, {van Dokkum}, {Feldmann}, {Fujimoto}, {Franx}, {Khullar}, {Nelson}, \& {Mowla}}]{Furtak23}
{Furtak}, L.~J., {Zitrin}, A., {Weaver}, J.~R., {et~al.} 2023, \bibinfo{title}{{UNCOVERing the extended strong lensing structures of Abell 2744 with the deepest JWST imaging},} \mnras, 523, 4568, \dodoi{10.1093/mnras/stad1627}

\bibitem[{B. {Greig} \& A. {Mesinger}(2018){Greig} \& {Mesinger}}]{Greig18}
{Greig}, B., \& {Mesinger}, A. 2018, \bibinfo{title}{{21CMMC with a 3D light-cone: the impact of the co-evolution approximation on the astrophysics of reionization and cosmic dawn},} \mnras, 477, 3217, \dodoi{10.1093/mnras/sty796}

\bibitem[{B. {Greig} {et~al.}(2019){Greig}, {Mesinger}, \& {Ba{\~n}ados}}]{2019MNRAS.484.5094G}
{Greig}, B., {Mesinger}, A., \& {Ba{\~n}ados}, E. 2019, \bibinfo{title}{{Constraints on reionization from the z = 7.5 QSO ULASJ1342+0928},} \mnras, 484, 5094, \dodoi{10.1093/mnras/stz230}

\bibitem[{M. {Gronke} {et~al.}(2021){Gronke}, {Ocvirk}, {Mason}, {Matthee}, {Bosman}, {Sorce}, {Lewis}, {Ahn}, {Aubert}, {Dawoodbhoy}, {Iliev}, {Shapiro}, \& {Yepes}}]{Gronke21}
{Gronke}, M., {Ocvirk}, P., {Mason}, C., {et~al.} 2021, \bibinfo{title}{{Lyman-{\ensuremath{\alpha}} transmission properties of the intergalactic medium in the CoDaII simulation},} \mnras, 508, 3697, \dodoi{10.1093/mnras/stab2762}

\bibitem[{J.~E. {Gunn} \& B.~A. {Peterson}(1965){Gunn} \& {Peterson}}]{GP65}
{Gunn}, J.~E., \& {Peterson}, B.~A. 1965, \bibinfo{title}{{On the Density of Neutral Hydrogen in Intergalactic Space.},} \apj, 142, 1633, \dodoi{10.1086/148444}

\bibitem[{Y. {Harikane} {et~al.}(2023){Harikane}, {Nakajima}, {Ouchi}, {Umeda}, {Isobe}, {Ono}, {Xu}, \& {Zhang}}]{Ha23}
{Harikane}, Y., {Nakajima}, K., {Ouchi}, M., {et~al.} 2023, \bibinfo{title}{{Pure Spectroscopic Constraints on UV Luminosity Functions and Cosmic Star Formation History From 25 Galaxies at $z_\mathrm{spec}=8.61-13.20$ Confirmed with JWST/NIRSpec},} arXiv e-prints, arXiv:2304.06658, \dodoi{10.48550/arXiv.2304.06658}

\bibitem[{Y. {Harikane} {et~al.}(2024){Harikane}, {Nakajima}, {Ouchi}, {Umeda}, {Isobe}, {Ono}, {Xu}, \& {Zhang}}]{Harikane24b}
{Harikane}, Y., {Nakajima}, K., {Ouchi}, M., {et~al.} 2024, \bibinfo{title}{{Pure Spectroscopic Constraints on UV Luminosity Functions and Cosmic Star Formation History from 25 Galaxies at z $_{spec}$ = 8.61-13.20 Confirmed with JWST/NIRSpec},} \apj, 960, 56, \dodoi{10.3847/1538-4357/ad0b7e}

\bibitem[{K.~E. {Heintz} {et~al.}(2023){Heintz}, {Watson}, {Brammer}, {Vejlgaard}, {Hutter}, {Strait}, {Matthee}, {Oesch}, {Jakobsson}, {Tanvir}, {Laursen}, {Naidu}, {Mason}, {Killi}, {Jung}, {Hsiao}, {Abdurro'uf}, {Coe}, {Arrabal Haro}, {Finkelstein}, \& {Toft}}]{Heintz23}
{Heintz}, K.~E., {Watson}, D., {Brammer}, G., {et~al.} 2023, \bibinfo{title}{{Extreme damped Lyman-$\alpha$ absorption in young star-forming galaxies at $z=9-11$},} arXiv e-prints, arXiv:2306.00647, \dodoi{10.48550/arXiv.2306.00647}

\bibitem[{K.~E. {Heintz} {et~al.}(2025){Heintz}, {Brammer}, {Watson}, {Oesch}, {Keating}, {Hayes}, {Abdurro'uf}, {Arellano-C{\'o}rdova}, {Carnall}, {Christiansen}, {Cullen}, {Dav{\'e}}, {Dayal}, {Ferrara}, {Finlator}, {Fynbo}, {Flury}, {Gelli}, {Gillman}, {Gottumukkala}, {Gould}, {Greve}, {Hardin}, {Hsiao}, {Hutter}, {Jakobsson}, {Killi}, {Khosravaninezhad}, {Laursen}, {Lee}, {Magdis}, {Matthee}, {Naidu}, {Narayanan}, {Pollock}, {Prescott}, {Rusakov}, {Shuntov}, {Sneppen}, {Smit}, {Tanvir}, {Terp}, {Toft}, {Valentino}, {Vijayan}, {Weaver}, {Wise}, \& {Witstok}}]{2025A&A...693A..60H}
{Heintz}, K.~E., {Brammer}, G.~B., {Watson}, D., {et~al.} 2025, \bibinfo{title}{{The JWST-PRIMAL archival survey: A JWST/NIRSpec reference sample for the physical properties and Lyman-{\ensuremath{\alpha}} absorption and emission of {\ensuremath{\sim}}600 galaxies at z = 5.0 ‑ 13.4},} \aap, 693, A60, \dodoi{10.1051/0004-6361/202450243}

\bibitem[{A. {Hoag} {et~al.}(2019){Hoag}, {Brada{\v{c}}}, {Huang}, {Mason}, {Treu}, {Schmidt}, {Trenti}, {Strait}, {Lemaux}, {Finney}, \& {Paddock}}]{2019ApJ...878...12H}
{Hoag}, A., {Brada{\v{c}}}, M., {Huang}, K., {et~al.} 2019, \bibinfo{title}{{Constraining the Neutral Fraction of Hydrogen in the IGM at Redshift 7.5},} \apj, 878, 12, \dodoi{10.3847/1538-4357/ab1de7}

\bibitem[{T.~Y.-Y. {Hsiao} {et~al.}(2023{\natexlab{a}}){Hsiao}, {Abdurro'uf}, {Coe}, {Larson}, {Jung}, {Mingozzi}, {Dayal}, {Kumari}, {Kokorev}, {Vikaeus}, {Brammer}, {Furtak}, {Adamo}, {Andrade-Santos}, {Antwi-Danso}, {Bradac}, {Bradley}, {Broadhurst}, {Carnall}, {Conselice}, {Diego}, {Donahue}, {Eldridge}, {Fujimoto}, {Henry}, {Hernandez}, {Hutchison}, {James}, {Norman}, {Park}, {Pirzkal}, {Postman}, {Ricotti}, {Rigby}, {Vanzella}, {Welch}, {Wilkins}, {Windhorst}, {Xu}, {Zackrisson}, \& {Zitrin}}]{Hsiao23}
{Hsiao}, T. Y.-Y., {Abdurro'uf}, {Coe}, D., {et~al.} 2023{\natexlab{a}}, \bibinfo{title}{{JWST NIRSpec spectroscopy of the triply-lensed $z = 10.17$ galaxy MACS0647$-$JD},} arXiv e-prints, arXiv:2305.03042, \dodoi{10.48550/arXiv.2305.03042}

\bibitem[{T.~Y.-Y. {Hsiao} {et~al.}(2023{\natexlab{b}}){Hsiao}, {Abdurro'uf}, {Coe}, {Larson}, {Jung}, {Mingozzi}, {Dayal}, {Kumari}, {Kokorev}, {Vikaeus}, {Brammer}, {Furtak}, {Adamo}, {Andrade-Santos}, {Antwi-Danso}, {Bradac}, {Bradley}, {Broadhurst}, {Carnall}, {Conselice}, {Diego}, {Donahue}, {Eldridge}, {Fujimoto}, {Henry}, {Hernandez}, {Hutchison}, {James}, {Norman}, {Park}, {Pirzkal}, {Postman}, {Ricotti}, {Rigby}, {Vanzella}, {Welch}, {Wilkins}, {Windhorst}, {Xu}, {Zackrisson}, \& {Zitrin}}]{H23}
{Hsiao}, T. Y.-Y., {Abdurro'uf}, {Coe}, D., {et~al.} 2023{\natexlab{b}}, \bibinfo{title}{{JWST NIRSpec spectroscopy of the triply-lensed $z = 10.17$ galaxy MACS0647$-$JD},} arXiv e-prints, arXiv:2305.03042, \dodoi{10.48550/arXiv.2305.03042}

\bibitem[{M. {Huberty} {et~al.}(2025){Huberty}, {Scarlata}, {Hayes}, \& {Gazagnes}}]{Huberty25}
{Huberty}, M., {Scarlata}, C., {Hayes}, M.~J., \& {Gazagnes}, S. 2025, \bibinfo{title}{{The Pitfalls of Using Lyman Alpha Damping Wings in High-z Galaxy Spectra to Measure the Intergalactic Neutral Hydrogen Fraction},} arXiv e-prints, arXiv:2501.13899, \dodoi{10.48550/arXiv.2501.13899}

\bibitem[{A.~K. {Inoue} {et~al.}(2018){Inoue}, {Hasegawa}, {Ishiyama}, {Yajima}, {Shimizu}, {Umemura}, {Konno}, {Harikane}, {Shibuya}, {Ouchi}, {Shimasaku}, {Ono}, {Kusakabe}, {Higuchi}, \& {Lee}}]{2018PASJ...70...55I}
{Inoue}, A.~K., {Hasegawa}, K., {Ishiyama}, T., {et~al.} 2018, \bibinfo{title}{{SILVERRUSH. VI. A simulation of Ly{\ensuremath{\alpha}} emitters in the reionization epoch and a comparison with Subaru Hyper Suprime-Cam survey early data},} \pasj, 70, 55, \dodoi{10.1093/pasj/psy048}

\bibitem[{R. {Ishimoto} {et~al.}(2022){Ishimoto}, {Kashikawa}, {Kashino}, {Ito}, {Liang}, {Cai}, {Yoshioka}, {Okoshi}, {Misawa}, {Onoue}, {Takeda}, \& {Uchiyama}}]{Ishimoto22}
{Ishimoto}, R., {Kashikawa}, N., {Kashino}, D., {et~al.} 2022, \bibinfo{title}{{The physical origin for spatially large scatter of IGM opacity at the end of reionization: The IGM Ly{\ensuremath{\alpha}} opacity-galaxy density relation},} \mnras, 515, 5914, \dodoi{10.1093/mnras/stac1972}

\bibitem[{P. {Jakobsen} {et~al.}(2022){Jakobsen}, {Ferruit}, {Alves de Oliveira}, {Arribas}, {Bagnasco}, {Barho}, {Beck}, {Birkmann}, {B{\"o}ker}, {Bunker}, {Charlot}, {de Jong}, {de Marchi}, {Ehrenwinkler}, {Falcolini}, {Fels}, {Franx}, {Franz}, {Funke}, {Giardino}, {Gnata}, {Holota}, {Honnen}, {Jensen}, {Jentsch}, {Johnson}, {Jollet}, {Karl}, {Kling}, {K{\"o}hler}, {Kolm}, {Kumari}, {Lander}, {Lemke}, {L{\'o}pez-Caniego}, {L{\"u}tzgendorf}, {Maiolino}, {Manjavacas}, {Marston}, {Maschmann}, {Maurer}, {Messerschmidt}, {Moseley}, {Mosner}, {Mott}, {Muzerolle}, {Pirzkal}, {Pittet}, {Plitzke}, {Posselt}, {Rapp}, {Rauscher}, {Rawle}, {Rix}, {R{\"o}del}, {Rumler}, {Sabbi}, {Salvignol}, {Schmid}, {Sirianni}, {Smith}, {Strada}, {te Plate}, {Valenti}, {Wettemann}, {Wiehe}, {Wiesmayer}, {Willott}, {Wright}, {Zeidler}, \& {Zincke}}]{2022A&A...661A..80J}
{Jakobsen}, P., {Ferruit}, P., {Alves de Oliveira}, C., {et~al.} 2022, \bibinfo{title}{{The Near-Infrared Spectrograph (NIRSpec) on the James Webb Space Telescope. I. Overview of the instrument and its capabilities},} \aap, 661, A80, \dodoi{10.1051/0004-6361/202142663}

\bibitem[{X. {Jin} {et~al.}(2023){Jin}, {Yang}, {Fan}, {Wang}, {Ba{\~n}ados}, {Bian}, {Davies}, {Eilers}, {Farina}, {Hennawi}, {Pacucci}, {Venemans}, \& {Walter}}]{2023ApJ...942...59J}
{Jin}, X., {Yang}, J., {Fan}, X., {et~al.} 2023, \bibinfo{title}{{(Nearly) Model-independent Constraints on the Neutral Hydrogen Fraction in the Intergalactic Medium at z 5-7 Using Dark Pixel Fractions in Ly{\ensuremath{\alpha}} and Ly{\ensuremath{\beta}} Forests},} \apj, 942, 59, \dodoi{10.3847/1538-4357/aca678}

\bibitem[{X. {Jin} {et~al.}(2024){Jin}, {Yang}, {Fan}, {Wang}, {Kakiichi}, {Meyer}, {Becker}, {Zou}, {Ba{\~n}ados}, {Champagne}, {D'Odorico}, {Yue}, {Bosman}, {Cai}, {Eilers}, {Hennawi}, {Jun}, {Li}, {Li}, {Liu}, {Pudoka}, {Satyavolu}, {Sun}, {Tee}, \& {Wu}}]{2024ApJ...976...93J}
{Jin}, X., {Yang}, J., {Fan}, X., {et~al.} 2024, \bibinfo{title}{{A SPectroscopic Survey of Biased Halos In the Reionization Era (ASPIRE): JWST Supports Earlier Reionization around [O III] Emitters},} \apj, 976, 93, \dodoi{10.3847/1538-4357/ad82de}

\bibitem[{G.~C. {Jones} {et~al.}(2024){Jones}, {Bunker}, {Saxena}, {Arribas}, {Bhatawdekar}, {Boyett}, {Carniani}, {Charlot}, {Curtis-Lake}, {Hainline}, {Johnson}, {Kumari}, {Maseda}, {Rix}, {Robertson}, {Tacchella}, {{\"U}bler}, {Williams}, {Willott}, {Witstok}, \& {Zhu}}]{Jones24}
{Jones}, G.~C., {Bunker}, A.~J., {Saxena}, A., {et~al.} 2024, \bibinfo{title}{{JADES: Measuring reionization properties using Lyman-alpha emission},} arXiv e-prints, arXiv:2409.06405, \dodoi{10.48550/arXiv.2409.06405}

\bibitem[{I. {Jung} {et~al.}(2020){Jung}, {Finkelstein}, {Dickinson}, {Hutchison}, {Larson}, {Papovich}, {Pentericci}, {Straughn}, {Guo}, {Malhotra}, {Rhoads}, {Song}, {Tilvi}, \& {Wold}}]{2020ApJ...904..144J}
{Jung}, I., {Finkelstein}, S.~L., {Dickinson}, M., {et~al.} 2020, \bibinfo{title}{{Texas Spectroscopic Search for Ly{\ensuremath{\alpha}} Emission at the End of Reionization. III. The Ly{\ensuremath{\alpha}} Equivalent-width Distribution and Ionized Structures at z > 7},} \apj, 904, 144, \dodoi{10.3847/1538-4357/abbd44}

\bibitem[{Y. {Kageura} {et~al.}(2025){Kageura}, {Ouchi}, {Nakane}, {Umeda}, {Harikane}, {Yoshiura}, {Nakajima}, {Yajima}, \& {Thai}}]{Kageura25}
{Kageura}, Y., {Ouchi}, M., {Nakane}, M., {et~al.} 2025, \bibinfo{title}{{Census of Ly$\alpha$ Emission from $\sim 600$ Galaxies at $z=5-14$: Evolution of the Ly$\alpha$ Luminosity Function and a Late Sharp Cosmic Reionization},} arXiv e-prints, arXiv:2501.05834, \dodoi{10.48550/arXiv.2501.05834}

\bibitem[{K. {Kakiichi} {et~al.}(2025){Kakiichi}, {Jin}, {Wang}, {Meyer}, {Garaldi}, {Bosman}, {Davies}, {Fan}, {Trebitsch}, {Yang}, {Ba{\~n}ados}, {Champagne}, {Eilers}, {Hennawi}, {Sun}, {Wu}, {Zou}, {Kannan}, {Smith}, {Becker}, {D'Odorico}, {Connor}, {Liu}, {Protu{\v{s}}ov{\'a}}, {Walter}, \& {Zhang}}]{2025arXiv250307074K}
{Kakiichi}, K., {Jin}, X., {Wang}, F., {et~al.} 2025, \bibinfo{title}{{JWST ASPIRE: How Did Galaxies Complete Reionization? Evidence for Excess IGM Transmission around ${\rm [O\,{\scriptstyle III}]}$ Emitters during Reionization},} arXiv e-prints, arXiv:2503.07074, \dodoi{10.48550/arXiv.2503.07074}

\bibitem[{D. {Kashino} {et~al.}(2023){Kashino}, {Lilly}, {Matthee}, {Eilers}, {Mackenzie}, {Bordoloi}, \& {Simcoe}}]{2023ApJ...950...66K}
{Kashino}, D., {Lilly}, S.~J., {Matthee}, J., {et~al.} 2023, \bibinfo{title}{{EIGER. I. A Large Sample of [O III]-emitting Galaxies at 5.3 < z < 6.9 and Direct Evidence for Local Reionization by Galaxies},} \apj, 950, 66, \dodoi{10.3847/1538-4357/acc588}

\bibitem[{H. {Katz} {et~al.}(2024){Katz}, {Cameron}, {Saxena}, {Barrufet}, {Choustikov}, {Cleri}, {de Graaff}, {Ellis}, {Fosbury}, {Heintz}, {Maseda}, {Matthee}, {McConchie}, \& {Oesch}}]{Katz24}
{Katz}, H., {Cameron}, A.~J., {Saxena}, A., {et~al.} 2024, \bibinfo{title}{{21 Balmer Jump Street: The Nebular Continuum at High Redshift and Implications for the Bright Galaxy Problem, UV Continuum Slopes, and Early Stellar Populations},} arXiv e-prints, arXiv:2408.03189, \dodoi{10.48550/arXiv.2408.03189}

\bibitem[{L.~C. {Keating} {et~al.}(2023){Keating}, {Bolton}, {Cullen}, {Haehnelt}, {Puchwein}, \& {Kulkarni}}]{2023arXiv230805800K}
{Keating}, L.~C., {Bolton}, J.~S., {Cullen}, F., {et~al.} 2023, \bibinfo{title}{{JWST observations of galaxy damping wings during reionization interpreted with cosmological simulations},} arXiv e-prints, arXiv:2308.05800, \dodoi{10.48550/arXiv.2308.05800}

\bibitem[{A. {Konno} {et~al.}(2014){Konno}, {Ouchi}, {Ono}, {Shimasaku}, {Shibuya}, {Furusawa}, {Nakajima}, {Naito}, {Momose}, {Yuma}, \& {Iye}}]{Konno14}
{Konno}, A., {Ouchi}, M., {Ono}, Y., {et~al.} 2014, \bibinfo{title}{{Accelerated Evolution of the Ly{\ensuremath{\alpha}} Luminosity Function at z >\raisebox{-0.5ex}\textasciitilde 7 Revealed by the Subaru Ultra-deep Survey for Ly{\ensuremath{\alpha}} Emitters at z = 7.3},} \apj, 797, 16, \dodoi{10.1088/0004-637X/797/1/16}

\bibitem[{P. {Laursen} {et~al.}(2011){Laursen}, {Sommer-Larsen}, \& {Razoumov}}]{Laursen11}
{Laursen}, P., {Sommer-Larsen}, J., \& {Razoumov}, A.~O. 2011, \bibinfo{title}{{Intergalactic Transmission and Its Impact on the Ly{\ensuremath{\alpha}} Line},} \apj, 728, 52, \dodoi{10.1088/0004-637X/728/1/52}

\bibitem[{T.-Y. {Lu} {et~al.}(2023){Lu}, {Mason}, {Hutter}, {Mesinger}, {Qin}, {Stark}, \& {Endsley}}]{Lu23}
{Lu}, T.-Y., {Mason}, C., {Hutter}, A., {et~al.} 2023, \bibinfo{title}{{The reionising bubble size distribution around galaxies},} arXiv e-prints, arXiv:2304.11192, \dodoi{10.48550/arXiv.2304.11192}

\bibitem[{P. {Madau}(1995){Madau}}]{Madau95}
{Madau}, P. 1995, \bibinfo{title}{{Radiative Transfer in a Clumpy Universe: The Colors of High-Redshift Galaxies},} \apj, 441, 18, \dodoi{10.1086/175332}

\bibitem[{P. {Madau} {et~al.}(2024){Madau}, {Giallongo}, {Grazian}, \& {Haardt}}]{Madau24}
{Madau}, P., {Giallongo}, E., {Grazian}, A., \& {Haardt}, F. 2024, \bibinfo{title}{{Cosmic Reionization in the JWST Era: Back to AGNs?},} \apj, 971, 75, \dodoi{10.3847/1538-4357/ad5ce8}

\bibitem[{C.~A. {Mason} {et~al.}(2025){Mason}, {Chen}, {Stark}, {Lu}, {Topping}, \& {Tang}}]{Mason25}
{Mason}, C.~A., {Chen}, Z., {Stark}, D.~P., {et~al.} 2025, \bibinfo{title}{{Constraints on the $z\sim6-13$ intergalactic medium from JWST spectroscopy of Lyman-alpha damping wings in galaxies},} arXiv e-prints, arXiv:2501.11702, \dodoi{10.48550/arXiv.2501.11702}

\bibitem[{C.~A. {Mason} {et~al.}(2015){Mason}, {Trenti}, \& {Treu}}]{Mason15}
{Mason}, C.~A., {Trenti}, M., \& {Treu}, T. 2015, \bibinfo{title}{{The Galaxy UV Luminosity Function before the Epoch of Reionization},} \apj, 813, 21, \dodoi{10.1088/0004-637X/813/1/21}

\bibitem[{C.~A. {Mason} {et~al.}(2018{\natexlab{a}}){Mason}, {Treu}, {Dijkstra}, {Mesinger}, {Trenti}, {Pentericci}, {de Barros}, \& {Vanzella}}]{Mason18}
{Mason}, C.~A., {Treu}, T., {Dijkstra}, M., {et~al.} 2018{\natexlab{a}}, \bibinfo{title}{{The Universe Is Reionizing at z {\ensuremath{\sim}} 7: Bayesian Inference of the IGM Neutral Fraction Using Ly{\ensuremath{\alpha}} Emission from Galaxies},} \apj, 856, 2, \dodoi{10.3847/1538-4357/aab0a7}

\bibitem[{C.~A. {Mason} {et~al.}(2018{\natexlab{b}}){Mason}, {Treu}, {de Barros}, {Dijkstra}, {Fontana}, {Mesinger}, {Pentericci}, {Trenti}, \& {Vanzella}}]{Mason18b}
{Mason}, C.~A., {Treu}, T., {de Barros}, S., {et~al.} 2018{\natexlab{b}}, \bibinfo{title}{{Beacons into the Cosmic Dark Ages: Boosted Transmission of Ly{\ensuremath{\alpha}} from UV Bright Galaxies at z {\ensuremath{\gtrsim}} 7},} \apjl, 857, L11, \dodoi{10.3847/2041-8213/aabbab}

\bibitem[{C.~A. {Mason} {et~al.}(2019){Mason}, {Fontana}, {Treu}, {Schmidt}, {Hoag}, {Abramson}, {Amorin}, {Brada{\v{c}}}, {Guaita}, {Jones}, {Henry}, {Malkan}, {Pentericci}, {Trenti}, \& {Vanzella}}]{2019MNRAS.485.3947M}
{Mason}, C.~A., {Fontana}, A., {Treu}, T., {et~al.} 2019, \bibinfo{title}{{Inferences on the timeline of reionization at z {\ensuremath{\sim}} 8 from the KMOS Lens-Amplified Spectroscopic Survey},} \mnras, 485, 3947, \dodoi{10.1093/mnras/stz632}

\bibitem[{J. {Matthee} {et~al.}(2022){Matthee}, {Naidu}, {Pezzulli}, {Gronke}, {Sobral}, {Oesch}, {Hayes}, {Erb}, {Schaerer}, {Amor{\'\i}n}, {Tacchella}, {Paulino-Afonso}, {Llerena}, {Calhau}, \& {R{\"o}ttgering}}]{Matthee22}
{Matthee}, J., {Naidu}, R.~P., {Pezzulli}, G., {et~al.} 2022, \bibinfo{title}{{(Re)Solving reionization with Ly{\ensuremath{\alpha}}: how bright Ly{\ensuremath{\alpha}} Emitters account for the z {\ensuremath{\approx}} 2-8 cosmic ionizing background},} \mnras, 512, 5960, \dodoi{10.1093/mnras/stac801}

\bibitem[{M. {McQuinn} {et~al.}(2007){McQuinn}, {Hernquist}, {Zaldarriaga}, \& {Dutta}}]{McQuinn07}
{McQuinn}, M., {Hernquist}, L., {Zaldarriaga}, M., \& {Dutta}, S. 2007, \bibinfo{title}{{Studying reionization with Ly{\ensuremath{\alpha}} emitters},} \mnras, 381, 75, \dodoi{10.1111/j.1365-2966.2007.12085.x}

\bibitem[{A. {Mesinger} {et~al.}(2015{\natexlab{a}}){Mesinger}, {Aykutalp}, {Vanzella}, {Pentericci}, {Ferrara}, \& {Dijkstra}}]{Mesinger15}
{Mesinger}, A., {Aykutalp}, A., {Vanzella}, E., {et~al.} 2015{\natexlab{a}}, \bibinfo{title}{{Can the intergalactic medium cause a rapid drop in Ly{\ensuremath{\alpha}} emission at z > 6?},} \mnras, 446, 566, \dodoi{10.1093/mnras/stu2089}

\bibitem[{A. {Mesinger} {et~al.}(2015{\natexlab{b}}){Mesinger}, {Aykutalp}, {Vanzella}, {Pentericci}, {Ferrara}, \& {Dijkstra}}]{2015MNRAS.446..566M}
{Mesinger}, A., {Aykutalp}, A., {Vanzella}, E., {et~al.} 2015{\natexlab{b}}, \bibinfo{title}{{Can the intergalactic medium cause a rapid drop in Ly{\ensuremath{\alpha}} emission at z > 6?},} \mnras, 446, 566, \dodoi{10.1093/mnras/stu2089}

\bibitem[{A. {Mesinger} \& S.~R. {Furlanetto}(2008{\natexlab{a}}){Mesinger} \& {Furlanetto}}]{MF08a}
{Mesinger}, A., \& {Furlanetto}, S.~R. 2008{\natexlab{a}}, \bibinfo{title}{{Ly{\ensuremath{\alpha}} damping wing constraints on inhomogeneous reionization},} \mnras, 385, 1348, \dodoi{10.1111/j.1365-2966.2007.12836.x}

\bibitem[{A. {Mesinger} \& S.~R. {Furlanetto}(2008{\natexlab{b}}){Mesinger} \& {Furlanetto}}]{Mesinger08}
{Mesinger}, A., \& {Furlanetto}, S.~R. 2008{\natexlab{b}}, \bibinfo{title}{{Ly{\ensuremath{\alpha}} damping wing constraints on inhomogeneous reionization},} \mnras, 385, 1348, \dodoi{10.1111/j.1365-2966.2007.12836.x}

\bibitem[{A. {Mesinger} {et~al.}(2016){Mesinger}, {Greig}, \& {Sobacchi}}]{Mesinger16}
{Mesinger}, A., {Greig}, B., \& {Sobacchi}, E. 2016, \bibinfo{title}{{The Evolution Of 21 cm Structure (EOS): public, large-scale simulations of Cosmic Dawn and reionization},} \mnras, 459, 2342, \dodoi{10.1093/mnras/stw831}

\bibitem[{A. {Mesinger} \& Z. {Haiman}(2004){Mesinger} \& {Haiman}}]{MesingerHaiman04}
{Mesinger}, A., \& {Haiman}, Z. 2004, \bibinfo{title}{{Evidence of a Cosmological Str{\"o}mgren Surface and of Significant Neutral Hydrogen Surrounding the Quasar SDSS J1030+0524},} \apjl, 611, L69, \dodoi{10.1086/423935}

\bibitem[{R.~A. {Meyer} {et~al.}(2025){Meyer}, {Roberts-Borsani}, {Oesch}, \& {Ellis}}]{Meyer25}
{Meyer}, R.~A., {Roberts-Borsani}, G., {Oesch}, P., \& {Ellis}, R.~S. 2025, \bibinfo{title}{{Probing patchy reionisation with JWST: IGM opacity constraints from the Lyman-$α$ forest of galaxies in legacy extragalactic fields},} arXiv e-prints, arXiv:2504.02683.
\newblock \doarXiv{2504.02683}

\bibitem[{J. {Miralda-Escud{\'e}}(1998){Miralda-Escud{\'e}}}]{ME98}
{Miralda-Escud{\'e}}, J. 1998, \bibinfo{title}{{Reionization of the Intergalactic Medium and the Damping Wing of the Gunn-Peterson Trough},} \apj, 501, 15, \dodoi{10.1086/305799}

\bibitem[{A.~M. {Morales} {et~al.}(2021){Morales}, {Mason}, {Bruton}, {Gronke}, {Haardt}, \& {Scarlata}}]{Morales21}
{Morales}, A.~M., {Mason}, C.~A., {Bruton}, S., {et~al.} 2021, \bibinfo{title}{{The Evolution of the Lyman-alpha Luminosity Function during Reionization},} \apj, 919, 120, \dodoi{10.3847/1538-4357/ac1104}

\bibitem[{T. {Morishita} {et~al.}(2023){Morishita}, {Roberts-Borsani}, {Treu}, {Brammer}, {Mason}, {Trenti}, {Vulcani}, {Wang}, {Acebron}, {Bah{\'e}}, {Bergamini}, {Boyett}, {Bradac}, {Calabr{\`o}}, {Castellano}, {Chen}, {De Lucia}, {Filippenko}, {Fontana}, {Glazebrook}, {Grillo}, {Henry}, {Jones}, {Kelly}, {Koekemoer}, {Leethochawalit}, {Lu}, {Marchesini}, {Mascia}, {Mercurio}, {Merlin}, {Metha}, {Nanayakkara}, {Nonino}, {Paris}, {Pentericci}, {Rosati}, {Santini}, {Strait}, {Vanzella}, {Windhorst}, \& {Xie}}]{Mo23}
{Morishita}, T., {Roberts-Borsani}, G., {Treu}, T., {et~al.} 2023, \bibinfo{title}{{Early Results from GLASS-JWST. XIV. A Spectroscopically Confirmed Protocluster 650 Million Years after the Big Bang},} \apjl, 947, L24, \dodoi{10.3847/2041-8213/acb99e}

\bibitem[{D.~J. {Mortlock} {et~al.}(2011){Mortlock}, {Warren}, {Venemans}, {Patel}, {Hewett}, {McMahon}, {Simpson}, {Theuns}, {Gonz{\'a}les-Solares}, {Adamson}, {Dye}, {Hambly}, {Hirst}, {Irwin}, {Kuiper}, {Lawrence}, \& {R{\"o}ttgering}}]{M11}
{Mortlock}, D.~J., {Warren}, S.~J., {Venemans}, B.~P., {et~al.} 2011, \bibinfo{title}{{A luminous quasar at a redshift of z = 7.085},} \nat, 474, 616, \dodoi{10.1038/nature10159}

\bibitem[{S. {Murray} {et~al.}(2020){Murray}, {Greig}, {Mesinger}, {Mu{\~n}oz}, {Qin}, {Park}, \& {Watkinson}}]{Murray20}
{Murray}, S., {Greig}, B., {Mesinger}, A., {et~al.} 2020, \bibinfo{title}{{21cmFAST v3: A Python-integrated C code for generating 3D realizations of the cosmic 21cm signal.},} The Journal of Open Source Software, 5, 2582, \dodoi{10.21105/joss.02582}

\bibitem[{R.~P. {Naidu} {et~al.}(2022){Naidu}, {Matthee}, {Oesch}, {Conroy}, {Sobral}, {Pezzulli}, {Hayes}, {Erb}, {Amor{\'\i}n}, {Gronke}, {Schaerer}, {Tacchella}, {Kerutt}, {Paulino-Afonso}, {Calhau}, {Llerena}, \& {R{\"o}ttgering}}]{2022MNRAS.510.4582N}
{Naidu}, R.~P., {Matthee}, J., {Oesch}, P.~A., {et~al.} 2022, \bibinfo{title}{{The synchrony of production and escape: half the bright Ly{\ensuremath{\alpha}} emitters at z {\ensuremath{\approx}} 2 have Lyman continuum escape fractions {\ensuremath{\approx}}50 per cent},} \mnras, 510, 4582, \dodoi{10.1093/mnras/stab3601}

\bibitem[{K. {Nakajima} {et~al.}(2023){Nakajima}, {Ouchi}, {Isobe}, {Harikane}, {Zhang}, {Ono}, {Umeda}, \& {Oguri}}]{Nk23}
{Nakajima}, K., {Ouchi}, M., {Isobe}, Y., {et~al.} 2023, \bibinfo{title}{{JWST Census for the Mass-Metallicity Star-Formation Relations at z=4-10 with Self-Consistent Flux Calibration and the Proper Metallicity Calibrators},} arXiv e-prints, arXiv:2301.12825, \dodoi{10.48550/arXiv.2301.12825}

\bibitem[{M. {Nakane} {et~al.}(2024){Nakane}, {Ouchi}, {Nakajima}, {Harikane}, {Ono}, {Umeda}, {Isobe}, {Zhang}, \& {Xu}}]{Nakane24}
{Nakane}, M., {Ouchi}, M., {Nakajima}, K., {et~al.} 2024, \bibinfo{title}{{Ly{\ensuremath{\alpha}} Emission at z = 7{\textendash}13: Clear Evolution of Ly{\ensuremath{\alpha}} Equivalent Width Indicating a Late Cosmic Reionization History},} \apj, 967, 28, \dodoi{10.3847/1538-4357/ad38c2}

\bibitem[{M. {Nakane} {et~al.}(2025){Nakane}, {Ouchi}, {Nakajima}, {Ono}, {Harikane}, {Isobe}, {Nomoto}, {Ishigaki}, {Yanagisawa}, {Kashino}, {Tominaga}, {Takahashi}, {Nishigaki}, {Takeda}, \& {Watanabe}}]{Nakane25}
{Nakane}, M., {Ouchi}, M., {Nakajima}, K., {et~al.} 2025, \bibinfo{title}{{Fe Abundances of Early Galaxies at $z=9-12$ Derived with Deep JWST Spectra},} arXiv e-prints, arXiv:2503.11457, \dodoi{10.48550/arXiv.2503.11457}

\bibitem[{L. {Napolitano} {et~al.}(2024){Napolitano}, {Pentericci}, {Santini}, {Calabr{\`o}}, {Mascia}, {Llerena}, {Castellano}, {Dickinson}, {Finkelstein}, {Amor{\'\i}n}, {Arrabal Haro}, {Bagley}, {Bhatawdekar}, {Cleri}, {Davis}, {Gardner}, {Gawiser}, {Giavalisco}, {Hathi}, {Holwerda}, {Hu}, {Jung}, {Kartaltepe}, {Koekemoer}, {Larson}, {Merlin}, {Mobasher}, {Papovich}, {Park}, {Pirzkal}, {Trump}, {Wilkins}, \& {Yung}}]{Napolitano24}
{Napolitano}, L., {Pentericci}, L., {Santini}, P., {et~al.} 2024, \bibinfo{title}{{Peering into cosmic reionization: Ly{\ensuremath{\alpha}} visibility evolution from galaxies at z = 4.5-8.5 with JWST},} \aap, 688, A106, \dodoi{10.1051/0004-6361/202449644}

\bibitem[{L. {Napolitano} {et~al.}(2025){Napolitano}, {Castellano}, {Pentericci}, {Arrabal Haro}, {Fontana}, {Treu}, {Bergamini}, {Calabr{\`o}}, {Mascia}, {Morishita}, {Roberts-Borsani}, {Santini}, {Vanzella}, {Vulcani}, {Zakharova}, {Bakx}, {Dickinson}, {Grillo}, {Leethochawalit}, {Llerena}, {Merlin}, {Paris}, {Rojas-Ruiz}, {Rosati}, {Wang}, {Yoon}, \& {Zavala}}]{Napolitano25}
{Napolitano}, L., {Castellano}, M., {Pentericci}, L., {et~al.} 2025, \bibinfo{title}{{Seven wonders of Cosmic Dawn: JWST confirms a high abundance of galaxies and AGN at z ≃ 9{\textendash}11 in the GLASS field},} \aap, 693, A50, \dodoi{10.1051/0004-6361/202452090}

\bibitem[{Y. {Ning} {et~al.}(2022){Ning}, {Jiang}, {Zheng}, \& {Wu}}]{Ning22}
{Ning}, Y., {Jiang}, L., {Zheng}, Z.-Y., \& {Wu}, J. 2022, \bibinfo{title}{{The Magellan M2FS Spectroscopic Survey of High-z Galaxies: Ly{\ensuremath{\alpha}} Emitters at z {\ensuremath{\approx}} 6.6 and the Evolution of Ly{\ensuremath{\alpha}} Luminosity Function over z {\ensuremath{\approx}} 5.7-6.6},} \apj, 926, 230, \dodoi{10.3847/1538-4357/ac4268}

\bibitem[{J.~B. {Oke} \& J.~E. {Gunn}(1983){Oke} \& {Gunn}}]{1983ApJ...266..713O}
{Oke}, J.~B., \& {Gunn}, J.~E. 1983, \bibinfo{title}{{Secondary standard stars for absolute spectrophotometry.},} \apj, 266, 713, \dodoi{10.1086/160817}

\bibitem[{M. {Ouchi} {et~al.}(2010){Ouchi}, {Shimasaku}, {Furusawa}, {Saito}, {Yoshida}, {Akiyama}, {Ono}, {Yamada}, {Ota}, {Kashikawa}, {Iye}, {Kodama}, {Okamura}, {Simpson}, \& {Yoshida}}]{Ouchi10}
{Ouchi}, M., {Shimasaku}, K., {Furusawa}, H., {et~al.} 2010, \bibinfo{title}{{Statistics of 207 Ly{\ensuremath{\alpha}} Emitters at a Redshift Near 7: Constraints on Reionization and Galaxy Formation Models},} \apj, 723, 869, \dodoi{10.1088/0004-637X/723/1/869}

\bibitem[{M. {Ouchi} {et~al.}(2018){Ouchi}, {Harikane}, {Shibuya}, {Shimasaku}, {Taniguchi}, {Konno}, {Kobayashi}, {Kajisawa}, {Nagao}, {Ono}, {Inoue}, {Umemura}, {Mori}, {Hasegawa}, {Higuchi}, {Komiyama}, {Matsuda}, {Nakajima}, {Saito}, \& {Wang}}]{Ouchi18}
{Ouchi}, M., {Harikane}, Y., {Shibuya}, T., {et~al.} 2018, \bibinfo{title}{{Systematic Identification of LAEs for Visible Exploration and Reionization Research Using Subaru HSC (SILVERRUSH). I. Program strategy and clustering properties of {\ensuremath{\sim}}2000 Ly{\ensuremath{\alpha}} emitters at z = 6-7 over the 0.3-0.5 Gpc$^{2}$ survey area},} \pasj, 70, S13, \dodoi{10.1093/pasj/psx074}

\bibitem[{H. {Park} {et~al.}(2024){Park}, {Jung}, {Yajima}, {Sorce}, {Shapiro}, {Ahn}, {Ocvirk}, {Teyssier}, {Yepes}, {Iliev}, \& {Lewis}}]{2024arXiv241007377P}
{Park}, H., {Jung}, I., {Yajima}, H., {et~al.} 2024, \bibinfo{title}{{Constraining Reionization with Ly{\ensuremath{\alpha}} Damping-Wing Absorption in Galaxy Spectra: A Machine Learning Model Based on Reionization Simulations},} arXiv e-prints, arXiv:2410.07377, \dodoi{10.48550/arXiv.2410.07377}

\bibitem[{ {Planck Collaboration} {et~al.}(2020{\natexlab{a}}){Planck Collaboration}, {Aghanim}, {Akrami}, {Ashdown}, {Aumont}, {Baccigalupi}, {Ballardini}, {Banday}, {Barreiro}, {Bartolo}, {Basak}, {Battye}, {Benabed}, {Bernard}, {Bersanelli}, {Bielewicz}, {Bock}, {Bond}, {Borrill}, {Bouchet}, {Boulanger}, {Bucher}, {Burigana}, {Butler}, {Calabrese}, {Cardoso}, {Carron}, {Challinor}, {Chiang}, {Chluba}, {Colombo}, {Combet}, {Contreras}, {Crill}, {Cuttaia}, {de Bernardis}, {de Zotti}, {Delabrouille}, {Delouis}, {Di Valentino}, {Diego}, {Dor{\'e}}, {Douspis}, {Ducout}, {Dupac}, {Dusini}, {Efstathiou}, {Elsner}, {En{\ss}lin}, {Eriksen}, {Fantaye}, {Farhang}, {Fergusson}, {Fernandez-Cobos}, {Finelli}, {Forastieri}, {Frailis}, {Fraisse}, {Franceschi}, {Frolov}, {Galeotta}, {Galli}, {Ganga}, {G{\'e}nova-Santos}, {Gerbino}, {Ghosh}, {Gonz{\'a}lez-Nuevo}, {G{\'o}rski}, {Gratton}, {Gruppuso}, {Gudmundsson}, {Hamann}, {Handley}, {Hansen}, {Herranz}, {Hildebrandt}, {Hivon}, {Huang}, {Jaffe}, {Jones}, {Karakci},
  {Keih{\"a}nen}, {Keskitalo}, {Kiiveri}, {Kim}, {Kisner}, {Knox}, {Krachmalnicoff}, {Kunz}, {Kurki-Suonio}, {Lagache}, {Lamarre}, {Lasenby}, {Lattanzi}, {Lawrence}, {Le Jeune}, {Lemos}, {Lesgourgues}, {Levrier}, {Lewis}, {Liguori}, {Lilje}, {Lilley}, {Lindholm}, {L{\'o}pez-Caniego}, {Lubin}, {Ma}, {Mac{\'\i}as-P{\'e}rez}, {Maggio}, {Maino}, {Mandolesi}, {Mangilli}, {Marcos-Caballero}, {Maris}, {Martin}, {Martinelli}, {Mart{\'\i}nez-Gonz{\'a}lez}, {Matarrese}, {Mauri}, {McEwen}, {Meinhold}, {Melchiorri}, {Mennella}, {Migliaccio}, {Millea}, {Mitra}, {Miville-Desch{\^e}nes}, {Molinari}, {Montier}, {Morgante}, {Moss}, {Natoli}, {N{\o}rgaard-Nielsen}, {Pagano}, {Paoletti}, {Partridge}, {Patanchon}, {Peiris}, {Perrotta}, {Pettorino}, {Piacentini}, {Polastri}, {Polenta}, {Puget}, {Rachen}, {Reinecke}, {Remazeilles}, {Renzi}, {Rocha}, {Rosset}, {Roudier}, {Rubi{\~n}o-Mart{\'\i}n}, {Ruiz-Granados}, {Salvati}, {Sandri}, {Savelainen}, {Scott}, {Shellard}, {Sirignano}, {Sirri}, {Spencer}, {Sunyaev}, {Suur-Uski},
  {Tauber}, {Tavagnacco}, {Tenti}, {Toffolatti}, {Tomasi}, {Trombetti}, {Valenziano}, {Valiviita}, {Van Tent}, {Vibert}, {Vielva}, {Villa}, {Vittorio}, {Wandelt}, {Wehus}, {White}, {White}, {Zacchei}, \& {Zonca}}]{Planck}
{Planck Collaboration}, {Aghanim}, N., {Akrami}, Y., {et~al.} 2020{\natexlab{a}}, \bibinfo{title}{{Planck 2018 results. VI. Cosmological parameters},} \aap, 641, A6, \dodoi{10.1051/0004-6361/201833910}

\bibitem[{ {Planck Collaboration} {et~al.}(2020{\natexlab{b}}){Planck Collaboration}, {Aghanim}, {Akrami}, {Ashdown}, {Aumont}, {Baccigalupi}, {Ballardini}, {Banday}, {Barreiro}, {Bartolo}, {Basak}, {Battye}, {Benabed}, {Bernard}, {Bersanelli}, {Bielewicz}, {Bock}, {Bond}, {Borrill}, {Bouchet}, {Boulanger}, {Bucher}, {Burigana}, {Butler}, {Calabrese}, {Cardoso}, {Carron}, {Challinor}, {Chiang}, {Chluba}, {Colombo}, {Combet}, {Contreras}, {Crill}, {Cuttaia}, {de Bernardis}, {de Zotti}, {Delabrouille}, {Delouis}, {Di Valentino}, {Diego}, {Dor{\'e}}, {Douspis}, {Ducout}, {Dupac}, {Dusini}, {Efstathiou}, {Elsner}, {En{\ss}lin}, {Eriksen}, {Fantaye}, {Farhang}, {Fergusson}, {Fernandez-Cobos}, {Finelli}, {Forastieri}, {Frailis}, {Fraisse}, {Franceschi}, {Frolov}, {Galeotta}, {Galli}, {Ganga}, {G{\'e}nova-Santos}, {Gerbino}, {Ghosh}, {Gonz{\'a}lez-Nuevo}, {G{\'o}rski}, {Gratton}, {Gruppuso}, {Gudmundsson}, {Hamann}, {Handley}, {Hansen}, {Herranz}, {Hildebrandt}, {Hivon}, {Huang}, {Jaffe}, {Jones}, {Karakci},
  {Keih{\"a}nen}, {Keskitalo}, {Kiiveri}, {Kim}, {Kisner}, {Knox}, {Krachmalnicoff}, {Kunz}, {Kurki-Suonio}, {Lagache}, {Lamarre}, {Lasenby}, {Lattanzi}, {Lawrence}, {Le Jeune}, {Lemos}, {Lesgourgues}, {Levrier}, {Lewis}, {Liguori}, {Lilje}, {Lilley}, {Lindholm}, {L{\'o}pez-Caniego}, {Lubin}, {Ma}, {Mac{\'\i}as-P{\'e}rez}, {Maggio}, {Maino}, {Mandolesi}, {Mangilli}, {Marcos-Caballero}, {Maris}, {Martin}, {Martinelli}, {Mart{\'\i}nez-Gonz{\'a}lez}, {Matarrese}, {Mauri}, {McEwen}, {Meinhold}, {Melchiorri}, {Mennella}, {Migliaccio}, {Millea}, {Mitra}, {Miville-Desch{\^e}nes}, {Molinari}, {Montier}, {Morgante}, {Moss}, {Natoli}, {N{\o}rgaard-Nielsen}, {Pagano}, {Paoletti}, {Partridge}, {Patanchon}, {Peiris}, {Perrotta}, {Pettorino}, {Piacentini}, {Polastri}, {Polenta}, {Puget}, {Rachen}, {Reinecke}, {Remazeilles}, {Renzi}, {Rocha}, {Rosset}, {Roudier}, {Rubi{\~n}o-Mart{\'\i}n}, {Ruiz-Granados}, {Salvati}, {Sandri}, {Savelainen}, {Scott}, {Shellard}, {Sirignano}, {Sirri}, {Spencer}, {Sunyaev}, {Suur-Uski},
  {Tauber}, {Tavagnacco}, {Tenti}, {Toffolatti}, {Tomasi}, {Trombetti}, {Valenziano}, {Valiviita}, {Van Tent}, {Vibert}, {Vielva}, {Villa}, {Vittorio}, {Wandelt}, {Wehus}, {White}, {White}, {Zacchei}, \& {Zonca}}]{Planck20}
{Planck Collaboration}, {Aghanim}, N., {Akrami}, Y., {et~al.} 2020{\natexlab{b}}, \bibinfo{title}{{Planck 2018 results. VI. Cosmological parameters},} \aap, 641, A6, \dodoi{10.1051/0004-6361/201833910}

\bibitem[{S.~H. {Price} {et~al.}(2024){Price}, {Bezanson}, {Labbe}, {Furtak}, {de Graaff}, {Greene}, {Kokorev}, {Setton}, {Suess}, {Brammer}, {Cutler}, {Leja}, {Pan}, {Wang}, {Weaver}, {Whitaker}, {Atek}, {Burgasser}, {Chemerynska}, {Dayal}, {Feldmann}, {F{\"o}rster Schreiber}, {Fudamoto}, {Fujimoto}, {Glazebrook}, {Goulding}, {Khullar}, {Kriek}, {Marchesini}, {Maseda}, {Miller}, {Muzzin}, {Nanayakkara}, {Nelson}, {Oesch}, {Shipley}, {Smit}, {Taylor}, {van Dokkum}, {Williams}, \& {Zitrin}}]{Price24}
{Price}, S.~H., {Bezanson}, R., {Labbe}, I., {et~al.} 2024, \bibinfo{title}{{The UNCOVER Survey: First Release of Ultradeep JWST/NIRSpec PRISM spectra for \raisebox{-0.5ex}\textasciitilde700 galaxies from z\raisebox{-0.5ex}\textasciitilde0.3-13 in Abell 2744},} arXiv e-prints, arXiv:2408.03920, \dodoi{10.48550/arXiv.2408.03920}

\bibitem[{N.~A. {Reddy} {et~al.}(2016){Reddy}, {Steidel}, {Pettini}, {Bogosavljevi{\'c}}, \& {Shapley}}]{Reddy2016}
{Reddy}, N.~A., {Steidel}, C.~C., {Pettini}, M., {Bogosavljevi{\'c}}, M., \& {Shapley}, A.~E. 2016, \bibinfo{title}{{The Connection Between Reddening, Gas Covering Fraction, and the Escape of Ionizing Radiation at High Redshift},} \apj, 828, 108, \dodoi{10.3847/0004-637X/828/2/108}

\bibitem[{A. {Runnholm} {et~al.}(2025){Runnholm}, {Hayes}, {Mehta}, {Malkan}, {Scarlata}, {Nedkova}, {Rafelski}, {Vulcani}, {Huberty}, {Herenz}, {Hutter}, {Bruton}, {Acharyya}, {Atek}, {Baronchelli}, {Battisti}, {Brada{\v{c}}}, {Bunker}, {Dai}, {Hannahs}, {Hasan}, {Kim}, {Leethochawalit}, {Lin}, {Rutkowski}, {Saldana-Lopez}, {Sattari}, \& {Wang}}]{2025arXiv250219174R}
{Runnholm}, A., {Hayes}, M.~J., {Mehta}, V., {et~al.} 2025, \bibinfo{title}{{The JWST/PASSAGE Survey: Testing Reionization Histories with JWST's First Unbiased Survey for Lyman alpha Emitters at Redshifts 7.5-9.5},} arXiv e-prints, arXiv:2502.19174, \dodoi{10.48550/arXiv.2502.19174}

\bibitem[{A. {Saxena} {et~al.}(2024){Saxena}, {Cameron}, {Katz}, {Bunker}, {Chevallard}, {D'Eugenio}, {Arribas}, {Bhatawdekar}, {Boyett}, {Cargile}, {Carniani}, {Charlot}, {Curti}, {Curtis-Lake}, {Hainline}, {Ji}, {Johnson}, {Jones}, {Kumari}, {Laseter}, {Maseda}, {Robertson}, {Simmonds}, {Tacchella}, {Ubler}, {Williams}, {Willott}, {Witstok}, \& {Zhu}}]{Saxena24}
{Saxena}, A., {Cameron}, A.~J., {Katz}, H., {et~al.} 2024, \bibinfo{title}{{Hitting the slopes: A spectroscopic view of UV continuum slopes of galaxies reveals a reddening at z > 9.5},} arXiv e-prints, arXiv:2411.14532, \dodoi{10.48550/arXiv.2411.14532}

\bibitem[{J. {Scholtz} {et~al.}(2023){Scholtz}, {Maiolino}, {D'Eugenio}, {Curtis-Lake}, {Carniani}, {Charlot}, {Curti}, {Silcock}, {Arribas}, {Baker}, {Bhatawdekar}, {Boyett}, {Bunker}, {Chevallard}, {Circosta}, {Eisenstein}, {Hainline}, {Hausen}, {Ji}, {Ji}, {Johnson}, {Kumari}, {Looser}, {Lyu}, {Maseda}, {Parlanti}, {Perna}, {Rieke}, {Robertson}, {Rodr{\'\i}guez Del Pino}, {Sun}, {Tacchella}, {{\"U}bler}, {Venturi}, {Williams}, {Willmer}, {Willott}, \& {Witstok}}]{Scholtz23}
{Scholtz}, J., {Maiolino}, R., {D'Eugenio}, F., {et~al.} 2023, \bibinfo{title}{{JADES: A large population of obscured, narrow line AGN at high redshift},} arXiv e-prints, arXiv:2311.18731, \dodoi{10.48550/arXiv.2311.18731}

\bibitem[{J. {Schroeder} {et~al.}(2013){Schroeder}, {Mesinger}, \& {Haiman}}]{2013MNRAS.428.3058S}
{Schroeder}, J., {Mesinger}, A., \& {Haiman}, Z. 2013, \bibinfo{title}{{Evidence of Gunn-Peterson damping wings in high-z quasar spectra: strengthening the case for incomplete reionization at z {\ensuremath{\sim}} 6-7},} \mnras, 428, 3058, \dodoi{10.1093/mnras/sts253}

\bibitem[{A. {Smith} {et~al.}(2022){Smith}, {Kannan}, {Garaldi}, {Vogelsberger}, {Pakmor}, {Springel}, \& {Hernquist}}]{Smith22}
{Smith}, A., {Kannan}, R., {Garaldi}, E., {et~al.} 2022, \bibinfo{title}{{The THESAN project: Lyman-{\ensuremath{\alpha}} emission and transmission during the Epoch of Reionization},} \mnras, 512, 3243, \dodoi{10.1093/mnras/stac713}

\bibitem[{E. {Sobacchi} \& A. {Mesinger}(2015){Sobacchi} \& {Mesinger}}]{Sobacchi15}
{Sobacchi}, E., \& {Mesinger}, A. 2015, \bibinfo{title}{{The clustering of Lyman {\ensuremath{\alpha}} emitters at z {\ensuremath{\approx}} 7: implications for reionization and host halo masses},} \mnras, 453, 1843, \dodoi{10.1093/mnras/stv1751}

\bibitem[{B. {Spina} {et~al.}(2024){Spina}, {Bosman}, {Davies}, {Gaikwad}, \& {Zhu}}]{Spina24}
{Spina}, B., {Bosman}, S. E.~I., {Davies}, F.~B., {Gaikwad}, P., \& {Zhu}, Y. 2024, \bibinfo{title}{{Damping wings in the Lyman-{\ensuremath{\alpha}} forest: A model-independent measurement of the neutral fraction at 5.4 < z < 6.1},} \aap, 688, L26, \dodoi{10.1051/0004-6361/202450798}

\bibitem[{M. {Tang} {et~al.}(2024){Tang}, {Stark}, {Topping}, {Mason}, \& {Ellis}}]{Tang24}
{Tang}, M., {Stark}, D.~P., {Topping}, M.~W., {Mason}, C., \& {Ellis}, R.~S. 2024, \bibinfo{title}{{JWST/NIRSpec Observations of Ly$\alpha$ Emission in Star Forming Galaxies at $6.5\lesssim z\lesssim13$},} arXiv e-prints, arXiv:2408.01507, \dodoi{10.48550/arXiv.2408.01507}

\bibitem[{T. {Totani} {et~al.}(2006){Totani}, {Kawai}, {Kosugi}, {Aoki}, {Yamada}, {Iye}, {Ohta}, \& {Hattori}}]{2006PASJ...58..485T}
{Totani}, T., {Kawai}, N., {Kosugi}, G., {et~al.} 2006, \bibinfo{title}{{Implications for Cosmic Reionization from the Optical Afterglow Spectrum of the Gamma-Ray Burst 050904 at z = 6.3$^{*}$},} \pasj, 58, 485, \dodoi{10.1093/pasj/58.3.485}

\bibitem[{T. {Totani} {et~al.}(2014){Totani}, {Aoki}, {Hattori}, {Kosugi}, {Niino}, {Hashimoto}, {Kawai}, {Ohta}, {Sakamoto}, \& {Yamada}}]{2014PASJ...66...63T}
{Totani}, T., {Aoki}, K., {Hattori}, T., {et~al.} 2014, \bibinfo{title}{{Probing intergalactic neutral hydrogen by the Lyman alpha red damping wing of gamma-ray burst 130606A afterglow spectrum at z = 5.913},} \pasj, 66, 63, \dodoi{10.1093/pasj/psu032}

\bibitem[{H. {Umeda} {et~al.}(2024){Umeda}, {Ouchi}, {Nakajima}, {Harikane}, {Ono}, {Xu}, {Isobe}, \& {Zhang}}]{Umeda24}
{Umeda}, H., {Ouchi}, M., {Nakajima}, K., {et~al.} 2024, \bibinfo{title}{{JWST Measurements of Neutral Hydrogen Fractions and Ionized Bubble Sizes at z = 7{\textendash}12 Obtained with Ly{\ensuremath{\alpha}} Damping Wing Absorptions in 27 Bright Continuum Galaxies},} \apj, 971, 124, \dodoi{10.3847/1538-4357/ad554e}

\bibitem[{H. {Umeda} {et~al.}(2025){Umeda}, {Ouchi}, {Kikuta}, {Harikane}, {Ono}, {Shibuya}, {Inoue}, {Shimasaku}, {Liang}, {Matsumoto}, {Saito}, {Kusakabe}, {Kageura}, \& {Nakane}}]{Umeda25}
{Umeda}, H., {Ouchi}, M., {Kikuta}, S., {et~al.} 2025, \bibinfo{title}{{SILVERRUSH. XIV. Ly{\ensuremath{\alpha}} Luminosity Functions and Angular Correlation Functions from 20,000 Ly{\ensuremath{\alpha}} Emitters at z {\ensuremath{\sim}} 2.2{\textendash}7.3 from up to 24 deg$^{2}$ HSC-SSP and CHORUS Surveys: Linking the Postreionization Epoch to the Heart of Reionization},} \apjs, 277, 37, \dodoi{10.3847/1538-4365/adb1c0}

\bibitem[{D. {{\v{D}}urov{\v{c}}{\'\i}kov{\'a}} {et~al.}(2024){{\v{D}}urov{\v{c}}{\'\i}kov{\'a}}, {Eilers}, {Chen}, {Satyavolu}, {Kulkarni}, {Simcoe}, {Keating}, {Haehnelt}, \& {Ba{\~n}ados}}]{Durovcikova23}
{{\v{D}}urov{\v{c}}{\'\i}kov{\'a}}, D., {Eilers}, A.-C., {Chen}, H., {et~al.} 2024, \bibinfo{title}{{Chronicling the Reionization History at 6 {\ensuremath{\lesssim}} z {\ensuremath{\lesssim}} 7 with Emergent Quasar Damping Wings},} \apj, 969, 162, \dodoi{10.3847/1538-4357/ad4888}

\bibitem[{F. {Wang} {et~al.}(2020){Wang}, {Davies}, {Yang}, {Hennawi}, {Fan}, {Barth}, {Jiang}, {Wu}, {Mudd}, {Ba{\~n}ados}, {Bian}, {Decarli}, {Eilers}, {Farina}, {Venemans}, {Walter}, \& {Yue}}]{2020ApJ...896...23W}
{Wang}, F., {Davies}, F.~B., {Yang}, J., {et~al.} 2020, \bibinfo{title}{{A Significantly Neutral Intergalactic Medium Around the Luminous z = 7 Quasar J0252-0503},} \apj, 896, 23, \dodoi{10.3847/1538-4357/ab8c45}

\bibitem[{L.~H. {Weinberger} {et~al.}(2018){Weinberger}, {Kulkarni}, {Haehnelt}, {Choudhury}, \& {Puchwein}}]{Weinberger18}
{Weinberger}, L.~H., {Kulkarni}, G., {Haehnelt}, M.~G., {Choudhury}, T.~R., \& {Puchwein}, E. 2018, \bibinfo{title}{{Lyman-{\ensuremath{\alpha}} emitters gone missing: the different evolution of the bright and faint populations},} \mnras, 479, 2564, \dodoi{10.1093/mnras/sty1563}

\bibitem[{L.~R. {Whitler} {et~al.}(2020){Whitler}, {Mason}, {Ren}, {Dijkstra}, {Mesinger}, {Pentericci}, {Trenti}, \& {Treu}}]{2020MNRAS.495.3602W}
{Whitler}, L.~R., {Mason}, C.~A., {Ren}, K., {et~al.} 2020, \bibinfo{title}{{The impact of scatter in the galaxy UV luminosity to halo mass relation on Ly {\ensuremath{\alpha}} visibility during the epoch of reionization},} \mnras, 495, 3602, \dodoi{10.1093/mnras/staa1178}

\bibitem[{H. {Williams} {et~al.}(2023){Williams}, {Kelly}, {Chen}, {Brammer}, {Zitrin}, {Treu}, {Scarlata}, {Koekemoer}, {Oguri}, {Lin}, {Diego}, {Nonino}, {Hjorth}, {Langeroodi}, {Broadhurst}, {Rogers}, {Perez-Fournon}, {Foley}, {Jha}, {Filippenko}, {Strolger}, {Pierel}, {Poidevin}, \& {Yang}}]{Williams23}
{Williams}, H., {Kelly}, P.~L., {Chen}, W., {et~al.} 2023, \bibinfo{title}{{A magnified compact galaxy at redshift 9.51 with strong nebular emission lines},} Science, 380, 416, \dodoi{10.1126/science.adf5307}

\bibitem[{J. {Witstok} {et~al.}(2025){Witstok}, {Maiolino}, {Smit}, {Jones}, {Bunker}, {Helton}, {Johnson}, {Tacchella}, {Saxena}, {Arribas}, {Bhatawdekar}, {Boyett}, {Cameron}, {Cargile}, {Carniani}, {Charlot}, {Chevallard}, {Curti}, {Curtis-Lake}, {D'Eugenio}, {Eisenstein}, {Hainline}, {Hausen}, {Kumari}, {Laseter}, {Maseda}, {Rieke}, {Robertson}, {Scholtz}, {Shivaei}, {Williams}, {Willmer}, \& {Willott}}]{Witstok25}
{Witstok}, J., {Maiolino}, R., {Smit}, R., {et~al.} 2025, \bibinfo{title}{{JADES: primaeval Lyman {\ensuremath{\alpha}} emitting galaxies reveal early sites of reionization out to redshift z \raisebox{-0.5ex}\textasciitilde 9},} \mnras, 536, 27, \dodoi{10.1093/mnras/stae2535}

\bibitem[{I.~G.~B. {Wold} {et~al.}(2022){Wold}, {Malhotra}, {Rhoads}, {Wang}, {Hu}, {Perez}, {Zheng}, {Khostovan}, {Walker}, {Barrientos}, {Gonz{\'a}lez-L{\'o}pez}, {Harish}, {Infante}, {Jiang}, {Pharo}, {Moya-Sierralta}, {Bauer}, {Galaz}, {Valdes}, \& {Yang}}]{Wold22}
{Wold}, I. G.~B., {Malhotra}, S., {Rhoads}, J., {et~al.} 2022, \bibinfo{title}{{LAGER Ly{\ensuremath{\alpha}} Luminosity Function at z 7: Implications for Reionization},} \apj, 927, 36, \dodoi{10.3847/1538-4357/ac4997}

\bibitem[{H. {Yanagisawa} {et~al.}(2024){Yanagisawa}, {Ouchi}, {Nakajima}, {Harikane}, {Fujimoto}, {Ono}, {Umeda}, {Nakane}, {Yajima}, {Fukushima}, \& {Xu}}]{Yanagisawa24}
{Yanagisawa}, H., {Ouchi}, M., {Nakajima}, K., {et~al.} 2024, \bibinfo{title}{{A Galaxy with an Extremely Blue UV Slope $\beta=-3$ at $z=9.25$ Identified by JWST Spectroscopy: Evidence for a Weak Nebular Continuum and Efficient Ionizing Photon Escape?},} arXiv e-prints, arXiv:2411.19893, \dodoi{10.48550/arXiv.2411.19893}

\bibitem[{E. {Zackrisson} {et~al.}(2017){Zackrisson}, {Binggeli}, {Finlator}, {Gnedin}, {Paardekooper}, {Shimizu}, {Inoue}, {Jensen}, {Micheva}, {Khochfar}, \& {Dalla Vecchia}}]{Zackrisson17}
{Zackrisson}, E., {Binggeli}, C., {Finlator}, K., {et~al.} 2017, \bibinfo{title}{{The Spectral Evolution of the First Galaxies. III. Simulated James Webb Space Telescope Spectra of Reionization-epoch Galaxies with Lyman-continuum Leakage},} \apj, 836, 78, \dodoi{10.3847/1538-4357/836/1/78}

\bibitem[{Z.-Y. {Zheng} {et~al.}(2017){Zheng}, {Wang}, {Rhoads}, {Infante}, {Malhotra}, {Hu}, {Walker}, {Jiang}, {Jiang}, {Hibon}, {Gonzalez}, {Kong}, {Zheng}, {Galaz}, \& {Barrientos}}]{Zheng17}
{Zheng}, Z.-Y., {Wang}, J., {Rhoads}, J., {et~al.} 2017, \bibinfo{title}{{First Results from the Lyman Alpha Galaxies in the Epoch of Reionization (LAGER) Survey: Cosmological Reionization at z {\ensuremath{\sim}} 7},} \apjl, 842, L22, \dodoi{10.3847/2041-8213/aa794f}

\bibitem[{Y. {Zhu} {et~al.}(2022){Zhu}, {Becker}, {Bosman}, {Keating}, {D'Odorico}, {Davies}, {Christenson}, {Ba{\~n}ados}, {Bian}, {Bischetti}, {Chen}, {Davies}, {Eilers}, {Fan}, {Gaikwad}, {Greig}, {Haehnelt}, {Kulkarni}, {Lai}, {Pallottini}, {Qin}, {Ryan-Weber}, {Walter}, {Wang}, \& {Yang}}]{2022ApJ...932...76Z}
{Zhu}, Y., {Becker}, G.~D., {Bosman}, S. E.~I., {et~al.} 2022, \bibinfo{title}{{Long Dark Gaps in the Ly{\ensuremath{\beta}} Forest at z < 6: Evidence of Ultra-late Reionization from XQR-30 Spectra},} \apj, 932, 76, \dodoi{10.3847/1538-4357/ac6e60}

\bibitem[{Y. {Zhu} {et~al.}(2024){Zhu}, {Becker}, {Bosman}, {Cain}, {Keating}, {Nasir}, {D'Odorico}, {Ba{\~n}ados}, {Bian}, {Bischetti}, {Bolton}, {Chen}, {D'Aloisio}, {Davies}, {Davies}, {Eilers}, {Fan}, {Gaikwad}, {Greig}, {Haehnelt}, {Kulkarni}, {Lai}, {Puchwein}, {Qin}, {Ryan-Weber}, {Satyavolu}, {Spina}, {Walter}, {Wang}, {Wolfson}, \& {Yang}}]{Zhu24}
{Zhu}, Y., {Becker}, G.~D., {Bosman}, S. E.~I., {et~al.} 2024, \bibinfo{title}{{Damping wing-like features in the stacked Ly {\ensuremath{\alpha}} forest: Potential neutral hydrogen islands at z < 6},} \mnras, 533, L49, \dodoi{10.1093/mnrasl/slae061}

\end{thebibliography}
\bibliographystyle{aasjournalv7}



\end{document}